\documentclass{tlp}

\usepackage[utf8]{inputenc} 
\usepackage[T1]{fontenc}

\usepackage{amssymb}
\usepackage{amsmath}
\usepackage{textcomp}
\usepackage{calc}
\usepackage[justification=centering]{caption}
\usepackage[hidelinks,pdftex]{hyperref}
\usepackage{relsize}

\usepackage{xr} 
\usepackage{mathptmx}

\usepackage{dsfont}

\usepackage{calrsfs}

\DeclareMathAlphabet{\pazocal}{OMS}{zplm}{m}{n}
\newcommand{\domain}{\ensuremath{\pazocal{D}}\xspace}
\newcommand{\mf}{\ensuremath{\pazocal{F}}\xspace}

\usepackage{xspace}  
\newcommand{\R}{$\mathds{R}$\xspace}
\newcommand{\Q}{$\mathds{Q}$\xspace}
\newcommand{\h}{$\mathds{H}$\xspace}
\newcommand{\D}{$\mathds{D}_{\leq}$\xspace}
\newcommand{\Lat}{$\mathds{L}at$\xspace}
\newcommand{\FD}{$\mathds{FD}$\xspace}

\newcommand{\lfp}{\ensuremath{\mathit{lfp}}\xspace}

\usepackage{enumitem}

\setlist[enumerate,2]{ref=\arabic{enumi}.\alph*}

\usepackage{multirow}
\usepackage{longtable}

\usepackage{adjustbox}
\usepackage{rotating}

\usepackage{relsize}

\usepackage{url}

\usepackage{graphicx}

\usepackage{xcolor}
\definecolor{PrologPredicate}{RGB}{0,0,200}
\definecolor{PrologVar}      {RGB}{145,032,039}
\definecolor{PrologComment}  {RGB}{169,082,044}
\definecolor{PrologOther}    {rgb}{0.2,0.2,0.2}
\definecolor{PrologString}   {rgb}{0.2,0.2,0.2}

\usepackage{listings} 
\lstdefinestyle{MyInline}
{
  basicstyle = \ttfamily\color{PrologOther},
  breaklines = true,
  breakatwhitespace=true,
  literate =
  {,}{}{0\discretionary{,}{}{,}}
  {|}{{\color{PrologOther}$\mid$}}1
  {\\\{}{{\color{PrologOther}\{}}1
  {\\\}}{{\color{PrologOther}\}}}1
  {[}{{\color{PrologOther}\small[}}1
  {]}{{\color{PrologOther}\small]}}1
  {\\$}{{\color{PrologOther}\$}}1
  {.=.}{{\color{PrologOther}\#=}}3
  {.<.}{{\color{PrologOther}\#<}}3
  {.>.}{{\color{PrologOther}\#>}}3
  {.=<.}{{\color{PrologOther}\#=<}}4
  {.>=.}{{\color{PrologOther}\#>=}}4
  {.\\=.}{{\color{PrologOther}{\char"5C}=}}3
  {\\=}{{\color{PrologOther}\char"5C=}}3
  {?-}{{\color{PrologOther}?-\ }}2
  {:-}{{\color{PrologOther}:-}}3
  {),}{{\color{PrologOther})\hspace{-.2em},}}1
}
\lstdefinestyle{MyProlog}
{
    xleftmargin=0.5cm,
    numberstyle=\tiny,
    numbers=left,
    stepnumber=1,
    belowskip=5pt,
  keywords = {},
  upquote = true,
  basicstyle = \small\ttfamily\color{PrologPredicate},
  basewidth = 0.56em,
  moredelim = {**[s][\color{PrologString}]{'}{'}},
  moredelim = {*[s][\color{PrologVar}]{(}{)}},
  moredelim = {*[s][\color{PrologOther}]{:-}{.}},
  moredelim = {*[s][\color{red}]{/*}{*/}},
  commentstyle = \mdseries\color{PrologComment},
  morecomment=[l]\%,
  morecomment=[s]{/*}{*/},
  literate     =
  {|}{{\color{PrologOther}$\mid$}}1
  {[}{{\color{PrologOther}\small[}}1
  {]}{{\color{PrologOther}\small]}}1
  {\\$}{{\color{PrologOther}\$}}1
  {&(}{{\color{PrologOther}(}}1
  {&)}{{\color{PrologOther})}}1
  {&.}{{.}}0
  {.=.}{{\color{PrologOther}\ \#=\ }}4
  {.<.}{{\color{PrologOther}\ \#<\ }}4
  {.>.}{{\color{PrologOther}\ \#>\ }}4
  {.=<.}{{\color{PrologOther}\ \#=<\ }}5
  {.>=.}{{\color{PrologOther}\ \#>=\ }}5
  {.\\=.}{{\color{PrologOther}\char"5C=}}3
  {\\=}{{\color{PrologOther}\char"5C=}}3
  {,}{{\color{PrologOther}\footnotesize,}}1
  {;}{{\color{PrologOther}\footnotesize;}}1,
}
\lstdefinestyle{MyASP}
{
  keywords = {},
  upquote = true,
  basicstyle = \small\ttfamily\color{PrologPredicate},
  basewidth = 0.52em,
  moredelim = {*[s][\color{PrologOther}]{:-}{.}},
  moredelim = {*[s][\color{PrologOther}]{\{}{\}}},
  moredelim = {[s][\color{PrologVar}]{(}{)}},
  commentstyle = \color{PrologComment},
  morecomment=[l]\%,
  literate     =
  {..}{..}2
  {,}{{\color{PrologOther}\footnotesize,}}1
}
\newcommand{\myvrule}[2]
{
    \begin{minipage}[cl]{#1}
    \begin{flushleft}
      \vrule height #2
    \end{flushleft}
    \end{minipage}%
}

\usepackage{tikz}
\usetikzlibrary{shapes.geometric, arrows, patterns}

\usepackage[edges]{forest}

\usepackage{morefloats}
\usepackage{xcolor}
\usepackage[textwidth=0.3\textwidth,textsize=footnotesize]{todonotes}
\newcommand{\redbefcapt}{\vspace*{0.0em}}
\newcommand{\redbefsec}{\vspace*{0em}}
\newcommand{\redbeffig}{\vspace*{0em}}

\newcommand{\redbeflst}{\vspace*{0em}}

\newcommand{\redaftersec}{\vspace*{0em}}
\newcommand{\redafterfig}{\vspace*{0em}}

\newcommand{\redafterlst}{\vspace*{0em}}

\newcommand{\ok}{\checkmark}    
\newcommand{\no}{\ensuremath{\mathbf{\times}}}

\let\proof\relax

\usepackage{amsthm}
\newtheorem{definition}{Definition}

\newtheorem{theorem}{Theorem}
\newtheorem{example}{Example}

\let\oldc\,
\renewcommand{\,}{\oldc \allowbreak}
\let\oldland\land
\renewcommand{\land}{\oldland \allowbreak}

\newcommand{\goal}[1]{\ensuremath{\langle #1 \rangle}}
\newcommand{\goaltt}[2]{\ensuremath{\langle \mathtt{#1},\, \mathtt{#2} \rangle}}
\newcommand{\mt}[1]{\ensuremath{\mathtt{#1}}}

\newcommand*{\pr}[1]{%
  \ifcase#1%
  \relax
  \or \ensuremath{\mathtt{_1}}\nobreak
  \or \ensuremath{\mathtt{_2}}\nobreak
  \or \ensuremath{\mathtt{_3}}\nobreak
  \or \ensuremath{\mathtt{_4}}\nobreak
  \or \ensuremath{\mathtt{_5}}\nobreak
  \else
  \fi
}

\newcommand{\mtclp}{\mbox{Mod~TCLP}}

\title[TCLP: Description, Implementation, Evaluation]
{Description, Implementation, and Evaluation of a Generic Design for Tabled CLP}

\author[Joaqu\'in Arias and Manuel Carro]{ 
Joaqu\'in Arias$^{1,2}$ \and Manuel Carro$^{1,2}$ \\
\texttt{\emph{joaquin.arias@imdea.org, manuel.carro@\{imdea.org,upm.es\}}} \\
\footnotesize
$^1$IMDEA Software Institute, $^2$Universidad Politécnica de Madrid
}

\volume{\textbf{10} (3):}
\jdate{March 2002}
\pubyear{2002}

\begin{document}

\maketitle

\redbefsec
\begin{abstract}
  \redaftersec
  \label{sec:abstract}
  Logic programming with tabling and constraints (TCLP, \emph{tabled
    constraint logic programming}) has been shown to be more
  expressive and in some cases more efficient than LP, CLP or LP +
  tabling.  Previous designs of TCLP systems did not fully use
  entailment to determine call / answer subsumption and did not
  provide a simple and well-documented interface to facilitate the
  integration of constraint solvers in existing tabling systems.  We
  study the role of projection and entailment in the termination,
  soundness and completeness of TCLP systems, and present the design
  and an experimental evaluation of \mtclp, a framework that eases the
  integration of additional constraint solvers.  \mtclp\ views
  constraint solvers as clients of the tabling system, which is
  generic w.r.t.\ the solver and only requires a clear interface from
  the latter.
  We validate our design by integrating four constraint solvers: a
  previously existing constraint solver for difference constraints,
  written in C; the standard versions of Holzbaur's CLP(\Q) and
  CLP(\R), written in Prolog; and a new constraint solver for
  equations over finite lattices.  We evaluate the performance of our
  framework in several benchmarks using the aforementioned constraint
  solvers.  \mtclp\ is developed in Ciao Prolog, a robust, mature,
  next-generation Prolog system.

  Under consideration in Theory and Practice of Logic Programming (TPLP).
\end{abstract}


\redbefsec
\section{Introduction}
\redaftersec
\label{sec:introduction}

Constraint Logic Programming (CLP)~\cite{survey94} extends
Logic Programming (LP)
with variables which can belong to arbitrary constraint domains and
whose associated constraint solvers can incrementally simplify equations
set up during program executioCLP brings additional expressive power to LP, since constraints can
very concisely capture complex relationships.  Also,
shifting from ``generate-and-test'' to ``constrain-and-generate''
patterns reduces the search tree and therefore brings additional
performance, even if constraint solving is in general more expensive
than unification.

Tabling~\cite{tamaki.iclp86,Warren92} is an execution
strategy for logic programs which suspends repeated calls which could
cause infinite loops.  Answers from non-looping branches are
used to resume suspended calls which can, in turn, generate more answers.
Only new answers are saved, and evaluation finishes when no new
answers can be generated.  Tabled evaluation always terminates for
calls / programs with the bounded term depth property\footnote{That
  is, programs which can only generate terms with a finite bound on
  their depth.} and can improve
efficiency for terminating programs which repeat computations, as it automatically
implements a variant of dynamic programming.
Tabling has been successfully applied in a variety of contexts, including
deductive databases, program analysis, semantic Web reasoning, and
model
checking~\cite{pracabsin,Dawson:pldi96,zou05:owl-tabling,ramakrishna97:model_checking_tabling,Charatonik02:model_checking_tabulation}.

The combination of CLP and
tabling~\cite{toman_theo_const_tabling,tchr2007,bao2000,chico-tclp-flops2012}
brings several advantages:  it enhances termination
properties, increases speed in a range of programs, and provides
additional expressiveness.
It has been applied in several areas, including constraint
databases~\cite{KanellakisKR95,toman_theo_const_tabling}, verification
of timed automata and infinite
systems~\cite{Charatonik02:model_checking_tabulation},and abstract interpretation~\cite{toman97:const_data_absint}.

The theoretical basis of TCLP~\cite{toman_theo_const_tabling}
was established in the framework of bottom-up evaluation of Datalog
systems
and presents the basic operations (projection and entailment checking)
that are necessary to ensure completeness w.r.t.\ the declarative
semantics.  However, some previous implementations did not fully
use these two operations~\cite{tchr2007,bao2000}, likely due to
performance issues and also to the implementation difficulty.

On the other hand, previous TCLP frameworks featuring a more complete
treatment of constraint projection and
entailment~\cite{chico-tclp-flops2012} focused on adapting the
implementation of a tabling algorithm
to be used with constraints.  As a result, and although
the ideas therein were generic,
they were not easily extensible.  Adding new constraint domains to
them is a difficult task that requires deep knowledge about the
particular tabling implementation and the constraint solver.  The
modifications done to the tabling implementation for one particular
constraint solver may very well be not useful for another constraint
solver; in turn, constraint solvers had to be modified in order to
make then aware of internal characteristics and capabilities of the
tabling algorithm.  These adaptations generate a \emph{technical debt}
that made using the full potential of TCLP very difficult.

In this work, we complete previous work on conditions for
termination of TCLP, we provide a richer, more flexible answer
management mechanism, 
we generalize the design of a tabling implementation so that it can
use the projection and entailment operations provided by a constraint
solver
presented to the tabling engine as a \emph{server}, and we define a
set of operations that the constraint solver has to provide to the
tabling engine.  These operations are natural to the constraint
solver, and when they are not already present, they should be easy to
implement by extending the solver.

We have validatedour design (termed \mtclp) with an implementation in Ciao
Prolog~\cite{hermenegildo11:ciao-design-tplp}
where we interfaced four
non-trivial constraint solvers to provide four different TCLP systems.
We have experimentally evaluated these implementations with several
benchmarks using TCLP.
Graphical step by step executions of TCLP programs and performance
comparison between TCLP interface are given in corresponding
appendices, included in the supplementary material accompanying the
paper at the TPLP archive.

\section{Motivation}
\label{sec:motivation}

In order to highlight some of the advantages of TCLP versus LP,
tabling, and CLP with respect to declarativeness and logical reading,
we will compare their behavior
using
different versions of a program to compute distances between nodes in
a graph.  Each version will be adapted to a different paradigm, but
trying to stay as close as possible to the original code, so that the
additional expressiveness can be attributed more to the
semantics of the programming language than to differences in the
code itself.

\begin{figure}
  \begin{center}
  \begin{minipage}{.42\linewidth}
    \redbeflst
\begin{lstlisting}[style=MyProlog]
dist(X, Y, D) :- 
    dist(X, Z, D1), 
    edge(Z, Y, D2), 
    D is D1 + D2.
dist(X, Y, D) :-
    edge(X, Y, D).

?- dist(a,Y,D), D < K.
\end{lstlisting}
    \redafterlst
  \end{minipage}
  \myvrule{.1\linewidth}{9em}
  \begin{minipage}{.37\linewidth}
    \redbeflst
\begin{lstlisting}[style=MyProlog]
:- use_package(clpq).

dist(X, Y, D) :- 
    D1 #> 0, D2 #> 0, 
    D  #= D1 + D2, 
    dist(X, Z, D1), 
    edge(Z, Y, D2).
dist(X, Y, D) :- 
    edge(X, Y, D).

?- D #< K, dist(a,Y,D).
\end{lstlisting}
    \redafterlst
  \end{minipage}
  \caption{Left-recursive  distance traversal in a graph: Prolog (left) / CLP
    (right). \\ Note: The symbols \texttt{\#>} and \texttt{\#=} are
    (in)equalities in CLP.}
  \label{fig:dist-graph}
  \redafterfig
  \end{center}
\end{figure}

\subsection{LP vs.\ CLP}
\label{sec:lp-vs-clp}

The code in Fig.~\ref{fig:dist-graph}, left, is the Prolog version of
a program used to find nodes in a graph within a distance \texttt{K} from
each other.\footnote{This is a typical query for the analysis of
  social networks~\cite{swift2010:subsumption}.}
Fig.~\ref{fig:dist-graph}, right, is the CLP version of the same code.
The queries used to find the nodes \texttt{Y} from the node \texttt{a} within
a maximum distance \texttt{K} appear in the figures themselves.

In the Prolog version, the distance between two nodes is calculated by
adding variables \texttt{D1} and \texttt{D2}, corresponding to
distances to and from an intermediate node, once they are
instantiated.  In the CLP version, addition is modeled as a constraint
and placed at the beginning of the clause.  Since the total distance
is bound, this constraint is expected to prune the search in case it tries to go
beyond the maximum distance \texttt{K}.
These checks are not added to the Prolog version, since they would not
be useful for termination: they would have to be placed after the
calls to \texttt{edge/3} and \texttt{dist/3}, where it is too late to
avoid infinite loops.  In fact, none of the queries shown before
terminates as left recursion makes the recursive clause enter an
infinite loop even for acyclic graphs.

\begin{figure}
  \begin{center}
  \begin{minipage}{.42\linewidth}
    \redbeflst
\begin{lstlisting}[style=MyProlog]
dist(X, Y, D) :- 
    edge(X, Z, D1), 
    dist(Z, Y, D2), 
    D is D1 + D2.
dist(X, Y, D) :-
    edge(X, Y, D).
\end{lstlisting}
    \redafterlst
  \end{minipage}
  \myvrule{.1\linewidth}{7em}
  \begin{minipage}{.37\linewidth}
    \redbeflst
\begin{lstlisting}[style=MyProlog]
:- use_package(clpq).

dist(X, Y, D) :- 
    D1 #> 0, D2 #> 0, 
    D  #= D1 + D2, 
    edge(X, Z, D1),
    dist(Z, Y, D2).
dist(X, Y, D) :- 
    edge(X, Y, D).
\end{lstlisting}
    \redafterlst
  \end{minipage}
  \caption{Right-recursive distance traversal in a graph: Prolog (left) / CLP
    (right). \\ Note: The symbols \texttt{\#>} and \texttt{\#=} are
    (in)equalities in CLP.}
  \label{fig:dist-graph-right}
  \redafterfig
\end{center}
\end{figure}

If we convert the program to a right-recursive version by 
swapping the calls to \texttt{edge/3} and \texttt{dist/3}
(Fig.~\ref{fig:dist-graph-right}), the LP 
execution will still not terminate in a  cyclic graph.  
The right-recursive version of the CLP program will however terminate
because the initial bound to the distance eventually causes the
constraint store to become inconsistent, which provokes a failure in
the search.
This behavior is summarized in columns ``LP'' and ``CLP'' of
Table~\ref{tab:distance-summary}.

Note that this transformation is easy in this case, but in other
cases, such as
language interpreters and tree / graph traversal algorithms,
left (or double) recursion is much more natural. 
While there are techniques to in some cases remove left / double
recursion, most Prolog compilers do not feature them.  Therefore,
we assume that the original source code is straightforwardly mapped to
the low-level runtime system, and, if necessary, left / double
recursion has to be manually removed by adding extra arguments and /
or stacks ---
precisely the kind of program transformation that we would like to
avoid due to the difficulties that it brings with respect to
maintenance and clarity.

\redbefsec
\subsection{LP vs.\ Tabling}
\redaftersec
\label{sec:lp-vs-tabling}

Tabling records the first occurrence of each call to a tabled
predicate (the \emph{generator}) and its answers.  In variant tabling,
the most usual form of tabling, when a call equal (up to variable
renaming) to a previous generator is found, its execution is
suspended, and it is marked as a \emph{consumer} of the generator.
For example, \texttt{dist(a,Y,D)} is a variant of \texttt{dist(a,Z,D)}
if \texttt{Y} and \texttt{Z} are free variables.
When a generator finitely finishes exploring all of its clauses and
its answers are collected, its consumers are resumed and are fed
the answers of the generator.  This may make consumers 
produce new answers which will in turn resume them.

Tabling is a complete strategy for all programs with the bounded
term-depth property, which in turn implies that the Herbrand model is
finite. 
Therefore, left- or right-recursive \emph{reachability}
terminates in finite graphs with or without cycles.  However, the
program in Fig.~\ref{fig:dist-graph}, left, has an infinite minimum Herbrand model for
cyclic graphs: every cycle can be traversed an unbounded number of
times, giving rise to an unlimited number of answers with a different
distance each.  The query \mbox{\texttt{?- dist(a,Y,D), D < K.}}
will therefore not terminate under variant
tabling.

\begin{table}[tb]
  \centering
      \setlength{\tabcolsep}{0.75em}
  \begin{tabular}{@{\extracolsep{0pt}}|l|c|c|c|c|c|}
    \cline{1-6}           & LP     & CLP    & TAB   & TCLP  & Graph\\ \cline{1-6}                             
    ~Left recursion       & \no    & \no    & \ok   & \ok~  & \multirow{2}{*}{Without cycles} \\ \cline{1-5}  
    ~Right recursion      & \ok    & \ok    & \ok   & \ok~  & \\ \cline{1-6}                                  
    \multicolumn{6}{c}{\vspace*{-0.95em}} \\ \cline{1-6}                                                      
    ~Left recursion       & \no    & \no    & \no   & \ok~  & \multirow{2}{*}{With cycles}    \\ \cline{1-5}  
    ~Right recursion      & \no    & \ok    & \no   & \ok~  & \\ \cline{1-6} %
  \end{tabular}
  \redbefcapt
  \caption{Comparison of termination properties.}
  \label{tab:distance-summary}
  \redafterfig
\end{table}

\redbefsec
\subsection{TCLP vs.\ Tabling and CLP}
\redaftersec
\label{sec:tclp-tabling-clp}

The program in Fig.~\ref{fig:dist-graph}, right, can be executed with
tabling and using constraint entailment to suspend calls which are
more particular than previous calls and, symmetrically, to keep only
the most general answers returned.

Entailment can be seen as a generalization of subsumption for the case
of general constraints; in turn, subsumption was shown to enhance
termination and performance in tabling~\cite{swift2010:subsumption}.
For example, the goal $G_0 \equiv$ \texttt{dist(a,Y,D)} is subsumed by
$G_1 \equiv$ \texttt{dist(X,Y,D)} because the former is an instance of
the latter ($G_0 \sqsubseteq G_1$).  All the answers for $G_1$ where
\mt{X=a} are valid answers for $G_0$; on the other hand, all the
answers for $G_0$ are also answers for $G_1$.

The main idea behind the use of entailment in TCLP is that more
particular calls (consumers) can suspend and later reuse the answers
collected by more general calls (generators).  In order to make this
entailment relationship explicit, we define a TCLP goal as
\mbox{\goal{g, c_{g}}} where $g$ is the call (a literal) and $c_g$ is
the projection of the current constraint store onto the variables of
the call.  Then, \goaltt{dist(a,Y,D)}{D<150} %
is entailed by the goal \goaltt{dist(a,Y,D)}{D>0 \land D<75} because
\mt{D>0 \land D<75 \sqsubseteq D<150}.  We also say that the former
(the generator) is more general than the latter (the consumer).  All
the solutions of the consumer are solutions of the generator or, in
other words, the space of solutions of the consumer is a subset of
that of the generator.  However, not all the answers from a generator
are valid for its consumers.  For example \mt{Y=b \land D>125 \land D<135} is a
solution for our generator, but not for our consumer, since the
consumer call was made under a constraint store more restrictive than
the generator.  Therefore, the tabling engine should check and filter,
via the constraint solver, that the answer from the generator is
consistent w.r.t.\ the constraint store of the consumer.

The use of entailment in calls and answers enhances termination
properties and can also increase speed (Section~\ref{sec:performance}).
The column ``TCLP'' in Table~\ref{tab:distance-summary} summarizes the
termination properties of \texttt{dist/3} under TCLP, and shows that a
full integration of tabling and CLP makes it possible to find all the
solutions and finitely terminate in all the cases.
Our TCLP framework not only facilitates the integration of constraint
solvers with the tabling engine thanks to its simple interface
(Section~\ref{sec:desig-gener-interf}), but also minimizes the effort
required to execute existing CLP programs under tabling
(Section~\ref{fig:transformation}), since the changes required to the
source code are minimal.

\redbefsec
\section{Background}
\redaftersec
\label{sec:foundations}

In this section we present the syntax and semantics of constraint
logic programs (Section~\ref{sec:tclp_syntax}),
and extend the semantics and termination, soundness, and completeness
proofs of~\cite{toman_theo_const_tabling} for a TCLP top-down
execution (Sections~\ref{sec:semantics} and
\ref{sec:soundn-compl-term}, resp.).

\subsection{Constraint Logic Programs}
\label{sec:tclp_syntax}

Constraint logic programming~\cite{survey94} introduces constraint
solving methods in logic-based systems. A constraint logic program
consists of clauses of the form:
$$ h\ \text{:-}\ c_h,\, l_1,\, \ \dots,\,\ l_k. $$
where
$h$ is an atom, $c_h$ is a constraint, and $l_i$
are literals.  The head of the clause is $h$ and the rest is called
the body. The clauses where the body is always true,
$ h\ \text{:-}\ true$, are called facts and usually written omitting
the body
($h.$).  We will use $L$ to denote the set of $l_i$ in a clause.  We
will assume throughout this paper that the program has been rewritten
so that clause heads are linearized (all the
variables are different) and all head unifications
take place in $c_h$.
We will assume that we are dealing with \emph{definite programs},
i.e., programs where the literals in the body are always positive
(non-negated) atoms.  \emph{Normal programs}require a different treatment.

A query to a CLP program is a clause without head
$\text{?-}\ c_q,\ q_1,\ \dots,\ q_k$, 
where $c_q$ is a constraint and $q_i$ 
are the
literals in the query.
We denote the set of $q_i$ as $Q$.

During the evaluation of a query to a CLP program, the \emph{inference
  engine} generates constraints whose consistency with respect to the
current constraint store is checked by the \emph{constraint solver}.
If the check fails, the engine backtracks to a previous state and
takes a pending branch of the search tree.

A constraint solver, denoted by CLP(\domain), is a (partial)
executable implementation of a constraint domain \domain. A valuation
$v$ over a set of variables $S = \{X_1, \dots, X_n\}$ maps each
variable $X_i$ to a value $d_i$ in \domain, denoted by $v(X_i) = d_i$.
$v$ is extended to expressions by substituting the variables in the
expressions by the value they are mapped onto.
A constraint can be a singleton constraint or a conjunction of simpler
constraints.  We denote constraints with lowercase letters and sets
of constraints with uppercase letters.
A solution of a constraint $c$ is a valuation $v$ over the variables
occurring in $c$ if $v(c)$
holds in the constraint domain.

The minimal set of operations that we expect a constraint solver to
support, in order to interface it successfully with our tabling
system, are:
\begin{itemize}
  
\item Test for consistence or satisfiability: A constraint $c$ is
  consistent in the domain $\domain$, denoted $\domain \vDash c$, if
  $c$ has a solution in $\domain$.

\item Test for entailment ($\sqsubseteq_{\domain}$):\footnote{We may
    omit the subscript \domain if there is no ambiguity.}  We say that
  a constraint $c_0$ entails another constraint $c_1$
  ($c_0 \sqsubseteq_{\domain} c_1$) if any solution of $c_0$ is also a
  solution of $c_1$.
  We extend the notion of constraint entailment to a set of constraints: a set of
  constraints $C_0$ entails another set of
  constraints $C_1$ (and we write it as  $C_0 \sqsubseteq_{\domain}
  C_1$) if
  $\forall c_{i}\in C_0 \, \exists c_{j}\in C_1 . c_{i}
  \sqsubseteq_{\domain} c_{j}$. 

\item An operation
  to compute the projection of a constraint $c$ onto a set of
  variables $S$ to obtain a constraint $c_S$ involving only variables
  in $S$ such that any solution of $c$ is also a solution of $c_S$,
  and a valuation $v$ over $S$ that is a solution of $c_S$ is a
  partial solution of $c$ (i.e., there exists an extension of $v$
  that is a solution of $c$).  We denote the projection as
  $Proj(S,c)$.

\end{itemize}

\subsection{Semantics of CLP and TCLP}
\label{sec:semantics}

The CLP fixpoint
S-semantics~\cite{falaschi-vars89,toman_theo_const_tabling} is defined
as usual as the least fixpoint of the immediate consequence operators
$S_P^{\domain}$ where all the operations behave as defined in the
constraint domain $\domain$:

\begin{definition}[Operator $S_P^{\domain}$~\cite{falaschi-vars89,toman_theo_const_tabling}]
\label{def:S_p}
  Let $P$ be a CLP program and $I$ an
  interpretation. The immediate consequence operator
  $S_P^{\domain}$ is defined as:

  \begin{tabular}{ll}
    $S_P^{\domain}(I) = I \ \cup \ \{\ \goal{h,c}\ \mid$
    & $h$ :- $c_h$, $l_1$, $\dots$, $l_k$ is a clause of $P,$ \\
    & $\goal{a_i, c_i} \in I,\ 0 < i \leq k,$ \\
    & $ c' = Proj(vars(h),\ c_h\ \land\ \bigwedge_{i=1}^k (a_i = l_i\ \land\ c_i)),$\\
    & $\domain \vDash c',$ \\
    & $ \text{if } c' \sqsubseteq c'' \text{ for some } \goal{h,c''} \in I
      \text{ then } c = c'' \text{ else } c = c'\ \} $\\
  \end{tabular}

\end{definition}

Note that $S_P^\domain$ may not add a pair \emph{\goal{literal,
  constraint}} when a more general constraint is already present in
the interpretation being enlarged.  However, to guarantee
monotonicity, it does not remove existing, more particular constraints.
The operational semantics of TCLP will however be able to do that.

The operational semantics of TCLP extends that of
CLP programs under a top-down execution scheme~\cite{survey94} that is
defined in terms of a transition system between states:

\begin{definition}
  \label{def:state}
  A \emph{state} is a tuple \goal{R,c} where:
  
  \begin{itemize}[wide=0.5em, leftmargin =*, nosep, before =
    \leavevmode\vspace{-\baselineskip}]
  \item[--] $R$, the resolvent, is a multiset of literals and
    constraints that contains the collection of as-yet-unseen literals and
    constraints of the program.  For brevity, when the set is a
    singleton we will write its only element using a lowercase letter
    instead of an uppercase letter, e.g.  $t$ instead of $\{t\}$.
  \item[--] $c$, the constraint store, is a (possibly empty) conjunction
    of constraints. It is acted upon by the constraint solver.
  \end{itemize}

\end{definition}

In~\cite{survey94} the constraint store $c$ is divided in a collection
of \emph{awake} constraints and a collection of \emph{asleep}
constraints.  This separation is ultimately motivated by
implementation issues. We do not need to make that distinction here.

Given a query \goal{Q,c_q}, the initial state of the evaluation is
\goal{Q,c_q}.
Every transition step between states resolves literals of the
resolvent against the clauses of the program and adds constraints to
the constraint store.  A derivation is \emph{successful} if it is
finite and the final state has the form \goal{\emptyset,c} (i.e., the
resolvent becomes empty). The answer to the query corresponding to
this derivation is $c' = Proj(vars(Q), c)$.

The transitions due to constraint handling are deterministic (there is
only one possible descendant in every node), while the transitions due to
literal matching are non-deterministic (there are as many descendants
as clauses match with the node literal).
We denote the set of tabled predicates in a TCLP program $P$ by
$Tab_P$.  The set of generators (calls to tabled predicates that do
not entail previous calls), is denoted by $Gen_{P}$.
The evaluation of a query to  a TCLP program is usually represented as a forest
of search trees, where each search tree
corresponds to the evaluation of a
generator and its nodes are the transitions states generated during
the evaluation (see \ref{sec:tree-forest} of the supplementary
material accompanying the paper at the TPLP archive, for some
examples).

The order in which the literals / constraints are selected is decided
by the computation rule.  During the computation of a TCLP program,
two main phases are interleaved for the evaluation of every tabled
goal: the call entailment phase (Def.~\ref{def:call_entail}) and the
answer entailment phase (Def.~\ref{def:ans_entail}).

\begin{definition}[Call entailment phase]
  \label{def:call_entail}
  The call entailment phase checks if a new goal \goal{t,c}, where $t$ is
  a tabled literal (i.e., $t \in Tab_P$), entails a previous goal
  (called its generator).\footnote{Note that this entailment check
    includes subsumption in the Herbrand domain.}  The new goal is
  resolved by answer resolution consuming the answers $c_i$ such that:

\begin{itemize}[wide=0.5em, leftmargin =*, nosep, before =
    \leavevmode\vspace{-\baselineskip}]
\item $c_i \in Ans(g, c_g)$, where $Ans(g, c_g)$ is the set of
  answers of the oldest generator $\goal{g,t} \in Gen_{P}$, such that
  $g$ and $t$ are equal upon variable renaming, and
  \mbox{$c \land (t = g) \sqsubseteq_{\domain} c_g$}, where $t = g$ is
  an abbreviation for the conjunction of equations between the
  corresponding arguments of $t$ and $g$, i.e., \goal{g,c_g} is more
  general than \goal{t,c}. In this case the goal \goal{t,c} is marked as a
  consumer of \goal{g,c_g}.
\item Or $c_i \in Ans(t, c')$, where $c' = Proj(vars(t),c)$ and
  $Ans(t, c')$ is the set of answers of a new generator \goal{t,c'}
  which is added to $Gen_{P}$, the set of generator.
\end{itemize}

\end{definition}

In TCLP, goals that match heads of tabled
predicates are not resolved against program clauses. Instead, they
are resolved consuming the answer constraints from a generator; this
is termed \emph{answer resolution}.

\begin{definition}[Answer entailment phase]
  \label{def:ans_entail}
  The answer constraints of a generator \goal{g,c_g} are collected in
  the answer entailment phase in such a way that an answer which
  entails another more general answer is discarded / removed. The set
  of answers, denoted by $Ans(g,c_g)$, is the set of more general
  (w.r.t. $\sqsubseteq_{\domain}$) answer constraints $c'_i$ obtained
  as the projection of $c_i$ onto $vars(g)$, where
  \goal{\emptyset,c_i} (resp. \goal{\emptyset,c_j}) is the final state of a successful derivation:

  \medskip
      
\begin{tabular}{ll}
  \hspace*{-2cm}
  $Ans(g,c_g) = \{ c'_{i}\ \mid$
  & $c'_{i} = Proj(vars(g),c_i),$ \\
  & $\forall j \neq i,\ \nexists\ c'_j = Proj(vars(g),c_j).\ \ 
    c'_{i} \sqsubseteq_{\domain} c'_{j}\ \}$\\
\end{tabular}\\

\end{definition}

We will assume, without loss of generality, that the subscripts $i, j$
correspond to the order in which answers are found.
The answer management strategy used in the answer entailment phase
aims at keeping only the more general answers by discarding / removing
more particular answers.  This is specified by the quantification
$\forall j \neq i$, where $i$ and $j$ are the indexes of the final
constraint store.
Simpler answer management strategies are possible: the implementations
in~\cite{bao2000,chico-tclp-flops2012},
following~\cite{toman_theo_const_tabling}, only discard answers which
are more particular than a previous one, i.e., they implement
$\forall j < i$, and keep previously saved answers.  A third
possibility is to remove previous answers that are more particular
than new ones, implementing $\forall j > i$.  The choice among them
does not impact soundness or completeness properties.  However,
discarding and removing redundant answers can greatly increase the
efficiency of the implementation, as we experimentally show in
section~\ref{sec:eval-management}.

The order in which we search in the TCLP forest for a previous
generator during the call entailment phase does not impact the
completeness, soundness, or termination properties of the execution,
but it can change its efficiency.  Generators are naturally sorted
from more particular ones (older) to more general (younger) ones ---
note that a younger, more particular call would be a consumer.
Searching for a generator for a call can be performed in any
direction.  Starting at older, more particular generators, may need to
examine several generators and perform potentially expensive
entailment checks before finding one that suits the needs of the call.
On the other hand, starting at younger, more general generators,
should allow us to locate a suitable generator faster.  However, this
more general generator would have more answers associated which need
to be filtered than what a more particular generator would have.
Therefore, there does not seem to be a clear general strategy:
either more generators have to be traversed, or more answers have to
be filtered.

\subsection{Soundness, completeness and termination properties of
  TCLP}
\label{sec:soundn-compl-term}

\cite{toman_theo_const_tabling} proves soundness and completeness of
$SLG^C$ for TCLP Datalog programs by reduction to soundness and
completeness of bottom-up evaluation. It is possible to extend these
results to prove soundness and completeness of our proposal: they only
differ in the answer management strategy and the construction of the
TCLP forest.  The strategy used in $SLG^C$ only discards answers which
are more particular than a previous answer, while in our proposal we
in addition remove previously existing more particular answers
(Def.~\ref{def:ans_entail}).  The result of this is that only the most
general answers are kept.  In $SLG^C$, the generation of the forest is
modeled as the application of rewriting rules.

\begin{theorem}[Soundness w.r.t.\ the fixpoint semantics]
  Let $P$ be a TCLP definite program and \goal{q,c_q} a query. Then for
  any answer $c'$
  \begin{displaymath}
    c' \in Ans(q,c_q)\ \Rightarrow \exists
    \goal{q,c} \in \lfp(S_P^{\domain}(\emptyset)).
    \ c' = c_q \land c
  \end{displaymath}
\noindent 
I.e., all the answers derived from the forest construction are also
derived from the bottom-up computation (and are therefore correct).
\end{theorem}

Answer resolution recovers constraints from the $Ans(g, c_g)$ sets
corresponding to a goal $g$ and an constraint $c_g$, instead of
repeating SLD resolution on $\goal{g, c_g}$.  These sets were
ultimately generated by saving SLD resolution results, possibly using
previously generated sets of answers for intermediate
goals.\footnote{Answers more particular than other answers may be removed, but
those which remain are still \emph{correct} answers.}
Therefore, we can substitute any point where answers are recovered from
$Ans(g, c_g)$ for the corresponding SLD resolution.  So, if
$c' \in Ans(q,c_q)$ then there is an SLD derivation
$\goal{q, c_q} \leadsto \goal{\emptyset, c'}$.  Moreover, if
$\goal{q, \emptyset} \leadsto \goal{\emptyset, c}$ then
$c' = c \land c_q$ and we can construct the answer to $\goal{q, c_q}$
from that of $\goal{q,\emptyset}$. So if $c' \in Ans(q,c_q)$, then
there is $\goal{q, \emptyset} \leadsto \goal{\emptyset, c}$  and $c' = c \land c_q$.
From the correctness of SLD resolution, we have that if there is a
derivation $\goal{q, \emptyset} \leadsto \goal{\emptyset, c}$ then
$\goal{q,c} \in \lfp(S_P^{\domain}(\emptyset))$ and, as we said
before, $c' = c \land c_q$.

\begin{theorem} [Completeness w.r.t.\ the fixpoint semantics]
  Let $P$ be a TCLP definite program and \goal{h,true} a query. Then for
  every \goal{h,c} in $\lfp(S_P^{\domain})$:
  \begin{displaymath}
    \goal{h,c} \in \lfp(S_P^{\domain}(\emptyset))\ \Rightarrow \ \exists c' \in Ans(h,true). \ c \sqsubseteq c'
  \end{displaymath}
  \noindent
  I.e., all the answers derived from the bottom-up computation 
  entail answers generated by the TCLP execution.
\end{theorem}

For any answer derived from the bottom-up computation
$\goal{h,c} \in \lfp(S_P^{\domain}(\emptyset))$ there exists a
successful SLD derivation $\goal{h,true} \leadsto \goal{\emptyset,c}$.
Since answer resolution may keep the most general answer when
generating comparable answers (Def.~\ref{def:ans_entail}), it is also
complete if entailment with this most general answer is used instead
of equality with the more particular answers (which were removed).
Therefore, it will always be the case that
$\exists c' \in Ans(h,true). \ c \sqsubseteq c'$.

\medskip

Termination of TCLP Datalog programs under a top-down strategy when
the domain is constraint-compact (Def.~\ref{def:constraint-compact})
is proven in~\cite{toman_theo_const_tabling}.

\begin{definition}[Constraint-compact]
  \label{def:constraint-compact}
  Let \domain be a constraint domain, and $D$ the set of all
  constraints expressible in \domain. Then $\domain$ is
  constraint-compact iff:
  
  \begin{itemize}[wide=0.5em, leftmargin =*, nosep, before =
    \leavevmode\vspace{-\baselineskip}]
  \item[--] for every finite set of variables $S$, and
  \item[--] for every subset $C \subseteq D$ such that
    $\forall c \in C.vars(c) \subseteq S$,
  \end{itemize}
  
  there is a finite subset $C_{fin} \subseteq C$ such that
  $\forall c \in C.\exists c' \in C_{fin}. c \sqsubseteq_{\domain} c'$
\end{definition}

In that case, the evaluation will suspend the exploration of a call
whose constraint store is less general or comparable to a previous
call.  Eventually, the program will generate a set of call constraint
stores which can cover any infinite set of constraints in the domain,
therefore finishing evaluation. That is because, intuitively speaking,
a domain \domain is constraint-compact if for any (potentially
infinite) set of constraints $C$ expressible in \domain, there is a
\emph{finite} set of constraints $C_{fin} \subseteq C$ that covers (in
the sense of $\sqsubseteq_{\domain}$) $C$.  In other words, $C_{fin}$
is as general as $C$.
 
Most TCLP applications require domains which are not
constraint-compact because constraint-compact domains are not very
expressive.  Therefore, we refined the termination theorem (Theorem~23
in~\cite{toman_theo_const_tabling}) for Datalog programs with
constraint-compact domains to cover cases where a program, during the
evaluation, generates only a constraint-compact subset of all
constraints expressible in the domain.

\newcommand{\Cs}{\ensuremath{C_{g}}\xspace}
\newcommand{\As}{\ensuremath{A_{\goal{g,c_g}}}\xspace}

\begin{theorem} [Termination]
  \label{th:termination}
  Let $P$ be a TCLP($\domain$) definite program and \goal{Q,c_q} a
  query. Then the TCLP execution terminates iff:

  \begin{itemize}[wide=0.5em, leftmargin =*, nosep, before =
    \leavevmode\vspace{-\baselineskip}]
  \item For every literal $g$, the set \Cs is
    constraint-compact, where \Cs is the set
    of all the constraint stores $c_i$, projected and renamed
    w.r.t.\ the arguments of $g$, s.t. \goal{g,c_i} is in the forest
    $\mf(Q,c_q)$.

  \item For every goal \goal{g,c_g}, the set \As is
    constraint-compact, where \As is the set of all the answer
    constraints $c'$, projected and renamed w.r.t.\ the arguments of
    $g$, s.t. $c'$ is a successful derivation in the forest
    $\mf(Q,c_q)$.
  \end{itemize}
\end{theorem}

The intuition is that for every subset $C$ of the set of all
possible constraint stores \Cs that can be generated when evaluating
a call to $P$, if there is
a finite subset $C_{fin} \subseteq C$ that covers (i.e., is as general
as) $C$, then, at some point, any call will entail previous
calls, thereby allowing its suspension to avoid loops. %
Similarly, for every subset $A$ from the set of all possible answer
constraints \As that can be generated by a call, if there is a finite
subset $A_{fin} \subseteq A$ that covers $A$, then, at some point, any
answer will entail a previous one, ensuring that the class of
answers $Ans(g,c_g)$ which is entailed by any other possible answer returned
by the program is finite.\footnote{Note that a finite answer set does
  not imply a finite domain for the answers: the set of answers
  $Ans(Q,c_q) = \{V > 5\}$ if finite, but the domain of $V$ is
  infinite.}

\begin{figure}
  \begin{center}
    \begin{minipage}{.29\linewidth}
      \redbeflst
\begin{lstlisting}[style=MyProlog]
nat(X) :-
    X #= Y+1, 
    nat(Y).
nat(0).

\end{lstlisting}~
      \redafterlst
    \end{minipage}
    \myvrule{.1\linewidth}{5em}
    \begin{minipage}{.28\linewidth}
      \redbeflst
\begin{lstlisting}[style=MyProlog]
nat(X) :-
    X #= Y+1, 
    nat(Y).
nat(0).
nat(X) :- X #> 1000.
\end{lstlisting}
      \redafterlst
    \end{minipage}
    \caption{Left: definition of natural numbers in TCLP(\Q).  Right:
      natural numbers with a clause describing infinitely many
      numbers.}
    \label{fig:terminate-example}
    \redafterfig
  \end{center}
\end{figure}

  \begin{example}
    Fig.~\ref{fig:terminate-example}, left, shows a program which
    generates all the natural numbers using TCLP(\Q). Although CLP(\Q)
    is not constraint-compact, the constraint stores generated by that
    program for the query {\normalfont\texttt{?- X \#< 10, nat(X)}} are
    constraint-compact and the program finitely finishes. Let us look at
    its behavior from two points of view:
    \\[-1.5em]
    \begin{description}
    [leftmargin=.7cm,labelindent=.7cm]
  \item[Compactness of call / answer constraint sets] %
    The set of all active constraint stores generated for the
    predicate {\normalfont\texttt{nat/1}} under the query \goaltt{nat(X)}{X < 10}
    is $C_{\mt{nat(V)}} = \{\mt{V<10,\, V<9,\, \dots,\, V < -1, V < -2,\, \dots}\} $. It
    is constraint-compact because every subset $C \in C_{\mt{nat(V)}}$ is
    covered by $C_{fin} = \{\mt{V<10}\}$. %
    The set of all possible answer constraints for the query,
    $A_{\goaltt{nat(V)}{V<10}} = \{\mt{V=0,\, \dots,\, V=9}\}$, is also
    constraint-compact because it is finite. Therefore, the program
    terminates.
  \item[Suspension due to entailment] The first recursive call is
    \goaltt{nat(Y_1)}{X<10 \land {X=Y_1+1}} and the projection of its
    constraint store after renaming entails the initial one
    since $\mt{V<9 \sqsubseteq V<10}$.  Therefore, TCLP evaluation suspends
    in the recursive call, shifts execution to the second clause and
    generates the answer $\mt{X = 0}$. This answer is given to the
    recursive call, which was suspended, produces the constraint store
    $\mt{X<10 \land X=Y_1+1 \land Y_1=0}$, and generates the answer
    $\mt{X = 1}$. Each new answer $\mt{X_n = n}$ is used to feed the recursive
    call. When the answer $\mt{X = 9}$ is given, it results in the
    (inconsistent) constraint store
    $\mt{X<10 \land X = Y_1 + 1 \land Y_1=9}$ and the execution
    terminates.
  \end{description}

\end{example}

  \begin{example}
    The program in Fig.~\ref{fig:terminate-example}, left, does not
    terminate 
    for the query {\normalfont\texttt{\mbox{?- X \#> 0},\, \mbox{X \#< 10},
      nat(X)}}.  Let us examine its behaviour: %
    \\[-1.5em]
    \begin{description}
      [leftmargin=.7cm,labelindent=.7cm]
    \item[The constraint store sets are not compact] The set of all
      constraint stores generated by the query \goaltt{nat(X)}{X>0
        \land X<10} is
      $C_{\mt{nat(V)}} = \{\mt{V>0 \land V<10},\, \mt{V>-1 \land
        V<9},\, \dots,\, \mt{V>-n \land V<(10-n)},\, \dots\}$, which
      it is not constraint-compact.  Note that $\mt{V}$ is, in
      successive calls, restricted to a sliding interval
      $\mt{[k, k+10]}$ which starts at $\mt{k=0}$ and decreases
      $\mt{k}$ in each recursive call.  No finite set of intervals can
      cover any subset of the possible intervals.

    \item[The evaluation loops] The first recursive call is
      \goaltt{nat(Y_1)}{X>0 \land X<10 \land {X=Y_1+1}} and the
      projection of its constraint store does not entail the
      initial one  after
      renaming since
      $(\mt{V>-1 \land V<9) \not\sqsubseteq  (X>0 \land X<10)}$. Then this
      call is evaluated and produces the second recursive call,
      \goaltt{nat(Y_2)}{X>0\land X<10 \land X=Y_1+1 \land
        Y_1=Y_2+1}. Again, the projection of its constraint store,
      $\mt{Y_2>-2 \land Y_2<8}$, does not entail any of the previous
      constraint stores, and so on.  The evaluation therefore loops.
    \end{description}
  \end{example}

  \begin{example}
    The program in Fig.~\ref{fig:terminate-example}, left, also does not terminate for the 
    query
    {\normalfont\texttt{?- nat(X)}}. Let us examine its behavior:%
    \\[-1.5em]
    \begin{description}
      [leftmargin=.7cm,labelindent=.7cm]
    \item[The answer constraint set is not compact] %
      The equation in the body of the clause $\mt{X=Y_1+1}$ defines a
      relation between the variables but, since the domain of $\mt{X}$
      is not restricted, its projection onto $\mt{Y_1}$ will return no
      constraints (i.e., $Proj(\mt{Y_1, X=Y_1+1}) =
      \mt{true}$). Therefore, the set of all call constraint stores
      generated by the query \goaltt{nat(X)}{true} is
      $C_{\mt{nat(V)}} = \{\mt{true}\}$ which is finite and
      constraint-compact. %
      However, the answer constraint set
      $A_{\goaltt{nat(V)}{true}} = \{\mt{V=0,\, V=1,\, \dots,\, V=n,\,
        \dots}\}$ is not constraint-compact. 
    \item[Call suspension with an infinite answer constraint set] %
      The first recursive call is \goaltt{nat(Y_1)}{X = Y_1+1} and the
      projection of its constraint store entails the initial
      store. Therefore, the TCLP evaluation suspends the recursive
      call, shifts execution to the second clause, and generates the
      answer $\mt{X = 0}$.  This answer is used to feed the suspended
      recursive call, resulting in the constraint store
      $\mt{X=Y_1+1 \land Y_1=0}$ which generates the answer
      $\mt{X = 1}$. Each new answer $\mt{X = n}$ is used to feed the
      suspended recursive call.  Since the projection of the
      constraint stores on the call variables is \mt{true}, the
      execution tries to generate infinitely many natural
      numbers. Therefore, the program does not terminate.
    \end{description}

  \end{example}

    \begin{example}
      Unlike the situation that happens in pure Prolog / variant tabling,
      adding new clauses to a program under TCLP can make it
      terminate.\footnote{This depends on the strategy used by the
        TCLP engine to resume suspended goals.  Our implementation
        gathers all the answers for goals that can produce results
        first, and then these answers are used to feed suspended
        goals.  This makes the exploration of the forests proceed in a
        breadth-first fashion.} As an example,
      Fig.~\ref{fig:terminate-example}, right, is the same as
      Fig.~\ref{fig:terminate-example}, left, with the addition of
      the clause {\normalfont\texttt{nat(X):- X \#> 1000}}.  Let us examine
      its behavior under the query {\normalfont\texttt{?- nat(X)}}:
   \begin{description}
      [leftmargin=.7cm,labelindent=.7cm]
    \item[Compactness of call / answer constraint sets] %
      The set of all constraint stores generated remains
      $C_{\mt{nat(V)}} = \{\mt{true}\}$. But the new clause makes  the
      answer constraint set becomes
      $A_{\goaltt{nat(V)}{true}} = \{\mt{V=0},\, \mt{V=1},\, \dots,\, \mt{V=n},\,
      \dots,\, \mt{V>1000},\, \mt{V>1001},\, \dots,\, \mt{V>n},\, \dots\}$, which is
      constraint-compact because a constraint of the form $\mt{V>n}$
      is entailed by infinitely many constraints, i.e.\ it covers the infinite
      set $\{\mt{V=n+1},\, \dots,\, \mt{V>n+1},\, \dots\}$. Therefore, since
      both sets are constraint-compact, the program terminates.
      
    \item[First search, then consume] %
      The first recursive call \goaltt{nat(Y_1)}{X = Y_1+1} is suspended
      and the TCLP evaluation shifts to the second clause which
      generates the answer $\mt{X = 0}$.  Then, instead of feeding the
      suspended call, the evaluation continues the search and shifts
      to the third clause, {\normalfont\texttt{nat(X):- X \#> 1000}}, and
      generates the answer $\mt{X>1000}$.  Since no more clauses remain 
      to be explored, the answer $\mt{X=0}$ is used, generating $\mt{X=1}$. Then
      $\mt{X>1000}$ is used, resulting in the constraint store
      $\mt{X = Y_1 + 1 \land Y_1 > 1000}$, which generates the answer
      $\mt{X>1001}$. However, during the answer entailment phase, $\mt{X>1001}$
      is discarded because $\mt{X>1001 \sqsubseteq X>1000}$. Then, one by
      one each answer $\mt{X=n}$ is used, generating $\mt{X=n+1}$. But when the
      answer $\mt{X=1000}$ is used, the resulting answer $\mt{X=1001}$ is
      discarded, during the answer entailment phase, because
      $\mt{X=1001 \sqsubseteq X>1000}$. At this point the evaluation
      terminates because there are no more answers to be consumed. The
      resulting set of answers is
      $Ans(\mt{nat(X),true}) = \{\mt{X=0},\, \mt{X>1000},\, \mt{X=1},\, \dots,\, \mt{X=1000}\}$
    \end{description}
    
  \end{example}

\redbefsec
\section{The \emph{\mtclp}\ Framework}
\redaftersec
\label{sec:generic-interface}

In this section we describe the \mtclp\ framework, the operations
required by the interface, and the program transformation that we use
to compile programs with tabled constraints 
(Section~\ref{sec:desig-gener-interf}).  We also provide a sketch of its
 implementation and we describe step-by-step some executions
at the level of the TCLP libraries (Section~\ref{sec:implementation}
and \ref{sec:execution-dist3} of the supplementary material). In
Section~\ref{sec:tclpq-int} we present the implementation of the TCLP
interface for Holzbaur's CLP(\Q) solver and in
Section~\ref{sec:improvementss} we present an optimization, the \emph{Two-Step}
projection.

\subsection{Design of the Generic Interface}
\label{sec:desig-gener-interf}

\begin{figure}
  \small
\begin{minipage}{0.95\linewidth}
  \fbox{
 \begin{minipage}{1.0\linewidth}
    \begin{description}

   \item[\texttt{store\_projection(+Vars, -ProjStore)}] Returns in
    \texttt{ProjStore} a representation of the projection of the current
    constraint store onto the list of variables \texttt{Vars}.
      \\[-0em]

    \item[\texttt{call\_entail(+ProjStore,
        +ProjStore$\mt{_{gen}}$)}] Succeeds if the projection of the current
      constraint store, \texttt{ProjStore}, entails the
      projected store, \texttt{ProjStore$\mt{_{gen}}$}, of a previous
      generator. It fails otherwise.
    
    \item[\texttt{answer\_compare(+ProjStore,
        +ProjStore$\mt{_{ans}}$, -Res)}] Returns \texttt{Res=`\texttt{=<}'} if
      the projected store of the current answer, \texttt{ProjStore},
      entails the projected store of a previous answer,
      \texttt{ProjStore$\mt{_{ans}}$}, or \texttt{Res=`\texttt{>}'} if
      \texttt{ProjStore} is entailed by \texttt{ProjStore$\mt{_{ans}}$} and they
      are not equal.  It fails otherwise.
      \\[-0em]

    \item[\texttt{apply\_answer(+Vars, +ProjStore)}] Adds the projected
      constraint store \texttt{ProjStore} of the answer to the current
      constraint store and succeeds if the resulting constraint store
      is consistent.
      \\[-0em]

    \end{description}
\end{minipage}
  }
\end{minipage}
\redbefcapt
  \caption{Generic interface specification.}
  \label{fig:interface}
\redafterfig
\end{figure}

\mtclp\ provides a generic interface (Fig.~\ref{fig:interface})
designed to facilitate the integration of different constraint
solvers. The predicates of the interface use extensively two objects:
\texttt{Vars}, the list of constrained variables, provided by the
tabling engine to the constraint solver, and \texttt{ProjStore}, a
representation of the projected constraint store, opaque to the
tabling engine, and which should be self-contained and independent
(e.g., with fresh variables) from the \emph{main} constraint store.
For example, the constraint solver CLP(\D)
(Section~\ref{sec:design-diff-constr}) is written in C and the
projection of a constraint store is a C structure whose representation
is its memory address and length.

To implement these predicates, the constraint solver has to support
the (minimal) set of operations defined in
Section~\ref{sec:tclp_syntax}: projection, test for entailment, and
test for consistence. The predicates that the constraint solver must
provide in order to enable its interaction with the tabling engine
are:
 
\begin{itemize}

\item \texttt{store\_projection(+Vars, -ProjStore)}, that  is invoked before
  the call and the answer entailment phases:

  \begin{itemize}
  \item[--] It is used before the call entailment phase to generate
    the representation of the goal as a tuple \goaltt{G}{ProjStore},
    where $\mt{ProjStore}$ represents the projection of the constraint
    store at the moment of the call onto $\mt{Vars}$, the variables in
    $\mt{G}$. Although a generic implementation should include the
    Herbrand constraints of the call in the constraint store, our
    implementation does not consider Herbrand constraints to be part
    of the constraint store by default. Instead, calls are
    syntactically compared using variant checking, but the programmer
    can also choose to use subsumption, if required, by using a
    package described below.
    There are some reasons for that decision: on the one hand, programmers
    (even using tabling) are used to this behavior; on the other hand,
    there are data structures highly
    optimized~\cite{rama95:efficient_tabling} to save and retrieve
    calls together with their input / output substitutions which perform
    variant checking on the fly while taking advantage of the
    WAM-level representation of substitutions.
    
  \item[--] Similarly, before the answer entailment phase, the
    projection of the Herbrand constraints onto the variables of the
    goal is directly taken care of by their WAM-level representation.
    We use variant checking to detect when the Herbrand constraints
    associated to two calls are equal.  Therefore, an answer
    constraint is internally represented by a tuple
    \goaltt{S}{ProjStore} where $\mt{S}$ captures the Herbrand
    constraints of the variables of the goal and $\mt{ProjStore}$
    represents the projection of the rest of the answer constraint
    onto $\mt{Vars}$, the variables of the answer.
  \end{itemize}

\item \texttt{call\_entail(+ProjStore,
    +ProjStore$\mt{_{gen}}$)} is invoked during the call entailment
  phase to check if a new call, represented by \goaltt{G}{ProjStore},
  entails a previous generator, represented by
  \goaltt{G_{gen}}{ProjStore_{gen}}, where $\mt{G}$ is a variant of
  $\mt{G_{gen}}$. The predicate succeeds if
  $\mt{ProjStore \sqsubseteq ProjStore_{gen}}$, and fails
  otherwise. If Herbrand subsumption checking is needed, our
  implementation provides a package which transforms calls to
  tabled predicates so that suspension is based on entailment in \h.
  This transformation moves Herbrand constraint handling away from the
  level of the WAM by 
  creating attributed variables~\cite{bao2000} that
  carry the constraints --- i.e., the unifications. Later on, a Herbrand
  constraint solver is used to check subsumption.\footnote{If there
    are several constraint domains involved, such as e.g.\ CLP(\h) and
    CLP(\Q), we assume that we can distinguish them appropriately at
    run-time and the entailment is determined as
    $\mt{variant(G, G_{gen})}\ \land \ \mt{ProjStore}
    \sqsubseteq_{\text{\h}} \mt{ProjStore_{gen}}\ \land \
    \mt{ProjStore} \sqsubseteq_{\text{\Q}} \mt{ProjStore_{gen}}$ }
  
\item \texttt{answer\_compare(+ProjStore,
    +ProjStore$\mt{_{ans}}$, -Res)} is invoked during the answer
  entailment phase to check a new answer, represented by
  \goaltt{S}{ProjStore}, against a previous one, represented by
  \goaltt{S_{ans}}{ProjStore_{ans}}, when the Herbrand constraints
  $\mt{S}$ and $\mt{S_{ans}}$ are equal. The predicate compares
  $\mt{ProjStore}$ and $\mt{ProjStore_{ans}}$ and returns
  `\texttt{=<}' in its last argument when
  \mbox{$\mt{ProjStore} \sqsubseteq \mt{ProjStore_{ans}}$},
  `\texttt{>}' when $\mt{ProjStore} \sqsupset \mt{ProjStore_{ans}}$,
  and fails otherwise. This bidirectional entailment check, which is
  used to discard / remove more particular answers, is a potentially
  costly process, but it brings considerable advantage from saved
  resumptions (Section~\ref{sec:eval-management}): when an answer is
  added to a generator, consumers are resumed by that answer.  These
  consumers in turn generate more answers and cause further
  resumptions in cascade.  Reducing the number of redundant answers
  reduces the number of redundant resumptions, and we have
  experimentally observed that it brings about important savings in
  execution time.

\item \texttt{apply\_answer(+Vars, +ProjStore)} is invoked to consume
  an answer from a generator. In variant tabling, since consumers are
  variants of generators, answer substitutions from generators can
  always be applied to consumers. That is not the case when using
  entailment in TCLP: consumers may be called in the realm of a
  constraint store more restrictive than that of their generators, and
  answers from the generator have to be filtered to discard those
  which are inconsistent with the constraint store at the time of the
  call to the consumer. In our implementation, an answer is
  represented by \goaltt{S}{ProjStore}, where $\mt{S}$, the set of Herbrand
  constraints, is applied by the tabling engine
  and $\mt{ProjStore}$, the projection of the constraint
  answer, is added to the constraint store of the consumer by
  \texttt{apply\_answer/2}, which succeeds iff the resulting
  constraint store is consistent.
\end{itemize}

The design of the interface assumes 
that external constraint
solvers 
are compatible with Prolog operational semantics so that when Prolog
backtracks to a previous state, the corresponding constraint store is
transparently restored.
That can be done by adding a Prolog layer which uses the trail to
store \texttt{undo} information which is used to reconstruct the
previous constraint store when Prolog backtracks (this is a
reasonable, minimal assumption for any integration of constraint
solving and logic programming).
The TCLP interface then can readily use any constraint solver which
follows this design, because the suspension and resumption mechanisms
of the tabling are based on the trailing mechanism of Prolog.
When a consumer suspends, backtracking takes place, the memory stacks
are frozen, and the variable bindings are saved on untrailing.  They are
reinstalled upon consumer resumption.

However, the entailment operations \texttt{call\_entail/2} and
\texttt{answer\_compare/3} need to know the correspondence among variables in
\texttt{ProjStore} and in
\texttt{ProjStore$_\mathtt{gen}$} (resp., 
\texttt{ProjStore$_\mathtt{ans}$}).  To this end, 
projections are (conceptually) a pair \texttt{(VarList, Store)},
where \texttt{VarList} is a list of fresh variables
in \texttt{Store} that correspond to \texttt{Vars}, the 
variables on which the projection was originally made. Different,
independent constraint stores can then be compared by means of these
lists.
This list is also necessary to apply the \texttt{ProjStore} of an
answer to the global store: it is used to determine the correspondence
of variables between the global and the projected store.

The actual implementation may differ among constraint solvers. For
example, the TCLP(\Q) interface (Fig.~\ref{fig:int-clpq}) uses
a list of fresh variables following the same order as those in
\texttt{Vars}.  However, the TCLP(\D) interface
(Section~\ref{sec:design-diff-constr}) uses a vector containing the
index in the matrix corresponding to every variable in \texttt{Vars},
again following the same order.

\subsection{Implementation Sketch}
\label{sec:implementation}

We summarily describe now the implementation of \mtclp, including the
global table where generators, consumers, and answers are saved.  We
also present the transformation performed to execute tabled predicates
and a (simplified) flowchart showing the interactions between the
tabling engine and the constraint solver through the generic
interface.

\subsubsection{Global Table}

Tries are the data structure of choice for the call / answer global
table~\cite{rama95:efficient_tabling}.  In variant tabling, every
generator $\mt{G_{gen}}$ is uniquely associated (modulo variable renaming)
to a leaf from where the Herbrand constraints for every answer hangs.
Generators are identified in \mtclp\ by the projection of the
constraint store on the variables of the generator, i.e., with a tuple
\mbox{\goaltt{G_{gen}}{ProjStore_{gen}}}.  We store generators in a trie
where each leaf is associated to a call pattern $\mt{G_{gen}}$ 
and a list with a frame for each projected constraint store
$\mt{ProjStore_{gen_i}}$.
Each frame identifies: (i) the projected constraint store
$\mt{ProjStore_{gen_i}}$, (ii) the answer table where the generator's
answers $Ans(\mt{G_{gen}}, \mt{ProjStore_{gen_i}})$ are stored, and (iii) the list
of its consumers.

Answers are represented by a tuple \goaltt{S_{ans}}{ProjStore_{ans}}
and stored in a trie where each leaf points to %
the Herbrand constraints $\mt{S_{ans}}$ and to a list with the
projected constraint stores $\mt{ProjStore_{ans_i}}$ corresponding to
answers whose 
Herbrand constraints are a variant of $\mt{S_{ans}}$.
The answers are stored in order of generation, since (as we will see
later) it is not clear that other orders eventually pay off in terms
of speeding up the entailment check of future answers.

\subsubsection{TCLP Directives and Program Transformation}

Executing a CLP program under the TCLP framework only needs to enable
tabling and import a package which implements the bridge CLP / tabling
instead of the regular constraint
solver. Fig.~\ref{fig:transformation}, left, shows the TCLP version of
the left recursive distance traversal program in
Fig.~\ref{fig:dist-graph}, right. The constraint interface remains
unchanged, and the program code does not need to be modified to be
executed under TCLP.
The directive \mbox{\texttt{:-~use\_package(tabling)}} initializes the
tabling engine and the directive
\mbox{\texttt{:-~use\_package(t\_clpq)}} imports TCLP(\Q), the TCLP
interface for the CLP(\Q) solver
(Section~\ref{sec:tclpq-int}). To select another TCLP interface (more
examples in Section~\ref{sec:other-tclp-interfaces}) we just have to
import the corresponding package. Finally, the directive
\mbox{\texttt{:-~table~dist/3.}}  specifies that the predicate
\texttt{dist/3} should be tabled.

Fig.~\ref{fig:transformation}, right, shows the transformation applied to
the predicate \texttt{dist/3}.  The original entry point to the
predicate is rewritten to call an auxiliary predicate through the
meta-predicate \texttt{tabled\_call/1} (Fig.~\ref{fig:tabled_call}).
The auxiliary predicate corresponds to the original one with a renamed
head and with an additional \texttt{new\_answer/0}
(Fig.~\ref{fig:new_answer}) at the end of the body to collect the
answers. 
An internal global stack, called PTCP, is used to identify the
generator under execution when \texttt{new\_answer/0} is executed.

\begin{figure}
  \begin{center}
    \begin{minipage}{.42\linewidth}
      \redbeflst
\begin{lstlisting}[style=MyProlog]
:- use_package(tabling).
:- use_package(t_clpq).

:- table dist/3.
dist(X, Y, D) :- 
    D1 #> 0, D2 #> 0, 
    D  #= D1 + D2, 
    dist(X, Z, D1),
    edge(Z, Y, D2).
dist(X, Y, D) :- 
    edge(X, Y, D).
\end{lstlisting}
      \redafterlst
    \end{minipage}
    \myvrule{.05\linewidth}{9em}
    \begin{minipage}{.45\linewidth}
      \redbeflst
\begin{lstlisting}[style=MyProlog]
dist(A, B, C) :-   
    tabled_call(dist_aux(A,B,C)).

dist_aux(X, Y, D) :-
    D1 #> 0, D2 #> 0, 
    D  #= D1 + D2, 
    dist(X, Z, D1),
    edge(Z, Y, D2), 
    new_answer.
dist_aux(X, Y, D) :-
    edge(X, Y, D),
    new_answer.
\end{lstlisting}
      \redafterlst
    \end{minipage}
    \redbefcapt
    \caption{\texttt{dist/3} with the directives to enable TCLP (left) and its transformation (right).}
    \label{fig:transformation}
    \redafterfig
  \end{center}
\end{figure}

\begin{figure}
  \begin{center}
    \begin{minipage}{.8\linewidth}
\begin{lstlisting}[style=MyProlog]
tabled_call(Call) :-
    call_lookup_table(Call, Vars, Gen),
    store_projection(Vars, ProjStore),
    &(  
       projstore_Gs(Gen, List_GenProjStore),
       member(ProjStore_G, List_GenProjStore),
       call_entail(ProjStore, ProjStore_G) ->
       suspend_consumer(Call)
    ;  
       save_generator(Gen, ProjStore_G, ProjStore),
       execute_generator(Gen, ProjStore_G),
    &),
    answers(Gen, ProjStore_G, List_Ans),
    member(Ans, List_Ans),
    projstore_As(Ans, List_AnsProjStore),
    member(ProjStore_A, List_AnsProjStore), 
    apply_answer(Vars, ProjStore_A).
\end{lstlisting}
    \end{minipage}
  \end{center}
  \redbefcapt
  \caption{Implementation of \texttt{tabled\_call/1}.}
  \label{fig:tabled_call}
  \redafterfig
\end{figure}

\begin{figure}
  \begin{center}
    \begin{minipage}{.8\linewidth}
\begin{lstlisting}[style=MyProlog]
new_answer :-
    answer_lookup_table(Vars, Ans),
    store_projection(Vars, ProjStore),
    &(   
      projstore_As(Ans, List_AnsProjStore),
      member(ProjStore_A, List_AnsProjStore),
      answer_compare(ProjStore, ProjStore_A, Res),
      &(   
          Res == `=<'
       ;  
          Res ==  `>',
          remove_answer(ProjStore_A), 
          fail   
      &), !
    ;                                   
      save_answer(Ans, ProjStore)
    &), !,
    fail.

new_answer :- 
    complete.
\end{lstlisting}
    \end{minipage}
  \end{center}
  \redbefcapt
  \caption{Implementation of \texttt{new\_answer/0}.}
  \label{fig:new_answer}
  \redafterfig
\end{figure}

\subsubsection{Execution Flow}
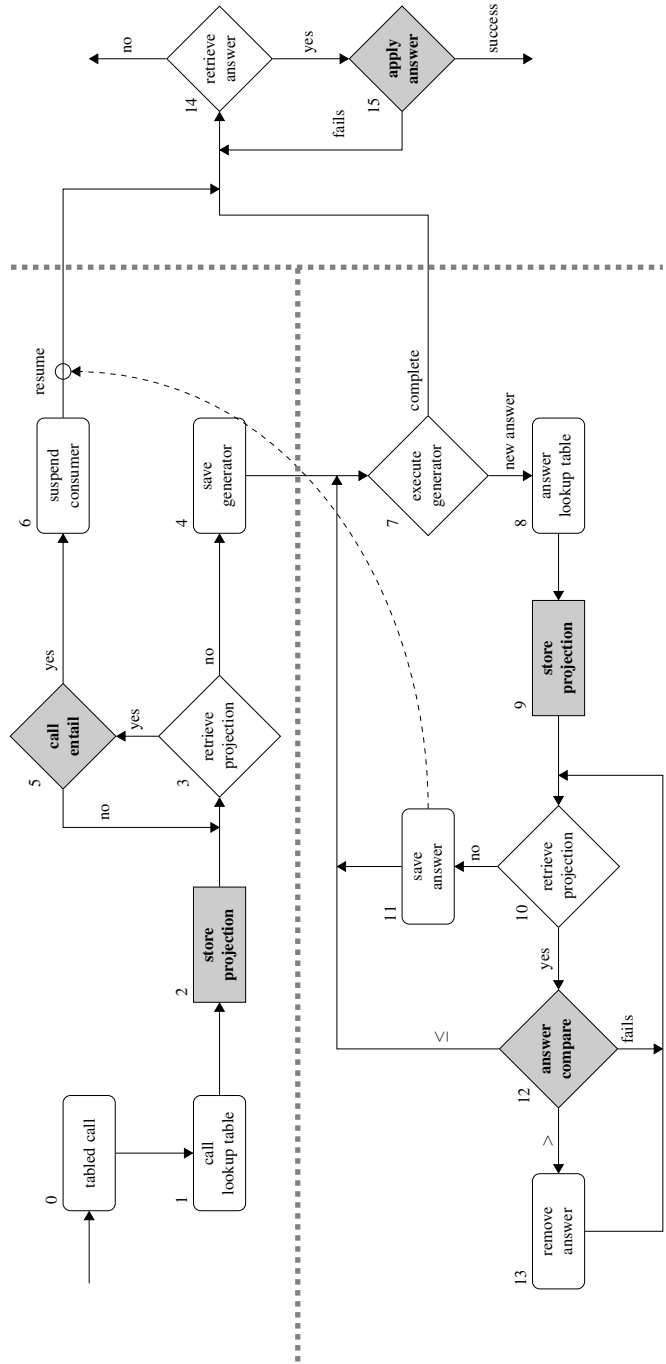
\begin{figure}
  \vspace*{-1cm}
  \begin{turn}{90}
    \begin{minipage}{1.8\textwidth}
      \begin{center}
  \resizebox{0.8\textwidth}{!} {%
\begin{tikzpicture}[node distance=2.5cm, transform shape]
\tikzstyle{rec} = [rectangle, rounded corners, minimum width=2cm, minimum height=1cm, text centered, text width=2cm, draw=black]
\tikzstyle{largerec} = [rectangle, rounded corners, minimum width=4cm, minimum height=1cm, text centered, text width=4cm, draw=black]
\tikzstyle{diam} = [diamond, minimum width=1cm, minimum height=1cm, text centered, text width=1cm, draw=black]
\tikzstyle{diamclp} = [diamond, minimum width=1cm, minimum height=1cm, text centered, text width=1cm, draw=black, fill=black!20]
\tikzstyle{clp} = [rectangle, minimum width=2cm, minimum height=1cm, text centered, text width=2cm, draw=black, fill=black!20]
\tikzstyle{largeclp} = [rectangle, minimum width=3cm, minimum height=1cm, text centered, text width=3cm, draw=black, fill=black!20]
\tikzstyle{flecha} = [draw,->, >=triangle 60,line width=0.01cm]
\tikzstyle{arrow} = [thick,->,>=triangle 60, line width=0.01cm]

\tikzset{L/.style args={[#1]#2}{label={[xshift=-0.9cm]#2}}}
\tikzset{D/.style args={[#1]#2}{label={[xshift=-0.9cm, yshift=-0.7cm]#2}}}

\node (0) [L={[-1]0}, rec, xshift=1cm] {tabled call};
\node (1) [L={[-1]1}, rec, below of=0] {call\\ lookup table};
\node (2) [L={[-1]2}, clp, right of=1, xshift=1.5cm] {\textbf{store\\ projection}};
\node (3) [D={[-1]3}, diam, right of=2, xshift=1.5cm, text width=1.25cm] {retrieve\\ projection};

\node (5) [L={[-1]4}, rec, right of=3, xshift=2.5cm] {save\\ generator};

\node (6) [D={[-1]5}, diamclp, above of=3, yshift=0.5cm] {\textbf{call\\ entail}};
\node (7) [L={[-1]6}, rec, right of=6, xshift=2.5cm] {suspend\\ consumer};

\node (8) [D={[-1]7}, diam, below of=5, yshift=-1.5cm, text width=1.25cm] {execute\\ generator};
\node (9) [L={[-1]8}, rec, below of=8] {answer\\ lookup table};
\node (10) [L={[-1]9}, clp,left of=9, xshift=-1cm] {\textbf{store\\ projection}};
\node (11) [D={[-1]10}, diam, left of=10, xshift=-1.5cm, text width=1.25cm] {retrieve\\ projection};
\node (12) [L={[-1]11}, rec, above of=11] {save\\ answer};
\node (13) [D={[-1]12}, diamclp,left of=11, xshift=-1cm, text width=1.25cm] {\textbf{answer\\ compare}};
\node (14) [L={[-1]13}, rec, left of=13, xshift=-1cm] {remove\\ answer};

\node (16) [D={[-1]14}, diam, right of=7, xshift=5.5cm, yshift=-3cm] {retrieve\\ answer};
\node (17) [D={[-1]15}, diamclp, below of=16, yshift=-1cm] {\textbf{apply\\ answer}};

\draw [flecha, <-] (0) -- +(-2.5,0);
\draw [arrow] (0) -- (1);
\draw [arrow] (1) -- (2);
\draw [arrow] (2) -- (3);
\draw [arrow] (3) -- node [anchor=south, xshift=-1cm] {no} (5);
\draw [arrow] (3) -- node [anchor=west] {yes} (6);
\draw [flecha, ->] (6) -| node [anchor=south, xshift=0.3cm, yshift=-1cm] {no} +(-1.75,-3);
\draw [arrow] (6) -- node [anchor=south, xshift=-1cm] {yes} (7);

\draw [arrow] (5) -- (8);
\draw [arrow] (8) -- node [anchor=west] {new answer} (9);
\draw [arrow] (9) -- (10);
\draw [arrow] (10) -- (11);
\draw [arrow] (11) -- node [anchor=west] {no} (12);
\draw [flecha, ->] (12) -- +(0,1.75);

\draw [arrow] (11) -- node [anchor=south] {yes} (13);
\draw [arrow] (13) -- node [anchor=south] {$>$} (14);
\draw [flecha, ->] (13) |- node [anchor=west, yshift=-2cm] {$\leq$} +(11,4.25);
\draw [flecha, ->] (13) -- node [anchor=west, yshift=0.25cm] {fails} +(0,-2) ;
\draw [flecha, ->] (14) |- +(8.75,-2) -- +(8.75,0);

\node (x) [circle, draw, minimum width=0.2cm, right of=7, xshift=-0.5cm] {};
\draw [flecha, ->, dashed]   (12) to[out=0,in=-90] (x);

\draw [flecha, ->] (7) -| node [anchor=south, xshift=-3.5cm, yshift=0.25cm] {resume} +(5.5,-3);
\draw [flecha] (8) -- node [anchor=south, xshift=-1.25cm] {complete} +(5,0) |- (16);

\draw [arrow] (16) -- node [anchor=west] {yes} (17);
\draw [flecha, ->] (16) -- node [anchor=west] {no} +(0,2.5);
\draw [flecha, ->] (17) -| node [anchor=south, yshift=1cm, xshift=0.5cm] {fails} +(-1.75,3.5);
\draw [flecha, ->] (17) -- node [anchor=west] {success} +(0,-2.5);

\draw [line width=0.1cm, gray, dashed] +(-3,-4) -- +(18,-4);
\draw [line width=0.1cm, gray, dashed] +(18,1.5) -- +(18,-11);

\end{tikzpicture}
}%
\end{center}
    \end{minipage}
  \end{turn}
  \vspace*{-1.5cm}
\caption{Flowchart of the execution algorithm of \mtclp.}
\label{fig:flow}
  \redafterfig
\end{figure}

Fig.~\ref{fig:flow} shows a (simplified) flowchart to illustrate how the
execution of a tabled call proceeds.  
The predicates of the interface to the constraint solver have a grey
background.
We explain next the steps of an execution, using the labels in the nodes.

\begin{enumerate}[leftmargin=*]
  \setcounter{enumi}{-1}

\item A call to a tabled predicate \texttt{Call} starts the tabled
  execution invoking \texttt{tabled\_call/1}, which takes the
  control of the execution. 
\item \texttt{call\_lookup\_table/3} returns in \texttt{Gen} a
  reference to the trie leaf 
  corresponding to the current call pattern \texttt{Call} 
  and in \texttt{Vars} a list with the constrained variables of
  \texttt{Call}.
\item The tabling engine calls \texttt{store\_projection/2}, which
  returns in \texttt{ProjStore} the projection onto \texttt{Vars} of
  the current constraint store.
\item The tabling engine uses \texttt{member/2} to retrieve in
  \mbox{\texttt{ProjStore\_G}} the projected constraint stores from
  the list of frames associated to \texttt{Gen}.  If it succeeds, the
  execution continues in step 5.  If it fails, it may be because
  \texttt{Gen} is the first occurrence of this call pattern, or
  because it does not entail any of the previous generators (and it
  is therefore a new generator).
\item The tabling engine calls \texttt{save\_generator/3} to add a new
  frame to \texttt{Gen}, identifying the new call as a generator.  The
  projected store \texttt{ProjStore} is saved in this new frame and the
  answer table and the consumer list are initialized.  From this point
  on, the generator is identified by \goaltt{Gen}{ProjStore_G} and the
  execution continues in step 7.
\item The constraint solver checks if the current store
  \texttt{ProjStore} entails the retrieved projected constraint
  store \texttt{ProjStore\_G} using \texttt{call\_entail/2}.  In that
  case, \texttt{Call} is suspended in step 6. Otherwise, the tabling
  engine tries to retrieve another projected constraint store in step
  3.
\item If the generator is not complete, the tabling engine suspends
  the execution of \texttt{Call} with \texttt{suspend\_consumer/1} and
  adds \texttt{Call} to the list of consumers of the generator.
  Execution then continues by backtracking over the youngest
  generator.  Otherwise, \texttt{Call} continues the execution in
  \mbox{step 14}.  A suspended consumer is resumed when its generator
  produces new answers, and also continues in step 14.
\item The generator \goaltt{Gen}{ProjStore\_G} is executed with
  \mbox{\texttt{execute\_generator/2}}, which calls the renamed tabled
  predicate, and its reference is pushed onto the PTCP stack.
  If the execution reaches the end of a clause,  a new
  answer has been found and \texttt{new\_answer/0} 
  continues the execution in step 8. %
\item This is the entry point for \texttt{new\_answer/0}. The tabling
  engine calls \mbox{\texttt{answer\_lookup\_table/2}}, which
  retrieves a reference to the generator in execution from the PTCP
  stack.  A reference to the Herbrand constraints of the current answer in the
  generator's answer table is returned in \texttt{Ans}, and the list of
  variables from the call that are now/still constrained is returned
  in \texttt{Vars}. %
\item The tabling engine invokes \texttt{store\_projection/2}.  This
  returns in \texttt{ProjStore} the projection of the current
  constraint store onto the constrained variables of the answer,
  \texttt{Vars}.
\item The tabling engine retrieves from \texttt{Ans} the list of
  projected constraint stores in \texttt{List\_AnsProjStore} and calls
  \texttt{member/2} to return the stores one at a time in
  \mbox{\texttt{ProjStore\_A}}.
  If it succeeds,
  the execution continues in step 12; otherwise,
  it %
  continues in step 11.  Failure can happen because all projected
  constraint stores were already retrieved from
  \texttt{List\_AnsProjStore} or because \texttt{Ans} is the first
  answer with these Herbrand constraints.
\item The tabling engine adds \texttt{ProjStore} to the list of
  projected constraint stores (\texttt{List\_AnsProjStore}) of the
  corresponding \texttt{Ans} with \texttt{save\_answer/2}, and
  resumes one by one the consumers of the current generator which
  were suspended in step 6. Since \texttt{new\_answer/0} always fails,
  the execution backtracks to complete the execution of the generator
  (\mbox{step 7}).
\item The constraint solver checks if the current store
  \texttt{ProjStore} entails the retrieved projected
  constraint store \texttt{ProjStore\_A} using
  \texttt{answer\_compare/3}.  If this is the case, it returns
  \texttt{Res = `$=<$'}, which makes \texttt{new\_answer/0} discard
  the current answer, and the generator is re-executed in step 7.  If
  \texttt{ProjStore} is entailed by \texttt{ProjStore\_A} and they are not
  equal, it returns \texttt{Res = `$>$'} and \texttt{ProjStore\_A} is
  removed in step 13.
  Otherwise, it fails and the execution continues in step 10, where
  the tabling engine tries to retrieve another projected constraint
  store.
\item The tabling engine marks the more particular answer as removed using
  \texttt{remove\_answer/1}.  Then the execution continues in step 10.
\item Once the generator has exhausted all the answers and does not
  have more dependencies, it is marked as complete using
  \mbox{\texttt{complete/0}} and the generator's reference is popped from the
  PTCP stack. %
  The tabling engine retrieves answers
  \texttt{\goaltt{Ans}{ProjStore\_A}} from the generator
  \texttt{\goaltt{Gen}{ProjStore\_G}} using \mbox{\texttt{member/2}}.
  If it succeeds and the answer is not marked as removed,
  the answer will be applied in step 15. Otherwise, the execution
  backtracks to retrieve another answer.
\item Applying the  Herbrand constraints \texttt{Ans} always succeeds,
  because the generator and its consumers have the same call pattern.
  Then the constraint solver adds the projected constraint store of
  the answer to the current constraint store with
  \texttt{apply\_answer/2}, and checks if the resulting constraint
  store is consistent. If so, execution continues; otherwise the
  execution goes back to \mbox{step 14}.
\end{enumerate}

\begin{figure}
  \begin{center}
  \begin{minipage}{.9\linewidth}
\begin{lstlisting}[style=MyProlog]
:- active_tclp.

store_projection(Vars, st(F,Proj) ) :-     
        clpqr_dump_constraints(Vars, F, Proj).

call_entail(st(F,Proj), st(FGen,ProjGen) ) :-    
        check_entailment(F, FGen, Proj, ProjGen).

answer_compare(st(F,Proj), st(FAns,ProjAns), =<) :-  
        check_entailment(F, FAns, Proj, ProjAns), !.
answer_compare(st(F,Proj), st(FAns,ProjAns),  >) :-  
        check_entailment(FAns, F, ProjAns, Proj).

apply_answer(Vars, st(FAns,ProjAns) ) :-         
        Vars = FAns, clpq_meta(ProjAns).

check_entailment(Vars1, Vars2, Proj1, Proj2) :-  
        Vars1 = Vars2, clpq_meta(Proj1), clpq_entailed(Proj2).
\end{lstlisting}
  \end{minipage}
\end{center}
  \redbefcapt
  \caption{The \mtclp\ interface for CLP(\Q) is a bridge to existing
    predicates.}
  \label{fig:int-clpq}
  \redafterfig
\end{figure}

In the \ref{sec:execution-dist3} of the supplementary material accompanying the paper
at the TPLP archive, we will walk through a step by step execution of
a program using the TCLP(\Q) interface described in
Section~\ref{sec:tclpq-int}.

\subsection{Implementation of the TCLP(\texorpdfstring{\Q}{Q}) Interface}
\label{sec:tclpq-int}

Fig.~\ref{fig:int-clpq} shows the interface for Holzbaur's CLP(\Q)
solver~\cite{holzbaur-clpqr} as an example of integration of a
constraint solver with \mtclp.
This CLP(\Q) implementation already provides most of the
functionality required by the tabling engine, and therefore the TCLP(\Q)
interface actually acts as a bridge to existing predicates.

A \mtclp\ constraint interface starts with the declaration
\mbox{\texttt{:-~active\_tclp.}}  It makes the compiler check which
interface predicates are available in order to adjust the program
transformation, and instructs the run-time to activate the TCLP
framework.
The functionality required by the interface is implemented as follows:

\begin{itemize}

\item \texttt{store\_projection(+Vars,-st(F,Proj))} calls the CLP(\Q)
  predicate \texttt{clpqr\_dump\_constraints(+Vars,-F,-Proj)} to perform
  the projection.  It returns in \texttt{Proj} the projection of the
  current store onto the list of variables \texttt{Vars}.  The
  variables in \texttt{Proj} are fresh and are contained in the list
  \texttt{F}, which follows the same order as those in \texttt{Vars} and
  can be used to restore the association between the variables in
  \texttt{Vars} and the constraints in \texttt{Proj}, as we said in
  Section~ \ref{sec:desig-gener-interf}.

\item \texttt{call\_entail(+st(F,Proj),+st(FGen,ProjGen))} calls the
  auxiliary predicate \texttt{check\_entailment(F, FGen, Proj,
    ProjGen)} which success if \texttt{Proj} $\sqsubseteq$
  \texttt{ProjGen}. First, \texttt{check\_entailment/4} unifies
  \texttt{F} and \texttt{FGen}, resp. the variables of the projection
  of the current store, and the variables of the generator's
  projection. Then, the CLP(\Q) predicate \texttt{clpq\_meta(+Proj)}
  makes \texttt{Proj} part of the current constraint store by
  executing it. This does not interact with the current store,
  because the variables in \texttt{F} and \texttt{FGen} are
  fresh. And finally, \texttt{clpq\_entailed(+ProjGen)} success if
  \texttt{ProjGen} is entailed by the current constraint store (i.e.,
  \texttt{Proj} $\sqsubseteq$ \texttt{ProjGen}).

\item \texttt{answer\_compare(+st(F,Proj),+st(FAns,ProjAns),Res)}
  calls the predicate \texttt{check\_entailment(F, FAns, Proj,
    ProjAns)} to check if \texttt{Proj} $\sqsubseteq$
  \texttt{ProjAns}. If it is the case, \textsf{answer\_compare/3}
  returns \texttt{=<} in \texttt{Res}. Otherwise, it calls
  \texttt{check\_entailment(FAns, F, ProjAns, Proj)} to check if
  \texttt{ProjAns} $\sqsubset$ \texttt{Proj}. If it is the case, it
  returns \texttt{>} in \texttt{Res}, otherwise it fails (i.e., there
  is no entailment in any direction).

\item \texttt{apply\_answer(+Vars,+st(FAns,ProjAns))} unifies
  \texttt{FAns}, the variables of \texttt{ProjAns} with \texttt{Vars},
  those in the pattern of the resumed call. Then, it uses the CLP(\Q)
  predicate \texttt{clpq\_meta(+ProjAns)} to add the answer constraint
  store \texttt{ProjAns} to the current constraint store.  If the
  resulting constraint store is consistent, execution continues, and
  it fails otherwise.  

\end{itemize}

The TCLP interface for CLP(\R) is similar to that  of CLP(\Q).
CLP(\R) uses floating-point numbers and its performance is better than
that of CLP(\Q), which uses exact fractions.  However, floating-point
rounding errors make CLP(\R) (and TCLP(\R)) inappropriate for some
applications, as entailment is unsound and therefore termination can
be compromised.

\subsection{Two-Step Projection} 
\label{sec:improvementss}

The design we have presented strives for simplicity.  There is however
an improvement that can be used to obtain more performance / reduce
memory usage,
at the cost of a slightly more complex design.  We present it now, with the
understanding that it does not change the general ideas we have
presented so far.

\texttt{store\_projection/2} is usually the most expensive operation
in the TCLP interface, but it is only mandatory when a call is a
generator, which we can determine from entailment
checking.\footnote{For efficiency, we can check entailment using
  the current constraint store $A$ instead of its projection onto a set of
  variables $S$ because
  \mbox{$A \sqsubseteq B \Longleftrightarrow Proj(S, A) \sqsubseteq
    B$, where $S=vars(B)$, as in our case}.}  We have however
placed \texttt{store\_projection/2} before entailment checking because
constraint solvers can often use the projection operation to compute
some information
needed by the entailment check.  Instead of recomputing this
information,
the projection is divided in two parts: an initial operation
\texttt{early\_call\_projection(+Vars, -EarlyProj)}, executed before the
entailment phase, that
returns in \texttt{EarlyProj} the information needed to check
entailment, and a second operation
\mbox{\texttt{final\_call\_projection(+Vars, +EarlyProj, -ProjStore)}}
that is executed after the entailment phase if the entailment check
fails (and the call would then be a generator).  If it is executed,
this operation returns the projected constraint store in
\texttt{ProjStore} using the information in \texttt{EarlyProj}.

For symmetry, a similar mechanism is used with answers.  Instead of
using \texttt{store\_projection/2} (step 10 in Fig.~\ref{fig:flow}),
two specialized versions are expected used:
\texttt{early\_ans\_projection/2},
is called before the answer entailment check and
\mbox{\texttt{final\_ans\_projection/3}}, called after the answer
entailment check.

\begin{example}
  The TCLP(\Q) interface in Fig.~\ref{fig:int-clpq} uses
  {\normalfont \texttt{store\_projection/2}} to project the constraint
  store of every new call.  But, since {\normalfont
    \texttt{clpq\_entailed/1}} does not need the projection of the
  current constraint store to check entailment w.r.t.\ the projected
  constraint store of a previous generator, the execution of the
  projection can be delayed:

\medskip
  {\normalfont
\begin{lstlisting}[style=MyProlog]
early_call_projection(Vars, st(Vars,_)).
call_entail(st(Vars,_), st(FGen,ProjGen)) :-        
                            Vars = FGen, clpq_entailed(ProjGen).
final_call_projection(_,st(Vars,_), st(F,Proj)) :-  
                          clpqr_dump_constraints(Vars, F, Proj).
\end{lstlisting}
}

\end{example}

The performance impact of implementing the \emph{Two-Step} projection
is evaluated in Section~\ref{sec:eval-two-step}, using the TCLP(\Q)
interface.

\section{Other TCLP Interfaces}
\label{sec:other-tclp-interfaces}

The design we presented brings more flexibility to a system with
tabled constraints at a reasonable cost in implementation effort.  To
support this claim we present the implementation of the TCLP interface
for a couple of additional solvers: a constraint solver for difference
constraints (Section~\ref{sec:design-diff-constr}) completely written
in C and ported from~\cite{chico-tclp-flops2012}, and a solver for
constraints over finite lattices
(Section~\ref{sec:design-constr-over-lat}).

\subsection{Difference Constraints}
\label{sec:design-diff-constr}

Difference constraints CLP(\D) is a simple but relatively powerful
constraint system whose constraints are generated from the set
$\pazocal{C}_{\text{\D}} = \{X - Y \leq d: X, Y, d \in \mathbb{Z}\}$, where $X$ and $Y$ are variables, and $d$ is a constant.

A system of difference constraints can be modeled with a weighted graph, and
it is satisfiable if there are no cycles with negative weight.  A
solver for this constraint system can be based on shortest-path
algorithms~\cite{FrigioniMN98} where
the constraint store is represented as an $n \times n$
matrix \texttt{A} of distances.  
The projection of a constraint store \texttt{A} onto a set of
variables \texttt{V} extracts 
a sub-matrix \texttt{A'} containing all pairs
\mbox{$(v_1,\, v_2)\ \textrm{s.t.}\ v_1,\, v_2\ \in\ V$}.  For
efficiency, a
projection can be represented as a vector of length $|V|$ containing
the index of each $v_i$ in \texttt{A}.  For example, if the indexes in
\texttt{A} of the variables \texttt{[X,Y,Z,T,W]} are
\texttt{(1,2,3,4,5)}, the projection onto the set of variables
\texttt{[T,X,Y]} is represented with the vector \texttt{(4,1,2)}.  The
implementation uses attributed variables to map Prolog variables onto
their representation in the matrix by having as attribute the index of each variable in
the matrix.  Therefore, calculating projection is fast.

The TCLP(\D) interface showcases that, as we mentioned in
Section~\ref{sec:desig-gener-interf}, the representation of the
projected constraint store depends on the constraint solver.  In this
case, the projected constraint store is represented by a triple
\texttt{st(Id,\,Ln,\,Proj)} where \texttt{Id} is the memory address of
the vector with the indexes of the constrained variables of the
call / answer, \texttt{Ln} is its length (the number of constrained
variables), and \texttt{Proj} is the memory address of a copy of the
sub-matrix which represents the projected constraint store. The
indexes of the vector \texttt{Id} follow the same order as the
variables in \texttt{Vars} and are used to restore the association
between \texttt{Vars} and \texttt{Proj} when they have to be compared or
applied.

CLP(\D) checks entailment using
\texttt{clpdiff\_entailed((Id,Ln),ProjGen)} and
\texttt{clpqdiff\_entails((Id,Ln),ProjGen)}, where \texttt{Id} and
\texttt{Ln} identify the position of the variables in the matrix
\texttt{A} (the current constraint store), and \texttt{ProjGen} is the
memory address of a sub-matrix which represent the projection of a
previous generator. Note that the indexes of the sub-matrix from $1$
to $n$ follows the order of the indexes in \texttt{Id}, i.e., the
$k$ column /row of the sub-matrix correspond to the variable
identified by the $k$ index in \texttt{Id}.

Therefore, the TCLP(\D) interface increases performance and reduces
memory footprint using the \emph{Two-Step} projection
(Section~\ref{sec:improvementss}) because it only makes a copy of the
sub-matrix \texttt{Proj} when the entailment phase fails and the
current call / answer becomes a generator / new answer. See below the
implementation of the projection and answer comparison operations using
the \emph{Two-Step} projection in the answer entailment check:

\medskip
  
  {\normalfont
\begin{lstlisting}[style=MyProlog]
early_ans_projection(Vars, st(Id,Ln,_) ) :-  
                               diff_project_index(Vars,(Id,Ln)).
answer_compare(st(Id,Ln,_), st(_,_,ProjAns), =<) :- 
                              diff_entailed((Id,Ln),ProjAns), !.
answer_compare(st(Id,Ln,_), st(_,_,ProjAns),  >) :- 
                                  diff_entails((Id,Ln),ProjAns).
final_ans_projection(_, st(Id,Ln,_), st(Id,Ln,Proj) ) :- 
                                  diff_projection((Id,Ln),Proj).
\end{lstlisting}
  }

\subsection{Constraints over Finite Lattices}
\label{sec:design-constr-over-lat}

A lattice is a triple {\sf ($\mathbb{S}$, $\sqcup$, $\sqcap$)} where
$\mathbb{S}$ is a set of points and {\sf join ($\sqcup$)} and {\sf meet
  ($\sqcap$)} are two internal operations that follow the commutative,
associative and absorption laws.
\textsf{($\mathbb{S}$,~$\sqsubseteq$)} is a poset where $\forall a,b
\in \mathbb{S}\ .\ a \sqsubseteq b\ \mathsf{if}\ a = a \sqcap b\
\mathsf{or}\ b = a \sqcup b $ and $\exists \bot, \top \in \mathbb{S}$
such that $\forall\ a \in \mathbb{S}\ .\ \bot \sqsubseteq a \sqsubseteq
\top$.

In the system of constraints over finite lattices \texttt{CLP(\Lat)},
the constraints between points in the lattice arise from (1) the
topological relationship of the lattice elements and (2) any
additional operations between the elements in the lattice.  These two
classes of constraints are handled by two different layers.

The external layer is concerned with the lattice topology and
implements the constraint \mbox{$Y \sqsubseteq X$} with
$X,Y \in \mathbb{S}$
and the projection operation for variable elimination using Fourier's
algorithm~\cite{intro_constraints_stuckey}: the projection of
$X \sqsubseteq d \land Y \sqsubseteq X$ onto $Y$ is
\mbox{$Y \sqsubseteq d$}.  This layer provides 
entailment checking and the operation to add a projected constraint
store to the current constraint store.

Further constraints on variables can be imposed by relationships
derived from internal operations other than those in the lattice.  
Compare, for example, $Y \sqsubseteq X$ with $Y
  \sqsubseteq X \land Y = X \oplus X$ for some operation
$\oplus$ among elements of the lattice: the additional information can
be helpful to simplify (or prove inconsistent) the constraint
store. In the lattice solver, a second layer implements these
additional operations (if they exist) and communicates with the
topology-related layer.  

We have used this solver to implement a constraint tabling-based
abstract interpreter (Section~\ref{sec:constr-over-lat}), where the points
of the lattice are the elements of the abstract domain.
The lattice implementation provides at least the operators $\sqcup$
and $\sqcap$ and 
the operations among the elements of the lattice, which are the
counterparts of the operations in the concrete
domain, as described above. 

\section{Experimental Evaluation}
\label{sec:exampl-constr-solv}

In this section we evaluate the performance of our framework using the
four constraint systems and interfaces we have summarily described
(\Q, \R, \D and \Lat).

In Section~\ref{sec:performance}, we quantify the performance benefits
of TCLP versus LP, tabling, and CLP using the \texttt{dist/3}
(Fig.~\ref{fig:dist-graph}) program presented in
section~\ref{sec:motivation} with the TCLP(\Q) interface.
Then we explore the impact and advantages of a more flexible modular
framework. On the one hand, in Section~\ref{sec:diff-constr} we
evaluate the performance impact of the increased overhead w.r.t.\
previous implementations with less flexibility (i.e., the previous
TCLP implementation of~\cite{chico-tclp-flops2012}) and, on the other
hand, in Section~\ref{sec:eval-management} we evaluate the benefits of
a more complete answer management strategy.

In Section~\ref{sec:eval-two-step}, we evaluate the performance
benefits of the \emph{Two Step} projection using TCLP(\Q). These
benefits are due to the reduction in the number of projections executed
during the evaluation.
In Section~\ref{sec:constr-over-lat}, we use tabling and the new
TCLP(\Lat) interface to implement a simple abstract interpreter, which
we benchmark. 

In \ref{sec:tclpd-tclpq-tclpr} of the supplementary material
accompanying the paper at the TPLP archive, we compare the
expressiveness and performance of the TCLP(\D), TCLP(\R), and TCLP(\Q)
interfaces. In this case, the expressiveness of CLP(\R/\Q) comes with
an overhead (which is higher in CLP(\Q) due to its higher precision),
but in certain problems this expressiveness can bring greats benefits
using TCLP (additionally, in some problems the precision could be
determinant).

The \mtclp\ framework presented in this paper is implemented in Ciao
Prolog.  The benchmarks and a Ciao Prolog distribution including the
libraries and interfaces presented in this paper are available at
\texttt{\url{http://www.cliplab.org/papers/tplp2018-tclp/}}.\footnote{Stable
  versions of Ciao Prolog are available at
  \texttt{\url{http://www.ciao-lang.org}}.  However, The libraries and
  interfaces are still in development, and they are not fully
  available yet in the stable versions.}  All the experiments were
performed on a Mac OS-X 10.9.5 machine with a 2.66 GHz Intel Core 2
Duo processor.  Times are given in milliseconds.%

\subsection{Absolute Performance of TCLP vs.\ LP vs.\ Tabling vs.\ CLP}
\label{sec:performance}
\redbefsec  

Let us recall Table~\ref{tab:distance-summary}, where we used the
\texttt{dist/3} program (Fig.~\ref{fig:dist-graph}) to support the use
of TCLP due to its better termination behavior.   We now
want to check whether, for those cases where LP or CLP also terminate,
the performance of TCLP is competitive and for those cases where only
TCLP terminates, whether its performance is reasonable.  We have used
a graph of 35 nodes without cycles (775 edges) and a graph of 49 nodes
with cycles (785 edges) and timed the results --- see
Table~\ref{tab:distance}.

As we already saw in Table~\ref{tab:distance-summary}, TCLP not only
terminates in all cases, but it is also faster than the rest of the
frameworks due to the combination of tabling (which avoids entering
loops and caches intermediate results) and constraint solving.
It also suggests, in line with the experience in tabling, that
left-recursive implementations are usually  faster and
preferable, as they avoid work by ``suspending first'' and reusing
answers when they are ready.

\begin{table}[tb]
  \setlength{\tabcolsep}{0.75em}
  \begin{tabular}{@{\extracolsep{0pt}}|l|r|r|r|r|c|}
    \cline{1-6}
                         & LP       & CLP(\Q)  & Tab    & \mtclp(\Q)          &  Graph\\ \cline{1-6}                             
    ~Left recursion      & --       & --       & 2311   & \textbf{1286}      &  \multirow{2}{*}{Without cycles} \\ 
    ~Right recursion     & > 5 min. &  5136    & 3672   & 2237              &  \\ \cline{1-6}                                  
    ~Left recursion      & --   & --      & --    & \textbf{742}     &  \multirow{2}{*}{With cycles}    \\ 
    ~Right recursion     & --   & 10992   & --    & 1776             &  \\ \cline{1-6} %
  \end{tabular}
  \redbefcapt
  \caption{Run time (ms) for \texttt{dist/3}. `--' means no termination.}
  \label{tab:distance} 
  \redafterfig
\end{table}

\subsection{The Cost of Modularity: \mtclp\ vs.\ Original TCLP}
\label{sec:diff-constr}

\begin{table}
    \setlength{\tabcolsep}{0.75em}
    \begin{tabular}{@{\extracolsep{0pt}}|l|r|r|r|}
      \cline{1-4}
      &  CLP(\D)     &  Orig.\ TCLP(\D) & \mtclp(\D)          \\[0.25em] \cline{1-4}
      \texttt{truckload(300)}~ & 40452    & \textbf{2903}          & 7268          \\ 
      \texttt{truckload(200)}  & 4179     & \textbf{1015}          & 2239          \\ 
      \texttt{truckload(100)}  & 145      & \textbf{140}           & 259           \\ \cline{1-4}
      %
      \texttt{step\_bound(30)}~  &  -     & 2657      & \textbf{1469}        \\ 
      \texttt{step\_bound(20)}   &  -     & 2170      & \textbf{1267}        \\ 
      \texttt{step\_bound(10)}   &  -     & 917       & \textbf{845}         \\ \cline{1-4}
    \end{tabular}
    \redbefcapt
    \caption{Performance comparison (ms) of CLP vs original TCLP vs \mtclp\ \\
      using \D for \texttt{truckload/4} and \texttt{step\_bound/4}.
      `--' means no termination.}
    \label{tab:bench-truck} 
    \redafterfig
\end{table}

The original TCLP implementation~\cite{chico-tclp-flops2012} was
deeply intertwined with the tabling engine and had a comparatively low
overhead.  Since it was done on the same platform as ours (Ciao
Prolog) and shares several components and low-level implementation
decisions, it seems a fair and adequate baseline to evaluate the
performance cost of the added modularity.  We will evaluate both
frameworks using exactly the same implementation of difference
constraints (Section~\ref{sec:design-diff-constr}) and two benchmarks:

\begin{description}
  \setlength{\itemsep}{0pt}%
  \setlength{\parskip}{0pt}%
\item [\mbox{\texttt{truckload(P, Load, Dest, Time)}}]
  \cite{bao2000,tchr2007}\textbf{:} it solves a \mbox{shipment} problem given a
  maximum \texttt{Load} for a truck, a destination \texttt{Dest}, and a list
  of packages to ship (1 to \texttt{P}.)  We set \texttt{P=30, Dest=chicago}
  and use \texttt{Load} as parameter to vary its
  complexity.
  \texttt{truckload/4} does not need tabling, but tabling speeds it up.

\item [\mbox{\texttt{step\_bound(Init, Dest, Steps,
      Limit)}}\textmd{:}] it is a left-recursive graph reachability
  program similar to \texttt{dist/3} that constrains the total number
  (\texttt{Limit}) of edge traversals. \texttt{step\_bound/4} needs
  tabling in the case of graphs with cycles, as it is the case of the
  graph we will use in this evaluation.
\end{description}

Table~\ref{tab:bench-truck} shows that \texttt{truckload/4} incurs a
nearly three-fold increase in execution time with respect to the
initial non-modular TCLP(\D) implementation.  This is mainly due to
the overhead of the control flow.  In the original implementation,
execution did not leave the level of C, as the tabling engine called
directly the constraint solver, also written in C.  However, in
\mtclp, the tabling engine (in C) calls the interface level (written
in Prolog), which calls back the constraint solver (in C).  The additional
overhead is the price we pay to make it much easier to plug in
additional constraint solvers, which in the original TCLP needed
ad-hoc, low level wiring.

However, \texttt{step\_bound/4} is less efficient in the original
TCLP(\D) implementation than in \mtclp, and cannot be executed in
CLP(\D) due to the cycles in the graph.  The reason behind this
improvement is the enhanced answer management strategy whose
implementation was made possible by our modular design.  We will
explore this point in the next section.

\subsection{Improved Answer Management Strategies}
\label{sec:eval-management}

The modular design of \mtclp\ makes it possible to implement more
easily hooks for internal operations.  In particular, the solver
interface can include the \texttt{answer\_compare/3} operation which
determines whether a new answer entails, is entailed by, or none of them,
some previous answer.  This can be used to decide whether to add or
not a new answer and remove or not an existing answer.
This is undoubtedly expensive in general, but as advanced in
Section~\ref{sec:desig-gener-interf}, it holds promise for improving
performance.
To validate this intuition, we executed again \texttt{truckload/4} and
\texttt{step\_bound/4} with TCLP(\D) under four different answer management
strategies:

\begin{description}%

\item[\ $\emptyset$\ ] all the answers are stored.

\item [$\leftarrow$\ ] checks if new answers entail previous
  answers.  If so, the new answer is discarded. That is the strategy
  used in the original TCLP framework.

\item [$\rightarrow$\ ] checks if new answers are entailed by previous
  answers.  If so, the previous answers are flagged as removed and
  ignored, and the new answer is stored.

\item [$\leftrightarrow$\ ]  checks entailment in both directions,
  discarding new answers and removing more particular answers.

\end{description}

\begin{table}[t]
  \setlength{\tabcolsep}{0.75em}
  \begin{tabular}{@{\extracolsep{0pt}}|l|r|r|r|r|}
    \cline{1-5}
    & \multicolumn{4}{c|}{\mtclp(\D)} \\[0.25em] \cline{2-5}
    &  \multicolumn{1}{c|}{$\emptyset$}& \multicolumn{1}{c|}{$\leftarrow$}  & \multicolumn{1}{c|}{$\rightarrow$}  & \multicolumn{1}{c|}{$\leftrightarrow$}  \\ \cline{1-5}
    \texttt{truckload(300)}   & 742039& 7806   & 7780  & \textbf{7268}   \\ 
    \texttt{truckload(200)}   & 11785 & 2314   & 2354  & \textbf{2239}   \\ 
    \texttt{truckload(100)}   & 300   & 263    &  263  &  \textbf{259}   \\ \cline{1-5}
    \texttt{step\_bound(30)}  & --    & 8450   &  --   & \textbf{1469}   \\ 
    \texttt{step\_bound(20)}  & --    & 6859   & 38107 & \textbf{1267}   \\ 
    \texttt{step\_bound(10)}  & --    & 2846   &  8879 &  \textbf{845}   \\ \cline{1-5}
  \end{tabular}
  \redbefcapt
  \caption{Run time (ms) comparison of answer management strategies \\
    using \mtclp(\D) for \texttt{truckload/4} and
    \texttt{step\_bound/4}.}
  \label{tab:bench-answer} 
    \redafterfig
\end{table}

The results in Table~\ref{tab:bench-answer} confirm that, in the
examples studied, and despite the cost of these strategies, the
computation time is reduced. The ``$\leftrightarrow$'' strategy proves
to be the best one, although by a small margin in some cases.

On the other hand, the worst strategy is `$\emptyset$', which for the
\texttt{truckload/4} program increases the runtime several  order of
magnitudes for large cases, while for the \texttt{step\_bound/4}
program the execution does not terminate because it runs out of memory
when trying to generate infinitely many repeated answers.
While \texttt{truckload/4} behaves similarly for the other strategies,
\texttt{step\_bound/4} varies drastically (i.e., `$\rightarrow$' runs
out of memory for the largest case).

Part of the reasons for these differences can be inferred from
Table~\ref{tab:bench-answer-analysis}, where, for each benchmark and
strategy, we show how many of the generated answers were saved,
discarded before being inserted, or removed after insertion. Note that
these results are independent of the constraint solver used (i.e.,
executing the same programs using CLP(\Q) or CLP(\R) instead of CLP(\D)
generates the same answers).

For \texttt{truckload/4}, the `$\rightarrow$' and the `$\leftarrow$'
strategies generate, discard / remove, and return a similar number of
answers, which means that their impact in execution time is not very
important. It is notwithstanding interesting to note that there is no
slowdown when using the more complex strategy, `$\leftrightarrow$'.
For \texttt{step\_bound/4}, `$\rightarrow$' generates many more
candidate answers than either of the other two --- in excess of one
million for \texttt{step\_bound(30)} --- but `$\leftarrow$' also
generates one order of magnitude more candidates answers than
`$\leftrightarrow$'. Note that the number of generated answers is not
always the same since, as discussed before, fewer saved answers wake
up fewer consumers.

\begin{table}
  \centering
  \setlength{\tabcolsep}{0.75em}
  \begin{tabular}{@{\extracolsep{0pt}}|c|l|@{\extracolsep{6pt}}r|r|r|r|}
\cline{1-6} 
    \multicolumn{1}{|p{1cm}|}{Answer \newline strategy}                   &                          & \# Sav. & \# Dis. & \# Rem.  & \# Ret.   \\ \cline{1-6}
\multirow{3}{*}{$\emptyset$ }   &\texttt{truckload(300)}~  & 448538   & 0            & 0           &  14999        \\
                                &\texttt{truckload(200)}   & 52349    & 0            & 0           &   1520        \\
                                &\texttt{truckload(100)}   &  2464    & 0            & 0           &     58        \\ \cline{1-6}
                                  %
                                  %
\multirow{6}{*}{$\leftarrow$ }  &\texttt{truckload(300)}~  & 67503    & 9971         & 0           &  41           \\ 
                                &\texttt{truckload(200)}   & 16456    & 1325         & 0           &  23           \\ 
                                &\texttt{truckload(100)}   &  1525    &   52         & 0           &   6           \\ \cline{2-6}
                                &\texttt{step\_bound(30)}~ & 44549    & 716826        & 0           & 252           \\ 
                                &\texttt{step\_bound(20)}  & 37548    & 599259        & 0           & 242           \\ 
                                &\texttt{step\_bound(10)}  & 15625    & 242351        & 0           & 165           \\ \cline{1-6}
                                  %
                                  %
\multirow{6}{*}{$\rightarrow$}  &\texttt{truckload(300)}~  & 75272    & 0         & 9460         &  30           \\ 
                                &\texttt{truckload(200)}   & 17568    & 0         & 1298         &  18           \\ 
                                &\texttt{truckload(100)}   &  1490    & 0         &   49         &   9           \\ \cline{2-6}
                                &\texttt{step\_bound(30)}~ &$>$1145690  & 0         &$>$1074071     & --        \\ 
                                &\texttt{step\_bound(20)}  &  946309  & 0         &  891078     & 441           \\ 
                                &\texttt{step\_bound(10)}  &  294728  & 0         &  276867     & 221           \\ \cline{1-6}
                                  %
                                  %
\multirow{6}{*}{$\leftrightarrow$ } &\texttt{truckload(300)}~   & 48524    & 6596         & 1740        &   5      \\ 
                                &\texttt{truckload(200)}   & 13550    & 1046         &  240        &   5           \\ 
                                &\texttt{truckload(100)}   &  1343    &   45         &   10        &   3           \\ \cline{2-6}
                                &\texttt{step\_bound(30)}~ &  9697    & 74528        & 4571        & 25           \\ 
                                &\texttt{step\_bound(20)}  &  9352    & 71658        & 4371        & 25           \\ 
                                &\texttt{step\_bound(10)}  &  6650    & 56935        & 3019        & 25           \\ \cline{1-6}
\end{tabular}
\redbefcapt
\caption{Number of answers: saved (\emph{Sav.}), discarded (\emph{Dis.}) , removed (\emph{Rem.}) and returned to the query (\emph{Ret.}) for each answer management strategy.}
\label{tab:bench-answer-analysis} 
\redafterfig
\end{table}

\begin{figure}
\redbeffig
\centering
\begin{minipage}{0.42\linewidth}
\begin{lstlisting}[style=MyProlog]
:- table sd/3.

sd(X,Y,D) :- 
     edge(X,Y,D0),
     D #>= D0.
sd(X,Y,D) :- 
     sd(X,Z,D1),
     edge(Z,Y,D2),
     D #>= D1+D2.
\end{lstlisting}
\end{minipage}
\begin{minipage}{0.25\linewidth}
\begin{tikzpicture}[node distance=1cm, transform shape]
\node (b) [circle, draw=black] {b};
\node (a) [circle, draw=black, below of=b, xshift=-1.25cm] {a};
\node (c) [circle, draw=black, below of=b, xshift=1.25cm] {c};
\node (d) [circle, draw=black, below of=a, xshift=0.8cm] {d};

\draw [draw, ->] (a) -- node [anchor=south, xshift=-0.25cm] {3.0} (b);
\draw [draw, ->] (b) -- node [anchor=south, xshift=0.25cm] {3.0} (c);
\draw [draw, ->] (a) -- node [anchor=north, xshift=-0.25cm] {1.0} (d);
\draw [draw, ->] (d) -- node [anchor=north, xshift=0.25cm] {2.0} (c);
\draw [draw, ->] (d) -- node [anchor=west] {1.0} (b);
\draw [draw, ->] (b) to[out=160, in=110] node [anchor=south, xshift=-0.25cm] {1.3} (a);
\end{tikzpicture}
\end{minipage}
\redbefcapt
\caption{\texttt{sd/3} -- a shortest-distance program.}
  \label{fig:sd}
\redafterfig
\end{figure}
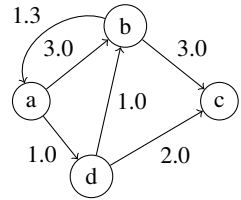

As an additional example of the usefulness of obtaining the most
general correct answer, Fig.~\ref{fig:sd} shows a graph and the
program \texttt{sd/3}, used in ~\cite{bao2000} to calculate the
``shortest distance'' between the nodes in the graph.
For a query such as \mbox{\texttt{?- sd(X,Y,Dist)}} the system
reported in~\cite{bao2000} returns a sequence of $n$ answers of the
form \texttt{Dist \#>= \mt{N_k}}. Each $\mt{N_k}$ is the current
achievable shortest distance from \texttt{X} to \texttt{Y}, such as
$\mt{N_1} > \dots > \mt{N_n}$, and the later $\mt{N_n}$ is the
shortest distance from \texttt{X} to \texttt{Y}. E.g., for the query
\texttt{?- sd(a,c,Dist)} it returns \texttt{Dist \#>= 6.0} and
\texttt{Dist \#>= 3.0}.
While using \mtclp\ under the `$\leftrightarrow$' strategy the
evaluation of the query \texttt{?- sd(a,c,Dist)} only returns the
answer \texttt{Dist \#>= 3.0} (the most general) which corresponds to
the tightest bound for the shortest distance between the nodes
\texttt{a} and \texttt{c}.%

\redbefsec
\subsection{Improved Two-Step Projection}
\redaftersec
\label{sec:eval-two-step}

The design of \mtclp\ makes it possible to postpone the projection
during the call / answer entailment phase using the \emph{Two-Step}
projection.  As we advanced in Section~\ref{sec:improvementss}, it
holds promise for performance improvements. To validate this
intuition, we use two benchmarks:
\begin{description}
  \setlength{\itemsep}{0pt}%
  \setlength{\parskip}{0pt}%
\item [\texttt{fib(N, F)}] the doubly recursive Fibonacci program run
  \emph{backwards}. It is well-known that tabling reduces
  \texttt{fib/2} complexity from exponential to linear.  In
  addition, CLP makes it possible to run exactly the same program
  \emph{backwards} to find the index of some Fibonacci number by
  generating a system of equations whose solution is the index of the
  given Fibonacci number (e.g., for the query \texttt{?- fib(N, 89)},
  the answer is \texttt{N=11}).  Under CLP, the size of this system of
  equations grows exponentially with the index of the Fibonacci
  number.  However, under TCLP, entailment makes redundant equations
  not to be added and solving them becomes less
  expensive. Additionally, entailment makes it possible to terminate
  (with failure) even when the query does not contain a non Fibonacci
  number, e.g., \texttt{fib(N,$10^{314}$)}.
\item [\texttt{dist(X, Y, D)}] the program already used in
  Section~\ref{sec:performance}.
\end{description}

We executed each of them with \mtclp(\Q) and the two designs for the
call projection we discussed earlier:

\begin{description}
\item [\emph{One-Step}:] The projection of the call is executed before
  the call entailment phase (Fig.~\ref{fig:int-clpq}). Note that
  CLP(\Q) does not need this projection to check entailment of the
  current call constraint store w.r.t.\ another constraint store.
\item [\emph{Two-Step}:] The projection of the call is executed using
  \texttt{final\_call\_projection/3} and, therefore, it is only
  executed when the call turns out to be a generator.
\end{description}

\begin{table}
  \setlength{\tabcolsep}{0.75em}
\begin{tabular}{@{\extracolsep{0pt}}|l|c|r|r|r|}
\cline{1-5}
  \multicolumn{2}{|l|}{Run-time (ms)}                              & \multicolumn{3}{c|}{\mtclp(\Q)}      \\ \cline{3-5} 
  \multicolumn{2}{|l|}{}                                           & One-Step   & Two-Step        & Ratio \\ \cline{1-5}
  \multicolumn{2}{|l|}{\texttt{fib(N,F$_{1500}$)}}                  & 206963     & \textbf{126461} & 1.63  \\ 
  \multicolumn{2}{|l|}{\texttt{fib(N,F$_{1000}$)}}                  & 89974      & \textbf{55183}  & 1.63  \\ 
  \multicolumn{2}{|l|}{\texttt{fib(N,F$_{500}$)}}                   & 22133      & \textbf{13612}  & 1.63  \\ 
  \multicolumn{2}{|l|}{\texttt{fib(N,$10^{314}$)}} & 205638     & \textbf{125670} & 1.63  \\ \cline{1-5}
  \multirow{2}{*}{\texttt{dist/3 right rec.}} & Without cycles & 2855       & \textbf{2506} & 1.14  \\ 
                                              & With cycles    & 2399       & \textbf{1850} & 1.30  \\ \cline{1-5}
  \multirow{2}{*}{\texttt{dist/3 left rec.}}  & Without cycles & 1436       & \textbf{1428} & 1.01   \\ 
                                              & With cycles    & 776        & \textbf{772}  & 1.01   \\ \cline{1-5}
\end{tabular}

\bigskip

\begin{tabular}{@{\extracolsep{0pt}}|l|c|r|r|r|}
\cline{1-5}
  \multicolumn{2}{|l|}{\# call projections}                    & \multicolumn{3}{c|}{\mtclp(\Q)} \\ \cline{3-5} 
  \multicolumn{2}{|l|}{}                                       & One-Step & Two-Step & Ratio \\ \cline{1-5}
  \multicolumn{2}{|l|}{\texttt{fib(N,F$_{1500}$)}}             & 1129497  &  \textbf{565500}  & 2.00 \\ 
  \multicolumn{2}{|l|}{\texttt{fib(N,F$_{1000}$)}}             &  502997  &  \textbf{252000}  & 2.00 \\ 
  \multicolumn{2}{|l|}{\texttt{fib(N,F$_{500}$)}}              &  126497  &   \textbf{63500}  & 1.99 \\ 
  \multicolumn{2}{|l|}{\texttt{fib(N,$10^{314}$)}}         & 1126499  &  \textbf{563252}  & 2.00 \\ \cline{1-5}
  \multirow{2}{*}{\texttt{dist/3 right rec.}} & Without cycles  & 1563 & \textbf{181} & 8.64 \\ 
                                          & With cycles     & 2144 & \textbf{443} & 4.84 \\ \cline{1-5}
  \multirow{2}{*}{\texttt{dist/3 left rec.}}  & Without cycles  & 2 & \textbf{1} & 2.00\\ 
                                          & With cycles     & 2 & \textbf{1} & 2.00\\ \cline{1-5}
\end{tabular}
  \redbefcapt
  \caption{Run time in ms (top) and number of call projections
    (bottom) \\ for each projection design, \emph{One-Step} and
    \emph{Two-Step}. \\ Note: $F_n$ is the $n^{th}$ Fibonacci number, and
    $10^{314}$ is not a Fibonacci number.}
  \label{tab:two_step_number} 
  \redafterfig
\end{table}

The results in Table~\ref{tab:two_step_number} (top) confirm that, in
the examples studied, the \emph{Two-Step} design reduces the
computation time, although only by a small margin in the case of
\texttt{dist/3} with left recursion. That is because, as we see in
Table~\ref{tab:two_step_number} (bottom), using the \emph{Two-Step}
projection, \texttt{dist/3} with left recursion executes the
projection of a call only once while using \emph{One-Step} it executes
the projection twice and therefore we only save the execution of one
projection. Since they are executed early in the
evaluation, the constraint store is small and their execution is
faster than in the case of \texttt{dist/3} with right recursion. Note
that using the \emph{Two-Step} projection, \texttt{dist/3} with right
recursion executes up to 8 times fewer call projections, and as
consequence its execution has better performance.

On the other hand, \texttt{fib/2} reduces drastically the computation
time using \emph{Two-Step} projection because, during the execution,
call entailment is checked many times (although the ratio of
\emph{useless} projection is similar to that of \texttt{dist/3} with
left recursion). Note that the Fibonacci number $F_{1500}$ and
$10^{314}$ have the same size (315 digits) and the run time / number
of projections for \texttt{fib(N,F$_{1500}$)} and for
\texttt{fib(N,$10^{314}$)} are similar. That is because the work
needed to find the index of a Fibonacci number is similar to the work
needed to confirm whether a number is or not a Fibonacci number.

\redbefsec
\subsection{Abstract Interpretation: Tabling vs.\ TCLP(\texorpdfstring{\Lat}{Lat})}
\redaftersec
\label{sec:constr-over-lat}

\begin{figure}
  \resizebox{0.4\textwidth}{!}{
    \begin{tikzpicture}[node distance=1.1cm]
      \tikzstyle{txt} = [ text centered, text width=1cm]
      \tikzstyle{sline} = [draw, ->, >=stealth]
      \node (top) [txt] {\textsf{top}};
      \node (num) [txt, below of=top] {\textsf{num}};
      \node (no-zero) [txt, below of=num] {$\mathsf{\pm}$};
      \node (neg+z) [txt, left of=no-zero] {$\mathsf{0^{-}}$};
      \node (pos+z) [txt, right of=no-zero] {$\mathsf{0^{+}}$};
      \node (zero) [txt, below of=no-zero] {\textsf{0}};
      \node (neg) [txt, left of=zero] {$\mathsf{-}$};
      \node (pos) [txt, right of=zero] {$\mathsf{+}$};
      \node (bottom) [txt, below of=zero] {\textsf{bottom}};
      \node (var) [txt, left of=neg+z, yshift=-0.7cm] {\textsf{var}};
      \node (str) [txt, right of=pos+z, yshift=-0.7cm] {\textsf{str}};
      \node (atom) [txt, right of=str] {\textsf{atom}};
      \draw [sline] (no-zero) -- (num);
      \draw [sline] (neg+z) -- (num);
      \draw [sline] (pos+z) -- (num);
      \draw [sline] (neg) -- (neg+z);
      \draw [sline] (zero) -- (neg+z);
      \draw [sline] (pos) -- (pos+z);
      \draw [sline] (zero) -- (pos+z);
      \draw [sline] (neg) -- (no-zero);
      \draw [sline] (pos) -- (no-zero);
      \draw [sline] (bottom) -- (neg);
      \draw [sline] (bottom) -- (zero);
      \draw [sline] (bottom) -- (pos);
      \draw [sline] (bottom) -- (var);
      \draw [sline] (bottom) -- (str);
      \draw [sline] (bottom) -- (atom);
      \draw [sline] (var) -- (top);
      \draw [sline] (str) -- (top);
      \draw [sline] (atom) -- (top);
      \draw [sline] (num) -- (top);
    \end{tikzpicture}
  }
  \redbefcapt
  \caption{\emph{Signs} abstract domain.}
  \label{fig:absdomain}
  \redafterfig
\end{figure}
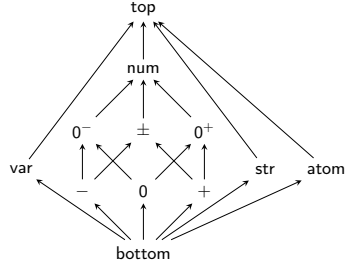

We compare here tabling and TCLP using two versions of a simple
abstract interpreter~\cite{Cousot77}.
The interpreter executes the programs to be analyzed on an abstract
domain, collecting the possible values at every point until a
fixpoint is reached.  The result of the execution is a safe
approximation of the run-time values of the variables in the concrete
domain.  The abstract domain we have used in this example is the
\emph{signs} abstract domain (Fig.~\ref{fig:absdomain}).

The two versions of the abstract interpreter we have used are:

\begin{description}
\item[Tabling] This version is a simple abstract interpreter written
  using tabling.  This ensures termination, as the abstract domain is
  finite.
\item[TCLP] This version is based on the previous abstract
  interpreter, but it uses the TCLP(\Lat) constraint solver interface
  (Section~\ref{sec:design-constr-over-lat}) to operate on the
  abstract domain and set up constraints over the variables. The main
  differences with the \emph{tabling} version is that TCLP uses
  constraint entailment instead of variant checking for loop detection
  and, therefore, it can also use the answers to more general goals to
  avoid computing more particular goals.
\end{description}

We applied our abstract interpreter to two programs:

\begin{description}
\item [\textbf{\texttt{takeuchi/m}}] (Fig.~\ref{fig:takeuchi}): a Prolog
implementation of the $n$-dimensional generalization of the Takeuchi
function~\cite{knuth1991textbook}.  
The program is
parametric on the number of input arguments $n$ and it returns the
 result in its last argument.

\item [\textbf{\texttt{sentinel/m}}] (Fig.~\ref{fig:sentinel}): a variant of a
synthetic program presented in~\cite{genaim2001worst}.  It
receives as input its first argument (the $\mt{Sentinel}$) and the next $n$
arguments $\mt{A_1, \dots, A_n}$ are a ring-ordered\footnote{I.e., there is
  a $j$ such that $\mt{A_j < A_{j+1},\  A_{j+1} < A_{j+2}},\  \ldots,\  \mt{A_{n} <
  A_1,\  A_1 < A_{2}},\  \ldots,\  \mt{A_{j-2} < A_{j-1}}$.} series of numbers.
The outputs are the arguments $\mt{B_1}, \ldots, \mt{B_n}$, which correspond to a
circular shift of $\mt{A_1}, \ldots, \mt{A_n}$ such that on success $\mt{B_i} <
\mt{B_{i+1}}$ for all $i < n$ and: if $\mt{Sentinel = 0}$, the first half of
$\mt{B_i}$ are negative and the second half are positive; if $\mt{Sentinel
< 0}$, $\mt{B_i < Sentinel}$ for all $i$; and if $\mt{Sentinel > 0}$, $\mt{B_i
> Sentinel}$ for all $i$.
\end{description}

\begin{figure}
\begin{lstlisting}
$t(x_1,\ x_2,\ \dots,\ x_n)\ = $ if $ x_1 \leq x_2 $ then $ x_2 $
                         else $ t(t(x_1 - 1,\ x_2,\ \dots,\ x_n),\ \dots,\ t(x_n - 1,\ x_1,\ \dots,\ x_{n-1})) $
\end{lstlisting}
$\centerline{\rule{5cm}{0.1mm}}$
\begin{center}
  \begin{minipage}{.8\linewidth}
\begin{lstlisting}[style=MyProlog, escapebegin=\tt\color{PrologOther}]
takeuchi(X$_1,$ X$_2,\ ...,$ _X$_n,$ R) :- X$_1$  <  X$_2,$ R = X$_2$.
takeuchi(X$_1,$ X$_2,\ ...,$ _X$_n,$ R) :- X$_1$ =:= X$_2,$ R = X$_2$.
takeuchi(X$_1,$ X$_2,\ ...,$  X$_n,$  R) :- X$_1$ > X$_2,$
	N$_{1}$ is X$_1$ - 1, takeuchi(N$_{1},$ X$_2,\ ...,$ X$_n,$ R$_1$),
	N$_{2}$ is X$_2$ - 1, takeuchi(N$_{2},$ X$_3,\ ...,$ X$_1,$ R$_2$),
        ...,
	N$_{n}$ is X$_n$ - 1, takeuchi(N$_{n},$ X$_1,\ ...,$ X$_{n-1},$ R$_n$),
	takeuchi(R$_1,$ R$_2,\ ...,$ R$_n,$ R).                
\end{lstlisting}
  \end{minipage}
\end{center}
\redbefcapt
\caption{$n$-dimensional Takeuchi function and its implementation.}
  \label{fig:takeuchi}
  \redafterfig
\end{figure}

\begin{figure}
  \centering
  \begin{minipage}{0.9\linewidth}
\begin{lstlisting}[style=MyProlog, escapebegin=\tt\color{PrologOther}]
sentinel(Sentinel$,$ A$_1,\ ...,$ A$_n,$ B$_1,\ ...,$ B$_n$) :- Sentinel =:= 0, 
        ring(A$_1,\ ...,$ A$_n, $ B$_1,\ ...,$ B$_n$), 
	B$_1$ < B$_2,\ ...,$ B$_{n-1}$ < B$_n,$ B$_{n/2}$ < Sentinel, Sentinel < B$_{n/2+1}$.
sentinel(Sentinel$,$ A$_1,\ ...,$ A$_n,$ B$_1,\ ...,$ B$_n$) :- Sentinel < 0, 
        ring(A$_1,\ ...,$ A$_n,$ B$_1,\ ...,$ B$_n$), 
	B$_1$ < B$_2,\ ...,$ B$_{n-1}$ < B$_n,$ B$_n$ < Sentinel.
sentinel(Sentinel$,$ A$_1,\ ...,$ A$_n,$ B$_1,\ ...,$ B$_n$) :- Sentinel > 0, 
        ring(A$_1,\ ...,$ A$_n,$ B$_1,\ ...,$ B$_n$), 
	B$_1$ < B$_2,\ ...,$ B$_{n-1}$ < B$_n,$ B$_1$ > Sentinel.

ring(A$_1,\ ...,$ A$_n,$ B$_1,\ ...,$ B$_n$) :- B$_1$ = A$_1,\ ...,$ B$_n$ = A$_n$.
ring(A$_1,\ ...,$ A$_n,$ B$_1,\ ...,$ B$_n$) :- A$_1$ > A$_2,$ 
	ring(A$_2,\ ...,$ A$_n,$ A$_1,$ B$_1,\ ...,$ B$_n$).
ring(A$_1,\ ...,$ A$_n,$ B$_1,\ ...,$ B$_n$) :- 
	ring(A$_n,$ A$_1,\ ...,$ A$_{n-1},$ B$_1,\ ...,$ B$_n$).
\end{lstlisting}
    \end{minipage}
\redbefcapt
  \caption{\texttt{sentinel/m} program.}
  \label{fig:sentinel}
\redafterfig
\end{figure}

Table~\ref{tab:abs_performance} shows the run time results of analyzing
\texttt{takeuchi/m} parameterized by the dimension of the function,
$n$ ($m = n + 1$), and \texttt{sentinel/m} parameterized by $n$, the
length of the ring ($m=2n+1$). In both examples the analysis with the
TCLP version of the interpreter is
faster than the analysis with the interpreter without constraints: the
latter has to evaluate each permutation completely in the recursive
predicates, while the former can suspend and save computation time
using results from a previous, more general, call.

Let us examine an example.  For a variable $\mt{A}$, let us write
$\mt{A}^{abs}$ to represent $\mt{A} \sqsubseteq abs$.
On the one hand, when an initial goal
\texttt{ring($\mt{A_1}^{top}, \dots, \mt{A_n}^{top}, \mt{B_1}^{top}, \dots, 
  \mt{B_n}^{top}$)} is interpreted by the TCLP analyzer, 
the first clause of \texttt{ring/2n} produces the first answer.  Then
the interpreter continues with the second clause, interprets the goal
$\mt{A_1 > A_2}$ and starts the evaluation of
\texttt{ring($\mt{A_2}^{num}, \dots, \mt{A_n}^{top}, \mt{A_1}^{num}, \mt{B_1}^{top}, \dots,
  \mt{B_n}^{top}$)}.  Since $num \sqsubseteq top$, this new call
entails the previous one and TCLP suspends this execution. Then
the interpreter continues with the third clause, which starts the
evaluation of \texttt{ring($\mt{A_n}^{top}, \mt{A_1}^{top}, \dots,
  \mt{A_{n-1}}^{top}, \mt{B_1}^{top}, \dots, \mt{B_n}^{top}$)}, and TCLP also
suspends the execution. Since the generator does not have more clauses to
evaluate, TCLP resumes the suspended execution with the
previously obtained answer. Each consumer produces a new answer but
since they are at least as particular as the previous one, they are
discarded.

On the other hand  when the initial
goal is evaluated in the tabling interpreter with $\mt{A_1}^{top}, \ldots, \mt{A_n}^{top}, \mt{B_1}^{top},
\ldots, \mt{B_n}^{top}$ as entry substitution, the first answer is also
produced. Then the interpreter continues with the second clause,
interprets the goal $\mt{A_1 > A_2}$ and starts the evaluation of
the recursive call with the entry substitution $\mt{A_1}^{num}, \ldots,
\mt{A_n}^{top}, \mt{B_1}^{num}, \ldots, \mt{B_n}^{top}$. However, tabling does not
suspend the execution because it is not a \emph{variant} call of the
previous one, which results in  increased computation time.

\begin{table}
  \centering
  \setlength{\tabcolsep}{0.75em}
  \begin{tabular}{@{\extracolsep{0pt}}|l|l|r|r|}
    \cline{1-4}
                                                               &      & Tabling        & \mtclp(\Lat)      \\ \cline{1-4}
    \multirow{3}{*}{\texttt{takeuchi/m {\small $(m=n+1)$}}}    & n=8  & 31.44          & \textbf{8.09}   \\ 
                                                               & n=6  & 13.75          & \textbf{5.85}   \\ 
                                                               & n=3  & \textbf{2.42}  & 3.12            \\ \cline{1-4}
    \multirow{3}{*}{\texttt{sentinel/m  {\small $(m=2n+1)$}}}  & n=8  & 1375.13        & \textbf{9.23}   \\ 
                                                               & n=6  & 218.93         & \textbf{6.53}   \\ 
                                                               & n=4  & 30.99          & \textbf{4.56}   \\ \cline{1-4}
  \end{tabular}
  \redbefcapt
  \caption{Run time (ms) for \texttt{analyze(takeuchi/m)} and for \texttt{analyze(sentinel/m)}. }
  \label{tab:abs_performance}
    \redafterfig
\end{table}

Table~\ref{tab:abs_restriction} shows the results of analyzing
\texttt{sentinel/m} in two different scenarios: without any
constraints in the abstract substitution of the variables
or adding the constraint \texttt{Sentinel $\sqsubseteq +$} before the
analysis.
Adding that domain restriction reduces analysis times by
approximately the same ratio in both cases.  Note that this compares two scenarios
which are possible both for TCLP and for tabling without constraints.
Additionally, the TCLP-based analyzer would be able to take into account
constraints \emph{among} variables, which would not be directly
possible using tabling without constraints.

\begin{table}
  \centering
  \setlength{\tabcolsep}{0.75em}
  \begin{tabular}{@{\extracolsep{0pt}}|l|l|r|r|r|r|}
    \cline{1-6}
    \multicolumn{2}{|c|}{}   &  \multicolumn{2}{c|}{Tabling} & \multicolumn{2}{c|}{\mtclp(\Lat)} \\ \cline{3-6}
   \multicolumn{2}{|c|}{}   & \multicolumn{1}{c|}{constraints}   & \multicolumn{1}{c|}{unconstrained}    & \multicolumn{1}{c|}{constraints}  &  \multicolumn{1}{c|}{unconstrained}     \\
   \multicolumn{2}{|c|}{}   & \multicolumn{1}{c|}{ before call}   & \multicolumn{1}{c|}{ call}    & \multicolumn{1}{c|}{ before call}  &  \multicolumn{1}{c|}{ call}     \\ \cline{1-6}
    \multirow{3}{1.8cm}{\texttt{sentinel/m} \\\texttt{{\small $(m=2n+1)$}}} & n=8  & \textbf{749.38} &  1375.13 & \textbf{5.29}  & 9.23  \\ 
                                                                            & n=6  & \textbf{ 98.80} &  218.93  & \textbf{3.31}  & 6.53  \\ 
                                                                            & n=4  & \textbf{  6.53} &  30.99   & \textbf{2.85}  & 4.56  \\ \cline{1-6}
  \end{tabular}
  \redbefcapt
  \caption{Run time (ms) for \texttt{analyze(sentinel/m)}. }
  \label{tab:abs_restriction}
  \redafterfig
\end{table}

\redbefsec
\section{Related Work}
\redaftersec
\label{sec:rel-work}

The initial ideas of tabling and constraints originate
in~\cite{KanellakisKR95}, where a variant of Datalog featuring
constraints was proposed.  The time and space properties associated
with the bottom-up evaluation of Datalog were studied
in~\cite{toman_theo_const_tabling}, where a top-down evaluation
strategy featuring tabling was proposed.

XSB~\cite{xsb-journal-2012} was the first logic programming system
that provided tabled CLP as a generic feature, instead of
resorting to ad-hoc adaptations.  This was done by extending XSB with
attributed variables~\cite{bao2000}, one of the most popular mechanism
to implement constraint solvers in Prolog.  
However, one of its drawbacks is that it only uses variant call
checking (even for goals with constraints), instead of entailment
checking of calls / answers.  This makes programs terminate in fewer
cases than using entailment
and takes longer in other cases.  This is similar to what happens in
tabled logic programs with and without
subsumption~\cite{swift2010:subsumption}.  From the point of view of
interfacing / adding additional CLP solvers to existing systems, the
framework in~\cite{bao2000} requires the constraint solver to provide
the predicates \texttt{projection/1} and \texttt{entail/2}, which are
used to discard more particular answers, but only in one direction.
It also requires the implementation of the predicate
\texttt{abstract/3}, which has to take care of the call abstraction.
However, it is not clear if this predicate is part of the constraint
solver or of the user program.

A general framework for CHR under tabled evaluation is described
in~\cite{tchr2007}. It takes advantage of the flexibility that
CHR provides for writing constraint solvers, but it also 
lacks call entailment checking
and enforces total call abstraction: all constraints are removed from
calls before executing them, which can result in non-termination
w.r.t.\ systems which use entailment.  Besides, the need to change the
representation between CHR and Herbrand terms takes a toll in
performance.  From the interface point of view, the framework provides
interesting hooks: 
\texttt{projection(PredName)} specifies that predicate
\texttt{PredName/1}  determines how projection is to be performed, which
makes it possible to, for example, ignore arguments;
\texttt{canonical\_form(PredName)} modifies the answer store to a
canonical form as defined by \texttt{PredName/2}, so that identical
answers can be detected (e.g. using \texttt{sort/2} the constraints
\texttt{[leq(1,X),leq(X,3)]} and \texttt{[leq(X,3),leq(1,X)]} are
reduced to the same canonical form); and
\texttt{answer\_combination(PredName)}, if specified, applies
\texttt{PredName/3} in such a way that two answers can be merged into
one.

Failure Tabled CLP~\cite{navas2013-FTCLP} implements a
verification-oriented system which has several points in common with
TCLP.  Interestingly, it can
learn from failed derivations and uses interpolants instead of
constraint projection to generate conditions for reuse.
It will however not terminate in some cases
even with the addition  of 
counters to implement a mechanism akin 
to iterative deepening.

Last, the original TCLP proposal~\cite{chico-tclp-flops2012} features
entailment checking for calls and (partially) for answers, executes
calls with all the constraints, and has good performance.  However,
from the interface point of view, it did not clearly state which
operations must be present in the constraint solver, which made it
difficult to extend, and was not focused on a modular design which, for
example, made implementing specific answer management
strategies cumbersome.

\redbefsec
\section{Conclusions and Further Work}
\redaftersec
\label{sec:conclusions} 

We have presented an approach to include constraint solvers in logic
programming systems with tabling.  Our main goal is making the addition of
new constraint solvers easier while taking full advantage of
entailment between constraint stores.
In order to achieve this, we determined the services that a constraint
solver should provide to a tabling engine.  This interface has been
designed to give
the constraint solver freedom to implement them.  To
validate our design, we have interfaced one solver previously written
in C (CLP(\D)), two existing classical solvers (CLP(\Q
/\R)), and a new solver (CLP(\Lat)), and we
have found the integration to be easy --- certainly easier than with
other designs ---, validating the usefulness of the capabilities that our system provides.

We evaluated its performance in a series of benchmarks.  In some of
them large savings are attained w.r.t.\ non-tabled/tabled executions,
even taking into account the penalty to pay for the additional
flexibility and modularity.  We are in any case confident that there
is still ample space to improve the efficiency of the implementation,
since in the current implementation we gave more importance to the
cleanliness of the code and the design.

The facilities that our framework provides to integrate constraint
solvers with tabling pave the way to new research directions:

\begin{itemize}
\item Explore richer, faster, and more flexible implementations of
  abstract interpretation-based analyzers.

\item Evolve CLP(\Lat) into a lattice domain that can capture
  reasoning in ontologies and explore its usage to implement constraint
  tabled-based reasoning systems featuring automatic reuse of more
  general concepts and combinations of answers into more general
  concepts.

\item Implement a TCLP interface for a constraint solver over finite
  domains, CLP(\FD)~\cite{VanHentenryck89,chip,wam-fd-iclp-diaz}.
  CLP(\FD) is widely used to model discrete problems such as
  scheduling, planning, packing, and timetabling. The implementation
  is not straightforward due to the difficulty of expressing
  projection and entailment inside
  CLP(\FD)~\cite{carlson1994entailment}.

\end{itemize}

\bibliographystyle{acmtrans}

\begin{thebibliography}{}

\end{thebibliography}


\begin{thebibliography}{}

\bibitem[\protect\citeauthoryear{Carlson, Carlsson, and Diaz}{Carlson
  et~al\mbox{.}}{1994}]{carlson1994entailment}
{\sc Carlson, B.}, {\sc Carlsson, M.}, {\sc and} {\sc Diaz, D.} 1994.
\newblock {Entailment of Finite Domain Constraints}.
\newblock In {\em International Conference on Logic Programming (ICLP'94)}. The
  MIT Press.

\bibitem[\protect\citeauthoryear{Charatonik, Mukhopadhyay, and
  Podelski}{Charatonik
  et~al\mbox{.}}{2002}]{Charatonik02:model_checking_tabulation}
{\sc Charatonik, W.}, {\sc Mukhopadhyay, S.}, {\sc and} {\sc Podelski, A.}
  2002.
\newblock {Constraint-Based Infinite Model Checking and Tabulation for
  Stratified CLP}.
\newblock In {\em ICLP'02}, {P.~J. Stuckey}, Ed. Lecture Notes in Computer
  Science, vol. 2401. Springer, 115--129.

\bibitem[\protect\citeauthoryear{{Chico~de~Guzm\'{a}n}, Carro, Hermenegildo,
  and Stuckey}{{Chico~de~Guzm\'{a}n}
  et~al\mbox{.}}{2012}]{chico-tclp-flops2012}
{\sc {Chico~de~Guzm\'{a}n}, P.}, {\sc Carro, M.}, {\sc Hermenegildo, M.~V.},
  {\sc and} {\sc Stuckey, P.} 2012.
\newblock {A} {G}eneral {I}mplementation {F}ramework for {T}abled {CLP}.
\newblock In {\em FLOPS'12}, {T.~Schrijvers} {and} {P.~Thiemann}, Eds. Number
  7294 in LNCS. Springer Verlag, 104--119.

\bibitem[\protect\citeauthoryear{Cousot and Cousot}{Cousot and
  Cousot}{1977}]{Cousot77}
{\sc Cousot, P.} {\sc and} {\sc Cousot, R.} 1977.
\newblock {A}bstract {I}nterpretation: a {U}nified {L}attice {M}odel for
  {S}tatic {A}nalysis of {P}rograms by {C}onstruction or {A}pproximation of
  {F}ixpoints.
\newblock In {\em {ACM} {S}ymposium on {P}rinciples of {P}rogramming
  {L}anguages (POPL'77)}. ACM Press, 238--252.

\bibitem[\protect\citeauthoryear{Cui and Warren}{Cui and
  Warren}{2000}]{bao2000}
{\sc Cui, B.} {\sc and} {\sc Warren, D.~S.} 2000.
\newblock {A} system for {T}abled {C}onstraint {L}ogic {P}rogramming.
\newblock In {\em Computational Logic}. 478--492.

\bibitem[\protect\citeauthoryear{Dawson, Ramakrishnan, and Warren}{Dawson
  et~al\mbox{.}}{1996}]{Dawson:pldi96}
{\sc Dawson, S.}, {\sc Ramakrishnan, C.~R.}, {\sc and} {\sc Warren, D.~S.}
  1996.
\newblock Practical {P}rogram {A}nalysis {U}sing {G}eneral {P}urpose {L}ogic
  {P}rogramming {S}ystems -- {A} {C}ase {S}tudy.
\newblock In {\em Proceedings of the ACM SIGPLAN'96 Conference on Programming
  Language Design and Implementation}. ACM Press, New York, USA, 117--126.

\bibitem[\protect\citeauthoryear{D\'{\i}az and Codognet}{D\'{\i}az and
  Codognet}{1993}]{wam-fd-iclp-diaz}
{\sc D\'{\i}az, D.} {\sc and} {\sc Codognet, P.} 1993.
\newblock {A} {M}inimal {E}xtension of the {WAM} for {\tt clp(fd)}.
\newblock In {\em Proceedings of the Tenth International Conference on Logic
  Programming}. Budapest, MIT press, 774--790.

\bibitem[\protect\citeauthoryear{Dincbas, Hentenryck, Simonis, and
  Aggoun}{Dincbas et~al\mbox{.}}{1988}]{chip}
{\sc Dincbas, M.}, {\sc Hentenryck, P.~V.}, {\sc Simonis, H.}, {\sc and} {\sc
  Aggoun, A.} 1988.
\newblock {T}he {C}onstraint {L}ogic {P}rogramming {L}anguage {CHIP}.
\newblock In {\em Proceedings of the 2nd International Conference on Fifth
  Generation Computer Systems}. 249--264.

\bibitem[\protect\citeauthoryear{Falaschi, Levi, Martelli, and
  Palamidessi}{Falaschi et~al\mbox{.}}{1989}]{falaschi-vars89}
{\sc Falaschi, M.}, {\sc Levi, G.}, {\sc Martelli, M.}, {\sc and} {\sc
  Palamidessi, C.} 1989.
\newblock Declarative {M}odeling of the {O}perational {B}ehaviour of {L}ogic
  {P}rograms.
\newblock {\em Theoretical Computer Science\/}~{\em 69}, 289--318.

\bibitem[\protect\citeauthoryear{Frigioni, Marchetti-Spaccamela, and
  Nanni}{Frigioni et~al\mbox{.}}{1998}]{FrigioniMN98}
{\sc Frigioni, D.}, {\sc Marchetti-Spaccamela, A.}, {\sc and} {\sc Nanni, U.}
  1998.
\newblock {F}ully {D}ynamic {S}hortest {P}aths and {N}egative {C}ycles
  {D}etection on {D}igraphs with {A}rbitrary {A}rc {W}eights.
\newblock In {\em ESA}. 320--331.

\bibitem[\protect\citeauthoryear{Gange, Navas, Schachte, S{\o}ndergaard, and
  Stuckey}{Gange et~al\mbox{.}}{2013}]{navas2013-FTCLP}
{\sc Gange, G.}, {\sc Navas, J.~A.}, {\sc Schachte, P.}, {\sc S{\o}ndergaard,
  H.}, {\sc and} {\sc Stuckey, P.~J.} 2013.
\newblock {Failure Tabled Constraint Logic Programming by Interpolation}.
\newblock {\em {TPLP}\/}~{\em 13,\/}~4-5, 593--607.

\bibitem[\protect\citeauthoryear{Genaim, Codish, and Howe}{Genaim
  et~al\mbox{.}}{2001}]{genaim2001worst}
{\sc Genaim, S.}, {\sc Codish, M.}, {\sc and} {\sc Howe, J.} 2001.
\newblock {Worst-Case Groundness Analysis Using Definite Boolean Functions}.
\newblock {\em Theory and Practice of Logic Programming\/}~{\em 1,\/}~05,
  611--615.

\bibitem[\protect\citeauthoryear{Hermenegildo, Bueno, Carro, L\'{o}pez, Mera,
  Morales, and Puebla}{Hermenegildo
  et~al\mbox{.}}{2012}]{hermenegildo11:ciao-design-tplp}
{\sc Hermenegildo, M.~V.}, {\sc Bueno, F.}, {\sc Carro, M.}, {\sc L\'{o}pez,
  P.}, {\sc Mera, E.}, {\sc Morales, J.}, {\sc and} {\sc Puebla, G.} 2012.
\newblock {A}n {O}verview of {C}iao and its {D}esign {P}hilosophy.
\newblock {\em Theory and Practice of Logic Programming\/}~{\em 12,\/}~1--2
  (January), 219--252.
\newblock http://arxiv.org/abs/1102.5497.

\bibitem[\protect\citeauthoryear{Holzbaur}{Holzbaur}{1995}]{holzbaur-clpqr}
{\sc Holzbaur, C.} 1995.
\newblock {{OFAI} CLP(Q,R) Manual, Edition 1.3.3}.
\newblock Tech. Rep. TR-95-09, Austrian Research Institute for Artificial
  Intelligence, Vienna.

\bibitem[\protect\citeauthoryear{Jaffar and Maher}{Jaffar and
  Maher}{1994}]{survey94}
{\sc Jaffar, J.} {\sc and} {\sc Maher, M.} 1994.
\newblock {C}onstraint {L}ogic {P}rogramming: {A} {S}urvey.
\newblock {\em Journal of Logic Programming\/}~{\em 19/20}, 503--581.

\bibitem[\protect\citeauthoryear{Kanellakis, Kuper, and Revesz}{Kanellakis
  et~al\mbox{.}}{1995}]{KanellakisKR95}
{\sc Kanellakis, P.~C.}, {\sc Kuper, G.~M.}, {\sc and} {\sc Revesz, P.~Z.}
  1995.
\newblock {C}onstraint {Q}uery {L}anguages.
\newblock {\em J. Comput. Syst. Sci.\/}~{\em 51,\/}~1, 26--52.

\bibitem[\protect\citeauthoryear{Knuth}{Knuth}{1991}]{knuth1991textbook}
{\sc Knuth, D.~E.} 1991.
\newblock {Textbook Examples of Recursion}.
\newblock {\em Artificial Intelligence and Mathematical Theory of Computation:
  Papers in Honor of John McCarthy\/}, 207--230.

\bibitem[\protect\citeauthoryear{Marriott and Stuckey}{Marriott and
  Stuckey}{1998}]{intro_constraints_stuckey}
{\sc Marriott, K.} {\sc and} {\sc Stuckey, P.~J.} 1998.
\newblock {\em {P}rogramming with {C}onstraints: an {I}ntroduction}.
\newblock MIT Press.

\bibitem[\protect\citeauthoryear{Ramakrishna, Ramakrishnan, Ramakrishnan,
  Smolka, Swift, and Warren}{Ramakrishna
  et~al\mbox{.}}{1997}]{ramakrishna97:model_checking_tabling}
{\sc Ramakrishna, Y.}, {\sc Ramakrishnan, C.}, {\sc Ramakrishnan, I.}, {\sc
  Smolka, S.}, {\sc Swift, T.}, {\sc and} {\sc Warren, D.} 1997.
\newblock {E}fficient {M}odel {C}hecking {U}sing {T}abled {R}esolution.
\newblock In {\em Computer Aided Verification}. Lecture Notes in Computer
  Science, vol. 1254. Springer Verlag, 143--154.

\bibitem[\protect\citeauthoryear{Ramakrishnan, Rao, Sagonas, Swift, and
  Warren}{Ramakrishnan et~al\mbox{.}}{1995}]{rama95:efficient_tabling}
{\sc Ramakrishnan, I.}, {\sc Rao, P.}, {\sc Sagonas, K.}, {\sc Swift, T.}, {\sc
  and} {\sc Warren, D.} 1995.
\newblock {E}fficient {T}abling {M}echanisms for {L}ogic {P}rograms.
\newblock In {\em ICLP'95}. 697--711.

\bibitem[\protect\citeauthoryear{Schrijvers, Demoen, and Warren}{Schrijvers
  et~al\mbox{.}}{2008}]{tchr2007}
{\sc Schrijvers, T.}, {\sc Demoen, B.}, {\sc and} {\sc Warren, D.~S.} 2008.
\newblock {TCHR}: a {F}ramework for {T}abled {CLP}.
\newblock {\em TPLP\/}~{\em 8,\/}~4, 491--526.

\bibitem[\protect\citeauthoryear{Swift and Warren}{Swift and
  Warren}{2010}]{swift2010:subsumption}
{\sc Swift, T.} {\sc and} {\sc Warren, D.~S.} 2010.
\newblock Tabling with answer subsumption: Implementation, applications and
  performance.
\newblock In {\em JELIA}, {T.~Janhunen} {and} {I.~Niemel{\"a}}, Eds. Lecture
  Notes in Computer Science, vol. 6341. Springer, 300--312.

\bibitem[\protect\citeauthoryear{Swift and Warren}{Swift and
  Warren}{2012}]{xsb-journal-2012}
{\sc Swift, T.} {\sc and} {\sc Warren, D.~S.} 2012.
\newblock {XSB}: {E}xtending {P}rolog with {T}abled {L}ogic {P}rogramming.
\newblock {\em TPLP\/}~{\em 12,\/}~1-2, 157--187.

\bibitem[\protect\citeauthoryear{Tamaki and Sato}{Tamaki and
  Sato}{1986}]{tamaki.iclp86}
{\sc Tamaki, H.} {\sc and} {\sc Sato, M.} 1986.
\newblock {OLD} {R}esolution with {T}abulation.
\newblock In {\em Third International Conference on Logic Programming}. Lecture
  Notes in Computer Science, Springer-Verlag, London, 84--98.

\bibitem[\protect\citeauthoryear{Toman}{Toman}{1997a}]{toman97:const_data_absint}
{\sc Toman, D.} 1997a.
\newblock {C}onstraint {D}atabases and {P}rogram {A}nalysis {U}sing {A}bstract
  {I}nterpretation.
\newblock In {\em Constraint Databases and Their Applications}. Lecture Notes
  in Computer Science, vol. 1191. 246--262.

\bibitem[\protect\citeauthoryear{Toman}{Toman}{1997b}]{toman_theo_const_tabling}
{\sc Toman, D.} 1997b.
\newblock {M}emoing {E}valuation for {C}onstraint {E}xtensions of {D}atalog.
\newblock {\em Constraints\/}~{\em 2,\/}~3/4, 337--359.

\bibitem[\protect\citeauthoryear{Van~Hentenryck}{Van~Hentenryck}{1989}]{VanHentenryck89}
{\sc Van~Hentenryck, P.} 1989.
\newblock {\em Constraint Satisfaction in Logic Programming}.
\newblock MIT Press.

\bibitem[\protect\citeauthoryear{Warren}{Warren}{1992}]{Warren92}
{\sc Warren, D.~S.} 1992.
\newblock {M}emoing for {L}ogic {P}rograms.
\newblock {\em Communications of the ACM\/}~{\em 35,\/}~3, 93--111.

\bibitem[\protect\citeauthoryear{Warren, Hermenegildo, and Debray}{Warren
  et~al\mbox{.}}{1988}]{pracabsin}
{\sc Warren, R.}, {\sc Hermenegildo, M.}, {\sc and} {\sc Debray, S.~K.} 1988.
\newblock {O}n the {P}racticality of {G}lobal {F}low {A}nalysis of {L}ogic
  {P}rograms.
\newblock In {\em Fifth International Conference and Symposium on Logic
  Programming}. {MIT} Press, 684--699.

\bibitem[\protect\citeauthoryear{Zou, Finin, and Chen}{Zou
  et~al\mbox{.}}{2005}]{zou05:owl-tabling}
{\sc Zou, Y.}, {\sc Finin, T.}, {\sc and} {\sc Chen, H.} 2005.
\newblock {F-OWL}: {A}n {I}nference {E}ngine for {S}emantic {W}eb.
\newblock In {\em Formal Approaches to Agent-Based Systems}. Lecture Notes in
  Computer Science, vol. 3228. Springer Verlag, 238--248.

\end{thebibliography}

\redbefsec

\redaftersec
\label{sec:bibliography}

\newpage

\end{document}


\maketitle

\appendix

\section{CLP Trees and TCLP Forests}
\label{sec:tree-forest}

Prolog and CLP follow a depth-first search strategy with chronological
backtracking.  The computation rule selects constraints and literals
from the resolvent from left to right. 
Literals are resolved against the clauses of the program, selected
from top to bottom.  When a literal unifies with a clause head, it is
substituted by the body of the clause after applying the unifier
obtained from the literal-head unification.
If a derivation branch fails because there are no more
matching clauses or the constraint store is inconsistent, the
evaluation backtracks to the youngest literal that has a candidate matching 
clause. 
Depth-first search is incomplete and in general not all answers can be
computed.  Moreover, there are programs with finite derivations for
which logically equivalent programs produce infinite derivations. The
use of TCLP can work around this issue in many cases.

\begin{figure}[h]
  \begin{center}
    \begin{minipage}{0.35\linewidth}
\begin{lstlisting}[style=MyProlog]
edge(a, b, 50).
edge(b, a,  D) :- 
     D #> 25, 
     D #< 35.
\end{lstlisting}
    \end{minipage}
    \begin{minipage}{0.2\linewidth}
      \begin{tikzpicture}[node distance=1cm, transform shape]
        \node (a) [circle, draw=black] {a};
        \node (b) [circle, draw=black, below of=a, xshift=1cm] {b};

        \draw [draw, ->] (a) to[out=0, in=90] node [anchor=south,xshift=2mm] {50} (b);
        \draw [draw, ->] (b) to[out=180, in=-90] node [anchor=north,
        xshift=-2mm, yshift=2.5mm, text width=2cm] {(25,35)} (a);
      \end{tikzpicture}
    \end{minipage}
    \redbefcapt
    \caption{Graph definition. (25,35) is the open interval from 25 to
      35. }
    \label{fig:edges}
    \redafterfig
  \end{center}
\end{figure}
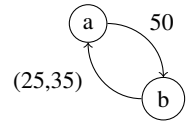

We will show the CLP trees and the TCLP forests for the query \texttt{?-
  D \#< 150, dist(a,Y,D).} for two logically equivalent versions of
the \texttt{dist/3} program: with left recursion 
(Fig.~\ref{fig:dist-graph}, right) and with
right recursion
(Fig.~\ref{fig:dist-graph-right}, right). We use the graph in
Fig.~\ref{fig:edges}, where the length of one of the edges is defined
with constraints.

Fig.~\ref{fig:tree02} and Fig.~\ref{fig:tree01-03} (top) are the CLP
trees of the right- / left-recursive programs
respectively. Fig.~\ref{fig:tree01-03} (bottom) and
Fig.~\ref{fig:tree04} are the TCLP forest of the left- /
right-recursive programs respectively. In these figures, the nodes of
the trees represent the states (Def.~\ref{def:state}) of the
computation. A state is a tuple \goal{R,c}, where $R$ is a sequence
of goals, $[g_1,g_2,\dots,g_n]$ and $c$ is a conjunction of
constraints.
The numbers attached to each state indicates the order in which they are
created.

On the one hand, Fig.~\ref{fig:tree02} shows a
finite CLP tree which finds all the answers and 
Fig.~\ref{fig:tree01-03} (top) shows an infinite CLP
tree caused by the left recursion.
On the other hand, Fig.~\ref{fig:tree01-03} (bottom) and
Fig.~\ref{fig:tree04} show that the TCLP forest for both programs are
finite and all the answers to the query are found, since the use of
tabling makes it terminate with left recursion as well.

\subsection{CLP Tree of \texttt{dist/3} with Right Recursion}
\label{sec:clp-tree-dist3-right}

\begin{figure}
  \vspace*{-.5cm}
  \begin{turn}{90}
    \begin{minipage}{1.7\textwidth}
      \begin{tabular}{|p{5cm}|p{14.5cm}|}\cline{1-2}
        {
  \renewcommand{\newstate}[3]{ \textbf{#1} }
  \renewcommand{\newstatelarge}[3]{ \newstate{#1}{}{} }

  \begin{forest}
    for tree={
      font=\sffamily\scriptsize,
      edge={->,>=stealth}
    }
    [ \newstate{s1}%
    {dist(a,Y,D)}%
    {D<150}
    [ \newstate{s2i}%
    {D1\pr1\#>0, D2\pr1\#>0, D\pr1\#=D1\pr1+D2\pr1, edge(a,Z\pr1,D1\pr1), dist(Z\pr1,Y\pr1,D2\pr1)}%
    {D<150,Y\pr1=Y,D\pr1=D}
    [ \newstate{s3}%
    {edge(a,Z\pr1,D1\pr1), dist(Z\pr1,Y,D2\pr1)}%
    {D<150,D1\pr1>0,D2\pr1>0,D=D1\pr1+D2\pr1}
    [ \newstate{s4}%
    {dist(b,Y,D2\pr1)}%
    {D<150,D1\pr1>0,D2\pr1>0,D=D1\pr1+D2\pr1,Z\pr1=b,D1\pr1=50}
    [ \newstatelarge{s5i}%
    {D1\pr2\#>0, D2\pr2\#>0, D\pr2\#=D1\pr2+D2\pr2, edge(b,Z\pr2,D1\pr2), dist(Z\pr2,Y\pr2,D2\pr2)}%
    {D<150,D2\pr1>0,D=50+D2\pr1,Y\pr2=Y,D\pr2=D2\pr1}
    [ \newstate{s6}%
    {edge(b,Z\pr2,D1\pr2), dist(Z\pr2,Y,D2\pr2)}%
    {D<150,D=50+D2\pr1,D1\pr2>0, D2\pr2>0, D2\pr1=D1\pr2+D2\pr2}
    [ \newstate{s7}%
    {dist(a,Y,D2\pr2)}%
    {D<150,D=50+D1\pr2+D2\pr2,D1\pr2>0, D2\pr2>0, Z\pr2=a,D1\pr2>25,D1\pr2<35}
    [ \newstatelarge{s8i}%
    {D1\pr3\#>0, D2\pr3\#>0, D\pr3\#=D1\pr3+D2\pr3, edge(a,Z\pr3,D1\pr3), dist(Z\pr3,Y\pr3,D2\pr3)}%
    {D<150,D>75+D2\pr2,D<85+D2\pr2,D2\pr2>0, Z\pr2=a,Y\pr3=Y,D\pr3=D2\pr2}
    [ \newstatelarge{s9}%
    {edge(a,Z\pr3,D1\pr3), dist(Z\pr3,Y,D2\pr3)}%
    {D<150, D>75+D2\pr2, D<85+D2\pr2, D2\pr2>0, D1\pr3>0, D2\pr3>0, D2\pr2=D1\pr3+D2\pr3}
    [ \newstatelarge{s10}%
    {dist(b,Y,D2\pr3)}%
    {D<150, D>75+D1\pr3+D2\pr3, D<85+D1\pr3+D2\pr3, D2\pr3>0, D1\pr3=50}
    [ \newstatelarge{s11i}%
    {D1\pr4\#>0, D2\pr4\#>0, D\pr4\#=D1\pr4+D2\pr4, edge(b,Z\pr4,D1\pr4), dist(Z\pr4,Y\pr4,D2\pr4)}%
    {D<150, D>125+D2\pr3, D<135+D2\pr3, D2\pr3>0, Y\pr4=Y, D\pr4=D2\pr3}
    [ \newstate{s12}%
    {edge(b,Z\pr4,D1\pr4), dist(Z\pr4,Y,D2\pr4)}%
    {D<150,D>125+D1\pr4+D2\pr4, D<135+D1\pr4+D2\pr4, D1\pr4>0, D2\pr4>0}
    [ \newstatelarge{s13}%
    {dist(a,Y,D2\pr4)}%
    {D<150,D>125+D1\pr4+D2\pr4, D<135+D1\pr4+D2\pr4, D2\pr4>0, D1\pr4>25,D1\pr4<35}
    [ {\bf fail} ]
    ]
    ]
    ]
    [ \newstate{s11ii}%
    {edge(b,Y,D2\pr3)}%
    {D<150, D>125+D2\pr3, D<135+D2\pr3, D2\pr3>0}
    [ \newstate{s14}%
    {\ }%
    {D<150, D>125+D2\pr3, D<135+D2\pr3, D2\pr3>0,Y=a,D2\pr3>25,D2\pr3<35}
    [ {\bf fail} ]
    ]
    ]
    ]
    ]
    ]
    [ \newstate{s8ii}%
    {edge(a,Y,D2\pr2)}%
    {D<150,D=50+D1\pr2+D2\pr2, D2\pr2>0, D1\pr2>25,D1\pr2<35}
    [ \newstatelarge{s15}%
    {\ }%
    {D<150,D=50+D2\pr1, D2\pr2>0, D2\pr1=D1\pr2+D2\pr2,D1\pr2>25,D1\pr2<35, Y=b, D2\pr2=50}
    [ {\textbf{a1} $\thickmuskip=0mu \mt{Y=b\ \land}$\\ $\thickmuskip=0mu \mt{D>125 \land  D<135}$}, delay={text width=2cm, text centered} ]
    ]
    ]
    ]
    ]
    ]
    [ \newstate{s5ii}%
    {edge(b,Y,D2\pr1)}%
    {D<150,D2\pr1>0,D=50+D2\pr1}
    [ \newstate{s16}%
    {\ }%
    {D<150,D=50+D2\pr1, Y=a, D2\pr1>25, D2\pr1<35}
    [ {{\bf a2} $\thickmuskip=0mu \mt{Y=a\ \land}$\\ $\thickmuskip=0mu \mt{D>75 \land  D<85}$}, delay={text width=2cm, text centered} ]
    ]
    ]
    ]
    ]
    ]
    [ \newstate{s2ii}%
    {edge(a,Y,D)}%
    {D<150}
    [ \newstate{s17}%
    {\ }%
    {Y=b, D=50}
    [ {{\bf a3} $\thickmuskip=0mu \mt{Y=b\ \land}$\\ $\thickmuskip=0mu \mt{D=50}$}, delay={text width=2cm, text centered} ]
    ]
    ]
    ]
  \end{forest}
}
        &
          {\vspace*{-11.5cm}\begin{forest}
  for tree={
    font=\sffamily\scriptsize,
    folder,
    grow'=0,
    fit=band,
    l sep'-=0pt,
    s sep+=-1pt,
    inner sep=1pt,
    text width=25cm,
    text height=.1cm,
    edge={->,>=stealth}
  }
  [ \newstate{s1}%
  {dist(a,Y,D)}%
  {D<150}
  [ \newstate{s2i}%
  {D1\pr1\#>0, D2\pr1\#>0, D\pr1\#=D1\pr1+D2\pr1, edge(a,Z\pr1,D1\pr1), dist(Z\pr1,Y\pr1,D2\pr1)}%
  {D<150 \land Y\pr1=Y \land D\pr1=D}
  [ \newstate{s3}%
  {edge(a,Z\pr1,D1\pr1), dist(Z\pr1,Y,D2\pr1)}%
  {D<150 \land D1\pr1>0 \land D2\pr1>0 \land D=D1\pr1+D2\pr1}
  [ \newstate{s4}%
  {dist(b,Y,D2\pr1)}%
  {D<150 \land D1\pr1>0 \land D2\pr1>0 \land D=D1\pr1+D2\pr1 \land Z\pr1=b \land D1\pr1=50}
  [ \newstatelarge{s5i}%
  {D1\pr2\#>0, D2\pr2\#>0, D\pr2\#=D1\pr2+D2\pr2, edge(b,Z\pr2,D1\pr2), dist(Z\pr2,Y\pr2,D2\pr2)}%
  {D<150 \land D2\pr1>0 \land D=50+D2\pr1 \land Y\pr2=Y \land D\pr2=D2\pr1}
  [ \newstate{s6}%
  {edge(b,Z\pr2,D1\pr2), dist(Z\pr2,Y,D2\pr2)}%
  {D<150 \land D=50+D2\pr1 \land D1\pr2>0 \land  D2\pr2>0 \land  D2\pr1=D1\pr2+D2\pr2}
  [ \newstate{s7}%
  {dist(a,Y,D2\pr2)}%
  {D<150 \land D=50+D1\pr2+D2\pr2 \land D1\pr2>0 \land  D2\pr2>0 \land  Z\pr2=a \land D1\pr2>25 \land D1\pr2<35}
  [ \newstatelarge{s8i}%
  {D1\pr3\#>0, D2\pr3\#>0, D\pr3\#=D1\pr3+D2\pr3, edge(a,Z\pr3,D1\pr3), dist(Z\pr3,Y\pr3,D2\pr3)}%
  {D<150 \land D>75+D2\pr2 \land D<85+D2\pr2 \land D2\pr2>0 \land  Z\pr2=a \land Y\pr3=Y \land D\pr3=D2\pr2}
  [ \newstatelarge{s9}%
  {edge(a,Z\pr3,D1\pr3), dist(Z\pr3,Y,D2\pr3)}%
  {D<150 \land  D>75+D2\pr2 \land  D<85+D2\pr2 \land  D2\pr2>0 \land  D1\pr3>0 \land  D2\pr3>0 \land  D2\pr2=D1\pr3+D2\pr3}
  [ \newstate{s10}%
  {dist(b,Y,D2\pr3)}%
  {D<150 \land  D>75+D1\pr3+D2\pr3 \land  D<85+D1\pr3+D2\pr3 \land  D2\pr3>0 \land  D1\pr3=50}
  [ \newstatelarge{s11i}%
  {D1\pr4\#>0, D2\pr4\#>0, D\pr4\#=D1\pr4+D2\pr4, edge(b,Z\pr4,D1\pr4), dist(Z\pr4,Y\pr4,D2\pr4)}%
  {D<150 \land  D>125+D2\pr3 \land  D<135+D2\pr3 \land  D2\pr3>0 \land  Y\pr4=Y \land  D\pr4=D2\pr3}
  [ \newstatelarge{s12}%
  {edge(b,Z\pr4,D1\pr4), dist(Z\pr4,Y,D2\pr4)}%
  {D<150 \land D>125+D1\pr4+D2\pr4 \land  D<135+D1\pr4+D2\pr4 \land  D1\pr4>0 \land  D2\pr4>0}
  [ \newstatelarge{s13}%
  {dist(a,Y,D2\pr4)}%
  {D<150 \land D>125+D1\pr4+D2\pr4 \land  D<135+D1\pr4+D2\pr4 \land  D2\pr4>0 \land  Z\pr4=a \land  D1\pr4>25 \land D1\pr4<35}
  [ {\bf fail} ]
  ]
  ]
  ]
  [ \newstate{s11ii}%
  {edge(b,Y,D2\pr3)}%
  {D<150 \land  D>125+D2\pr3 \land  D<135+D2\pr3 \land  D2\pr3>0}
  [ \newstate{s14}%
  {\,}%
  {D<150 \land  D>125+D2\pr3 \land  D<135+D2\pr3 \land  D2\pr3>0 \land Y=a \land D2\pr3>25 \land D2\pr3<35}
  [ {\bf fail} ]
  ]
  ]
  ]
  ]
  ]
  [ \newstate{s8ii}%
  {edge(a,Y,D2\pr2)}%
  {D<150 \land D=50+D1\pr2+D2\pr2 \land  D2\pr2>0 \land  D1\pr2>25 \land D1\pr2<35}
  [ \newstate{s15}%
  {\,}%
  {D<150 \land D=50+D2\pr1 \land  D2\pr2>0 \land  D2\pr1=D1\pr2+D2\pr2 \land D1\pr2>25 \land D1\pr2<35 \land  Y=b \land  D2\pr2=50}
  [ {\textbf{a1} $\thickmuskip=0mu \mt{Y=b \land  D>125 \land  D<135}$} ]
  ]
  ]
  ]
  ]
  ]
  [ \newstate{s5ii}%
  {edge(b,Y,D2\pr1)}%
  {D<150 \land D2\pr1>0 \land D=50+D2\pr1}
  [ \newstate{s16}%
  {\,}%
  {D<150 \land D=50+D2\pr1 \land  Y=a \land  D2\pr1>25 \land  D2\pr1<35}
  [ {{\bf a2} $\thickmuskip=0mu \mt{Y=a \land  D>75 \land  D<85}$} ]
  ]
  ]
  ]
  ]
  ]
  [ \newstate{s2ii}%
  {edge(a,Y,D)}%
  {D<150}
  [ \newstate{s17}%
  {\,}%
  {Y=b \land  D=50}
  [ {{\bf a3} $\thickmuskip=0mu \mt{Y=b \land  D=50}$} ]
  ]
  ]
  ]
\end{forest}
}\\ \cline{1-2}
      \end{tabular}
    \end{minipage}
  \end{turn}
  \vspace*{-.8cm}
  \caption{CLP tree of the query \texttt{?- D \#< 150, dist(a,Y,D)} \\
    for \texttt{dist/3} with right recursion}
  \label{fig:tree02}
  \redafterfig
\end{figure}

Fig.~\ref{fig:tree02} shows the CLP tree of the query using the
version of \texttt{dist/3} with right recursion
(Fig.~\ref{fig:dist-graph-right}, right). We see that the evaluation
of the recursive clause generates similar states (\textsf{s1, s4, s7}
and \textsf{s10}), but in each iteration the domain of the constrained
variable \texttt{D2$_i$} is reduced.  As a consequence, the constraint
store in state \textsf{s13} is inconsistent and the evaluation of this
derivation fails.  The pending branches are evaluated upon
backtracking. We explain now how we obtain some of the states; the
rest are obtained similarly, so we will skip them:

\begin{description}[leftmargin=!,labelwidth=\widthof{\bfseries s11a/b
  }, labelindent=0cm, align=right]
  
\item[s1] the initial state is the representation of the query.

\item[s2i/ii] are obtained by resolving the literal
  \texttt{dist(a,Y,D)} against the two clauses of the program. The
  constraints $\mt{Y\pr1=Y \land D\pr1=D}$ are added to the constraint
  store.

\item[s3] is obtained from the leftmost state \textsf{s2i} by adding
  the constraints of the resolvent \texttt{\mbox{[D1\pr1\#>0,} D2\pr1\#>0,\,
    D\pr1\#=D1\pr1+D2\pr1]} to the constraint store.  
\item[s4] is obtained by resolving the literal
  \texttt{edge(a,Z\pr1,D1\pr1)}. The constraint
  $\mt{Z\pr1=b \land {D1\pr1=50}}$ reduces the domain\footnote{We are
    considering a linear constraint solver over the rational numbers
    which from $\mt{D<150 \land D=D1\pr1+D2\pr1 \land D1\pr1=50}$ it
    infers that $\mt{D2\pr1<100}$} of \texttt{D2\pr1} to
  $\mt{D2\pr1 > 0 \land D2\pr1 < 100}$.

\item[s7] is obtained by resolving the literal
  \texttt{edge(b,Z\pr2,D1\pr2)}. The constraint
  $\mt{Z\pr2=a \land {D1\pr2>25} \land D1\pr2<35}$ reduces the domain of
  \texttt{D2\pr2} to $\mt{D2\pr2>0 \land D2\pr2<75}$.

\item[s10] is obtained by resolving the literal
  \texttt{edge(a,Z\pr3,D1\pr3)}. The constraint
  $\mt{Z\pr3=b \land {D1\pr3=50}}$ reduces the domain of \texttt{D2\pr3}
  to $\mt{D2\pr3>0 \land D2\pr3<25}$.

\item[s13] is obtained by resolving the literal
  \texttt{edge(b,Z\pr4,D1\pr4)}. The constraint $\mt{{Z\pr4=a} \land
    D1\pr4>25 \land D1\pr4<35}$ is inconsistent with the current
  constraint store, $\mt{D<150} \land \mt{D>125+D1\pr4+D2\pr4} \land
    \mt{D2\pr4>0} \land \dots$. Its child is a \textsf{fail} node.

\item[s14] is obtained, upon backtracking to the state \textsf{s11b}
  by resolving the literal \texttt{edge(b,Y,D2\pr3)}. However, it is
  also a failed derivation because the resulting constraint store is
  inconsistent.

\item[s15] is a final state of a
  successful derivation, obtained upon backtracking to the state
  \textsf{s8b} by resolving the literal
  \texttt{edge(a,Y,D2\pr2)}. The constraint $\mt{Y=a} \land \mt{D2\pr3>25} \land
    \mt{D2\pr3<35}$ is consistent with the constraint store.

  \item[a1] is the first answer
    $\mt{Y=a} \land \mt{D>125} \land \mt{D<315}$, projected
  onto the variables of the query ($vars(Q)=\{\mt{Y, D}\}$).

\item[s16] is a final state obtained upon backtracking to the state
  \textsf{s5b}.
  
\item[a2] is the second answer, $\mt{Y=a \land D>75 \land D<85}$.
  
\item[s17] is a final state obtained upon backtracking to the state
  \textsf{s2ii}.
  
\item[a3] is the third and last answer, $\mt{Y=b \land D=50}$

  .
\end{description}

\subsection{CLP Tree of \texttt{dist/3} with Left Recursion}
\label{sec:clp-tree-dist3-left}

Fig.~\ref{fig:tree01-03} (top) shows the CLP tree of the query to
\texttt{dist/3} with left recursion (Fig.~\ref{fig:dist-graph},
right). We see that the recursive clause also generates similar states
(\textsf{s1, s3, s5, $\dots$}) but in this example the domain of the
constrained variable $D1_i$ remains unchanged, and the evaluation
therefore enters a loop. As before, we only explain how we obtain some
of the states:

\begin{figure}
  \vspace*{-.8cm}
  \begin{turn}{90}
    \begin{minipage}{1.7\textwidth}
      \begin{tabular}{|p{6.75cm}|p{12.5cm}|}\cline{1-2}
        {
  \renewcommand{\newstate}[3]{ \textbf{#1} }
  \renewcommand{\newstatelarge}[3]{ \newstate{#1}{}{} }
  \hspace*{1cm}
  \begin{forest}
    for tree={
      font=\sffamily\scriptsize,
      l sep=0ex,
      s sep+=20pt,
      edge={->,>=stealth}     %
    }
    [ \newstate{s1}{dist(a,Y,D)}{D<150}
    [ \newstate{s2i}%
    {D1\pr1\#>0, D2\pr1\#>0, D\#=D1\pr1+D2\pr1, dist(a,Z\pr1,D1\pr1), edge(Z\pr1,Y,D2\pr1)}%
    {D<150,Y\pr1=Y,D\pr1=D}
    [ \newstate{s3}%
    {dist(a,Z\pr1,D1\pr1), edge(Z\pr1,Y,D2\pr1)}%
    {D<150,D1\pr1>0,D2\pr1>0,D=D1\pr1+D2\pr1}
    [ \newstatelarge{s4i}%
    {D1\pr2\#>0, D2\pr2\#>0, D1\pr1\#=D1\pr2+D2\pr2, dist(a,Z\pr2,D1\pr2), edge(Z\pr2,Z\pr1,D2\pr2),  edge(Z\pr1,Y,D2\pr1)}%
    {D<150,D1\pr1>0,D2\pr1>0,D=D1\pr1+D2\pr1, Y\pr2=Z\pr1, D\pr2=D1\pr1}
    [ \newstatelarge{s5}%
    {dist(a,Z\pr2,D1\pr2), edge(Z\pr2,Z\pr1,D2\pr2),  edge(Z\pr1,Y,D2\pr1)}%
    {D<150,D1\pr1>0,D2\pr1>0,D=D1\pr1+D2\pr1,D1\pr2>0, D2\pr2>0, D1\pr1=D1\pr2+D2\pr2}
    [ {\bf s6i }
    [ $\dots$, delay={edge={-}, yshift=1em} ]
    ]
    [ {\bf s6ii} ]
    ]
    ]
    [ \newstate{s4ii}{edge(a,Z\pr1,D1\pr1), edge(Z\pr1,Y,D2\pr1)}{D<150,D1\pr1>0,D2\pr1>0,D=D1\pr1+D2\pr1,Y\pr2=Z\pr1,D\pr2=D1\pr1} ]
    ]
    ]
    [ \newstate{s2ii}{edge(a,Y,D)}{D<150,Y\pr1=Y,D\pr1=D} ]
    ]
  \end{forest}
}
        &
          {\vspace{-5cm}\begin{forest}
  for tree={
    font=\sffamily\scriptsize,
    folder,
    grow'=0,
    fit=band,
    l sep'-=0pt,
    s sep+=-1pt,
    inner sep=1pt,
    text width=20cm,
    text height=.1cm,
    edge={->,>=stealth}
  }
  [ \newstate{s1}{dist(a,Y,D)}{D<150}
  [ \newstate{s2i}%
  {D1\pr1\#>0, D2\pr1\#>0, D\#=D1\pr1+D2\pr1, dist(a,Z\pr1,D1\pr1), edge(Z\pr1,Y,D2\pr1)}%
  {D<150 \land Y\pr1=Y \land D\pr1=D}
  [ \newstate{s3}%
  {dist(a,Z\pr1,D1\pr1), edge(Z\pr1,Y,D2\pr1)}%
  {D<150 \land D1\pr1>0 \land D2\pr1>0 \land D=D1\pr1+D2\pr1}
  [ \newstatelarge{s4i}%
  {D1\pr2\#>0, D2\pr2\#>0, D1\pr1\#=D1\pr2+D2\pr2, dist(a,Z\pr2,D1\pr2), edge(Z\pr2,Z\pr1,D2\pr2),  edge(Z\pr1,Y,D2\pr1)}%
  {D<150 \land D1\pr1>0 \land D2\pr1>0 \land D=D1\pr1+D2\pr1 \land  Y\pr2=Z\pr1 \land  D\pr2=D1\pr1}
  [ \newstatelarge{s5}%
  {dist(a,Z\pr2,D1\pr2), edge(Z\pr2,Z\pr1,D2\pr2),  edge(Z\pr1,Y,D2\pr1)}%
  {D<150 \land D1\pr1>0 \land D2\pr1>0 \land D=D1\pr1+D2\pr1 \land D1\pr2>0 \land  D2\pr2>0 \land  D1\pr1=D1\pr2+D2\pr2}
  [ {\bf s6i}
  [ \ $\dots$ ] 
  ]
  [ {\bf s6ii} ]
  ]
  ]
  [ \newstate{s4ii}{edge(a,Z\pr1,D1\pr1), edge(Z\pr1,Y,D2\pr1)}%
  {D<150 \land D1\pr1>0 \land D2\pr1>0 \land D=D1\pr1+D2\pr1 \land Y\pr2=Z\pr1 \land D\pr2=D1\pr1} ]
  ]
  ]
  [ \newstate{s2ii}{edge(a,Y,D)}%
  {D<150 \land Y\pr1=Y \land D\pr1=D} ]
  ]
\end{forest}
}\\
        \cline{1-2}
        \multicolumn{2}{c}{}\\[-0.5em]
        \cline{1-2}
        {
  \renewcommand{\newstate}[3]{ \textbf{#1} }
  \renewcommand{\newstatelarge}[3]{ \newstate{#1}{}{} }

  \begin{forest}
    for tree={
      font=\sffamily\scriptsize,
      delay={
        edge label/.wrap value={node[midway,font=\sffamily\scriptsize, left]{#1}},
      },
      edge={->,>=stealth}
    }
    [ \newstate{s1}%
    {dist(a,Y,D)}%
    {D<150}
    [ \newstate{s2i}%
    {D1\pr1\#>0, D2\pr1\#>0, D\pr1\#=D1\pr1+D2\pr1, dist(a,Z\pr1,D1\pr1), edge(Z\pr1,Y\pr1,D2\pr1)}%
    {D<150,Y\pr1=Y,D\pr1=D}
    [ \newstate{s3}%
    {dist(a,Z\pr1,D1\pr1), edge(Z\pr1,Y,D2\pr1)}%
    {D<150,D1\pr1>0,D2\pr1>0,D=D1\pr1+D2\pr1}
    [ {Ans({\bf s1})}, delay={text width=3cm, text centered, edge={-}} 
    [ \newstate{s5}%
    {edge(b,Y,D2\pr1)}%
    {D<150,D1\pr1>0,D2\pr1>0,D=D1\pr1+D2\pr1, Z\pr1=b, D1\pr1=50}, edge label=(a1)
    [ \newstate{s6}%
    {\ }%
    { D<150,D2\pr1>0,D=50+D2\pr1, Y=a, D2\pr1>25, D2\pr1<35}
    [ {\textbf{a2} $\thickmuskip=0mu \mt{V0=a\ \land}$\\ $\thickmuskip=0mu \mt{V1>75 \land  V1<85}$}, delay={text width=2cm, text centered}  ]
    ]
    ]
    [ \newstate{s7}%
    {edge(a,Y,D2\pr1)}%
    {D<150,D1\pr1>0,D2\pr1>0,D=D1\pr1+D2\pr1, Z\pr1=a, D1\pr1>75, D2\pr1<85}, edge label=(a2)
    [ \newstate{s8}%
    {\ }%
    { D<150,D2\pr1>0,D>75+D2\pr1, D<85+D2\pr1, Y=b, D2\pr1=50}
    [ {\textbf{a3} $\thickmuskip=0mu \mt{V0=b\ \land}$\\ $\thickmuskip=0mu \mt{V1>125 \land  V1<135}$}, delay={text width=2.2cm, text centered}  ]
    ]
    ]
    [ \newstate{s9}%
    {edge(b,Y,D2\pr1)}%
    {D<150,D1\pr1>0,D2\pr1>0,D=D1\pr1+D2\pr1, Z\pr1=b, D1\pr1>125, D2\pr1<135}, edge label=(a3)
    [ \newstate{s10}%
    {\ }%
    { D<150,D2\pr1>0,D>125+D2\pr1, D<135+D2\pr1, Y=a, D2\pr1>25, D2\pr1<35}
    [ \textbf{fail} ]
    ]
    ]
    ]
    ]
    ]
    [ \newstate{s2ii}%
    {edge(a,Y,D)}%
    {D<150,Y\pr1=Y,D\pr1=D}
    [ \newstate{s4}%
    {\ }%
    {D<150,Y=b,D=50}
    [ {\textbf{a1} $\thickmuskip=0mu \mt{V0=b\ \land}$\\ $\thickmuskip=0mu \mt{V1=50}$}, delay={text width=2cm, text centered}  ]
    ]     
    ]
    ]
  \end{forest}
}
        &
          {\vspace{-5.8cm}\begin{forest}
  for tree={
    font=\sffamily\scriptsize,
    delay={
      edge label/.wrap value={node[midway,font=\sffamily\scriptsize, left]{#1}},
    },
    folder,
    grow'=0,
    fit=band,
    l sep'-=0pt,
    s sep+=-1pt,
    inner sep=1pt,
    text width=20cm,
    text height=.1cm,
    edge={->,>=stealth}
  }
  [ \newstate{s1}%
  {dist(a,V0,V1)}%
  {V1<150}
  [ \newstate{s2i}%
  {D1\pr1\#>0, D2\pr1\#>0, D\pr1\#=D1\pr1+D2\pr1, dist(a,Z\pr1,D1\pr1), edge(Z\pr1,Y\pr1,D2\pr1)}%
  {V1<150 \land Y\pr1=V0 \land D\pr1=V1}
  [ \newstate{s3}%
  {dist(a,Z\pr1,D1\pr1), edge(Z\pr1,V0,D2\pr1)}%
  {V1<150 \land D1\pr1>0 \land D2\pr1>0 \land V1=D1\pr1+D2\pr1}, delay={l sep+=10pt,s sep+=7pt}
  [ {Ans(dist(a,V0\pr1,V1\pr1), $V1\pr1<150$) \smaller%
    is entailed because $V1\pr1>0 \land V1\pr1<150 \sqsubseteq V1\pr1<150$},%
  delay={%
    edge={-},%
    edge label={node[midway,font=\sffamily\scriptsize, right, yshift=10pt]%
      {\smaller with renaming $Z\pr1=V0\pr1\ \land\ D1\pr1=V1\pr1$}}}
  [ \newstate{s5}%
  {edge(b,V0,D2\pr1)}%
  {V1<150 \land D1\pr1>0 \land D2\pr1>0 \land V1=D1\pr1+D2\pr1 \land  Z\pr1=b \land  D1\pr1=50}, edge label=(a1)
  [ \newstate{s6}%
  {\,}%
  { V1<150 \land D2\pr1>0 \land V1=50+D2\pr1 \land  V0=a \land  D2\pr1>25 \land  D2\pr1<35}
  [ {\textbf{a2} $\thickmuskip=0mu \mt{V0=a \land  V1>75 \land  V1<85}$} ]
  ]
  ]
  [ \newstate{s7}%
  {edge(a,V0,D2\pr1)}%
  {V1<150 \land D1\pr1>0 \land D2\pr1>0 \land V1=D1\pr1+D2\pr1 \land  Z\pr1=a \land  D1\pr1>75 \land  D1\pr1<85}, edge label=(a2)
  [ \newstate{s8}%
  {\,}%
  { V1<150 \land D2\pr1>0 \land V1>75+D2\pr1 \land  V1<85+D2\pr1 \land  V0=b \land  D2\pr1=50}
  [ {\textbf{a3} $\thickmuskip=0mu \mt{V0=b \land  V1>125 \land  V1<135}$} ]
  ]
  ]
  [ \newstate{s9}%
  {edge(b,V0,D2\pr1)}%
  {V1<150 \land D1\pr1>0 \land D2\pr1>0 \land V1=D1\pr1+D2\pr1 \land  Z\pr1=b \land  D1\pr1>125 \land  D1\pr1<135}, edge label=(a3)
  [ \newstate{s10}%
  {\,}%
  { V1<150 \land D2\pr1>0 \land V1>125+D2\pr1 \land  V1<135+D2\pr1 \land  V0=a \land  D2\pr1>25 \land  D2\pr1<35}
  [ \textbf{fail} ]
  ]
  ]
  ]
  ]
  ]
  [ \newstate{s2ii}%
  {edge(a,V0,V1)}%
  {V1<150 \land Y\pr1=V0 \land D\pr1=V1}
  [ \newstate{s4}%
  {\,}%
  {V1<150 \land V0=b \land V1=50}
  [ {\textbf{a1} $\thickmuskip=0mu \mt{V0=b \land  V1=50}$} ]
  ]     
  ]
  ]
\end{forest}
}\\ \cline{1-2}
      \end{tabular}
    \end{minipage}
  \end{turn}
  \vspace*{-.9cm}
  \caption{CLP tree (top) and TCLP-forest (bottom) of the query \\
    \texttt{?- D \#< 150, dist(a,Y,D)} for \texttt{dist/3} with
    left recursion}
  \label{fig:tree01-03}
  \redafterfig
\end{figure}

\begin{description}[leftmargin=!,labelwidth=\widthof{\bfseries s2i/ii
  }, labelindent=0cm, align=right]

\item[s3] is obtained from the leftmost state \textsf{s2i}. The domain
  of \texttt{D1\pr1} is $\mt{D1\pr1>0 \land D1\pr1<150}$.

\item[s5] is obtained from the leftmost node \textsf{s4i}. The
  domain of \texttt{D1\pr2} is $\mt{D1\pr2>0 \land D1\pr2<150}$.

\item[$\dots$] the evaluation enters a loop.

\end{description}

Although the program that generates this CLP tree is
logically equivalent to the previous one, this tree is infinite and
 no answers are found.

\subsection{TCLP Forest of \texttt{dist/3} with Left Recursion}
\label{sec:tclp-tree-dist3-left}

Fig.~\ref{fig:tree01-03} (bottom) shows the TCLP forest for the query
we have been using to the \texttt{dist/3} program written using left
recursion (Fig.~\ref{fig:dist-graph}, right) where the set of tabled
predicates is $Tab_P=\{\mt{dist/3}\}$. The main point is that at state
\textsf{s3} the tabling engine detects that the evaluation of
\goaltt{dist(a,Z\pr1,D1\pr1)}{D1\pr1>0 \land D1\pr1<150} entails the
generator \goaltt{dist(a,V0,V1)}{V1<150} and therefore it suspends the
execution and waits until another generator feeds the suspended goal
with answers. The evaluation of the state \textsf{s2ii} generates the
first answer \textsf{a1} upon backtracking. Then, the tabling engine
resumes the consumer with \textsf{a1} and generates \textsf{a2} which
is used to generate \textsf{a3}. Finally, the evaluation fails after
consuming \textsf{a3} and, since all the clauses have been evaluated
and there are no more consumers to be resumed or answers to be
consumed, the generator is marked as complete and all the answers are
returned.  We explain below how some of the states are obtained. The
rest of the states are obtained similarly, so we skip them for
brevity:

\begin{description}[leftmargin=!,labelwidth=\widthof{\bfseries Ans(xx)
  }, labelindent=0cm, align=right]

\item[s0] We omit the representation of the TCLP tree for the query
  $\bigsymbol{\tau}_P(\mt{dist(a,Y,D),D<150})$ and its answer resolution.
  
\item[s1] the initial state of the TCLP tree
  $\bigsymbol{\tau}_P(\mt{dist(a,V0_1,V1_1),V1_1<150})$ is the renamed
  generator. (Def.~\ref{def:call_entail}).

\item[s2i/ii] are obtained by resolving the literal
  \texttt{dist(a,V0,V1)} against the two clauses of the program.

\item[s3] is obtained from the leftmost state \textsf{s2i} by adding
  the constraints to the constraint store as in the CLP tree.
  
\item[Ans(s1)] the tabled literal \texttt{dist(a,Z\pr1,D1\pr1)} has
  to be resolved by answer resolution
  (Def.~\ref{def:call_entail}) using the answer
  from the current TCLP tree
  $\bigsymbol{\tau}_P(\mt{dist(a,V0_1,V1_1),V1_1<150})$ because, after
  renaming, the projection of the current constraint store onto the
  variables of the literal entails the projected constraint
  store of the generator:
  $\mt{V1_1>0} \land \mt{V1_1<150} \sqsubseteq \mt{V1_1<150}$. Since the current TCLP
  tree is under construction and depends on itself, this branch
  derivation is \textsf{suspended}.
  
\item[s4] is a final state of a successful derivation. It is obtained,
  upon backtracking to the state \textsf{s2ii}, by resolving with
  \texttt{edge(a,V0,V1)}. The equations $\mt{V0=b} \land \mt{V1=50}$
  are consistent with the constraint store.
  
\item[a1] is the first answer, $\mt{V0=b \land V1=50}$
  (Def.~\ref{def:ans_entail}). Since is the first one,
  it is also the more general one.

\item[s5] is obtained from the state \textsf{s3} (because there are
  no more branches) by answer resolution consuming \textsf{a1}
  (Def.~\ref{def:ans_entail}).
  
\item[s6] is a final state obtained by resolving the literal
  \texttt{edge(b,V0,D2\pr1)}.
  
\item[a2] is the second answer, $\mt{V0=a \land V1>75 \land V1<85}$. It is
  neither more particular nor more general than \textsf{a1}.
  
\item[s7] is obtained from the state \textsf{s3} by consuming
  \textsf{a2}.
  
\item[s8] is a final state.
  
\item[a3] is the third answer, $\mt{V0=b \land V1>125 \land V1<135}$. It
  is neither more particular nor more general than \textsf{a1} or
  \textsf{a2}.
  
\item[s9] is obtained from the state \textsf{s3} by consuming,
  \textsf{a3}.
  
\item[s10] is a failed derivation because the resulting constraint
  store is inconsistent,
  $\mt{V1<150 \land \dots \land V1>125+D2\pr1 \land D2\pr1>25}$. Its child is a \textsf{fail} node.
  
\end{description}

Note that CLP execution entered a loop resolving the state
\textsf{s3}. Under TCLP, answer resolution avoids looping and the
resulting TCLP forest is finite and complete (i.e., the leaves of the
trees are either \textsf{fail} nodes or answers).

\subsection{TCLP Forest of \texttt{dist/3} with Right Recursion}
\label{sec:tclp-tree-dist3-right}

\begin{figure}
  \begin{tabular}{|p{.97\textwidth}|}\cline{1-1}
    \\[.1cm]
    {

  \renewcommand{\newstate}[3]{ \textbf{#1} }
  \renewcommand{\newstatelarge}[3]{ \newstate{#1}{}{} }
  
  \begin{minipage}{.45\textwidth}

    \begin{forest}
      for tree={
        font=\sffamily\scriptsize,
        delay={
          edge label/.wrap value={node[midway,font=\sffamily\scriptsize, left]{#1}},
        },
        text width=2cm,
        text centered,
        edge={->,>=stealth}
      }
      [ \newstate{s1}%
      {dist(a,Y,D)}%
      {D<150}
      [ \newstate{s2i}%
      {D1\pr1\#>0, D2\pr1\#>0, D\pr1\#=D1\pr1+D2\pr1, edge(a,Z\pr1,D1\pr1), dist(Z\pr1,Y\pr1,D2\pr1)}%
      {D<150,Y\pr1=Y,D\pr1=D}
      [ \newstate{s3}%
      {edge(a,Z\pr1,D1\pr1), dist(Z\pr1,Y,D2\pr1)}%
      {D<150,D1\pr1>0,D2\pr1>0,D=D1\pr1+D2\pr1}
      [ \newstate{s4}%
      {dist(b,Y,D2\pr1)}%
      {D<150,D1\pr1>0,D2\pr1>0,D=D1\pr1+D2\pr1,Z\pr1=b,D1\pr1=50}
      [ {Ans({\bf s5})}, delay={text width=3cm, text centered, edge={-}} 
      [ \newstate{s11}%
      {\ }%
      {D<150,D2\pr1>0,D=50+D2\pr1,V0=a,D2\pr1>25,D2\pr1<35}, edge label=(b1)
      [ {\textbf{a2} $\thickmuskip=0mu \mt{V0=a\ \land}$\\ $\thickmuskip=0mu \mt{V1>75 \land V1<85}$}, delay={text width=2cm, text centered}  ]
      ]
      [ \newstate{s14}%
      {\ }%
      {D<150,D2\pr1>0,D=50+D2\pr1,V0=b,D2\pr1>75,D2\pr1<85}, edge label=(b2)
      [ {\textbf{a3} $\thickmuskip=0mu \mt{V0=b\ \land}$\\ $\thickmuskip=0mu \mt{V1>125 \land  V1<135}$}, delay={text width=2.2cm, text centered}  ]
      ]
      ]
      ]
      ]
      ]
      [ \newstate{s2ii}%
      {edge(a,V0,D)}%
      {D<150,Y\pr1=V0,D\pr1=D}
      [ \newstate{s10}%
      {\ }%
      {D<150,V0=b,D=50}
      [  {\textbf{a1} $\thickmuskip=0mu \mt{V0=b\ \land}$\\ $\thickmuskip=0mu \mt{V1=50}$}, delay={text width=2cm, text centered}  ]
      ]
      ]
      ]
    \end{forest}

  \end{minipage}
  \begin{minipage}{.45\textwidth}

    \begin{forest}
      for tree={
        font=\sffamily\scriptsize,
        delay={
          edge label/.wrap value={node[midway,font=\sffamily\scriptsize, left]{#1}},
        },
        edge={->,>=stealth}
      }
      [ \newstate{s5}%
      {dist(b,Y,D)}%
      {D>0, D<100}
      [ \newstate{s6i}%
      {D1\pr1\#>0, D2\pr1\#>0, D\pr1\#=D1\pr1+D2\pr1, edge(b,Z\pr1,D1\pr1), dist(Z\pr1,Y\pr1,D2\pr1)}%
      {D>0, D<100,Y\pr1=Y,D\pr1=D}
      [ \newstate{s7}%
      {edge(b,Z\pr1,D1\pr1), dist(Z\pr1,Y,D2\pr1)}%
      {D>0, D<100,D1\pr1>0,D2\pr1>0,D=D1\pr1+D2\pr1}
      [ \newstate{s8}%
      {dist(a,Y,D2\pr1)}%
      {D>0, D<100,D1\pr1>0,D2\pr1>0,D=D1\pr1+D2\pr1,Z\pr1=a,D1\pr1>25,D1\pr1<35}
      [ {Ans({\bf s1})}, delay={text width=2cm, text centered, edge={-}} 
      [ \newstate{s12}%
      {\ }%
      {D>0, D<100,D2\pr1>0,D>25+D2\pr1,D<35+D2\pr1,Y=b,D2=50}, edge label=(a1)
      [  {\textbf{b2} $\thickmuskip=0mu \mt{V2=b\ \land}$\\ $\thickmuskip=0mu \mt{V3>75 \land  V3<85}$}, delay={text width=2cm, text centered}  ]
      ]
      [ \newstate{s13}%
      {\ }%
      {D>0, D<100,D2\pr1>0,D>25+D2\pr1,D<35+D2\pr1,Y=a,D2>75,D2<85}, edge label=(a2)
      [  {\bf fail} ]
      ]
      [ \newstate{s15}%
      {\ }%
      {D>0, D<100,D2\pr1>0,D>25+D2\pr1,D<35+D2\pr1,Y=b,D2>125,D2<135}, edge label=(a3), delay={xshift=1cm}
      [  {\bf fail}, delay={xshift=1cm} ]
      ]
      ]
      ]]]
      [ \newstate{s6ii}%
      { edge(b,Y,D)}%
      { D>0, D<100, Y\pr1=Y, D\pr1=D }
      [ \newstate{s9}%
      {\ }%
      { D>0, D<100, Y=a, D>25, D<35 }
      [ {\textbf{b1} $\thickmuskip=0mu \mt{V2=a\ \land}$\\ $\thickmuskip=0mu \mt{V3>25 \land  V3<35}$}, delay={text width=3cm, text centered}  ]
      ]
      ]]
    \end{forest}

  \end{minipage}
}
 \\
    \\[.2cm]
    \cline{1-1}
    {\vspace{0.5em}\begin{forest}
  for tree={
    font=\sffamily\scriptsize,
    delay={
      edge label/.wrap value={node[midway,font=\sffamily\scriptsize, left]{#1}},
    },
    folder,
    grow'=0,
    fit=band,
    l sep'-=1pt,
    s sep+=-1pt,
    inner sep=1pt,
    text width=20cm,
    text height=.1cm,
    edge={->,>=stealth}
  }
  [ \newstate{s1}%
  {dist(a,V0,V1)}%
  {V1<150}
  [ \newstatelarge{s2i}%
  {D1\pr1\#>0, D2\pr1\#>0, V1\pr1\#=D1\pr1+D2\pr1, edge(a,Z\pr1,D1\pr1), dist(Z\pr1,Y\pr1,D2\pr1)}%
  {V1<150 \land Y\pr1=V0 \land D\pr1=V1}
  [ \newstate{s3}%
  {edge(a,Z\pr1,D1\pr1), dist(Z\pr1,V0,D2\pr1)}%
  {V1<150 \land D1\pr1>0 \land D2\pr1>0 \land V1=D1\pr1+D2\pr1}
  [ \newstate{s4}%
  {dist(b,V0,D2\pr1)}%
  {V1<150 \land D1\pr1>0 \land D2\pr1>0 \land V1=D1\pr1+D2\pr1 \land Z\pr1=b \land D1\pr1=50}, delay={l sep+=5pt,s sep+=7pt}
  [ {Ans(dist(b,V2,V3), $V3>0 \land  V3<100$)},%
  delay={%
    edge={-},%
    edge label={node[midway,font=\sffamily\scriptsize, right, yshift=10pt]%
      {\smaller with renaming $V0=V2\ \land\  D2\pr1=V3$}}}
  [ \newstate{s11}%
  {\,}%
  {V1<150 \land D2\pr1>0 \land V1=50+D2\pr1 \land V0=a \land D2\pr1>25 \land D2\pr1<35}, edge label=(b1)
  [ {\textbf{a2} $\thickmuskip=0mu \mt{V0=a \land  V1>75 \land  V1<85}$} ]
  ]
  [ \newstate{s14}%
  {\,}%
  {V1<150 \land D2\pr1>0 \land V1=50+D2\pr1 \land V0=b \land D2\pr1>75 \land D2\pr1<85}, edge label=(b2)
  [ {\textbf{a3} $\thickmuskip=0mu \mt{V0=b \land  V1>125 \land  V1<135}$} ]
  ]
  ]
  ]
  ]
  ]
  [ \newstate{s2ii}%
  {edge(a,V0,V1)}%
  {V1<150 \land Y\pr1=V0 \land D\pr1=V1}
  [ \newstate{s10}%
  {\,}%
  {V1<150 \land V0=b \land V1=50}
  [  {\textbf{a1} $\thickmuskip=0mu \mt{V0=b \land  V1=50}$} ]
  ]
  ]
  ]
\end{forest}
} \\
    {\vspace{0.5em}\begin{forest}
  for tree={
    font=\sffamily\scriptsize,
    delay={
      edge label/.wrap value={node[midway,font=\sffamily\scriptsize, left]{#1}},
    },
    folder,
    grow'=0,
    fit=band,
    l sep'-=1pt,
    s sep+=-1pt,
    inner sep=1pt,
    text width=20cm,
    text height=.1cm,
    edge={->,>=stealth}
  }
  [ \newstate{s5}%
  {dist(b,V2,V3)}%
  {V3>0 \land  V3<100}
  [ \newstatelarge{s6i}%
  {D1\pr1\#>0, D2\pr1\#>0, D\pr1\#=D1\pr1+D2\pr1, edge(b,Z\pr1,D1\pr1), dist(Z\pr1,Y\pr1,D2\pr1)}%
  {V3>0 \land  V3<100 \land Y\pr1=V2 \land D\pr1=V3}
  [ \newstate{s7}%
  {edge(b,Z\pr1,D1\pr1), dist(Z\pr1,V2,D2\pr1)}%
  {V3>0 \land  V3<100 \land D1\pr1>0 \land D2\pr1>0 \land V3=D1\pr1+D2\pr1}
  [ \newstate{s8}%
  {dist(a,V2,D2\pr1)}%
  {V3>0 \land  V3<100 \land D1\pr1>0 \land D2\pr1>0 \land V3=D1\pr1+D2\pr1 \land Z\pr1=a \land D1\pr1>25 \land D1\pr1<35}, delay={l sep+=5pt,s sep+=7pt}
  [ {Ans(dist(a,V0\pr1,V1\pr1), $V1\pr1<150$) \smaller%
    is entailed  because $V1\pr1>0 \land  V1\pr1<75$ $\sqsubseteq$ $V1\pr1<150$},%
  delay={%
    edge={-},%
    edge label={node[midway,font=\sffamily\scriptsize, right, yshift=10pt]%
      {\smaller with renaming $V2=V0\pr1\ \land\  D2\pr1=V1\pr1$}}}
  [ \newstate{s12}%
  {\,}%
  {V3>0 \land  V3<100 \land D2\pr1>0 \land V3>25+D2\pr1 \land V3<35+D2\pr1 \land V2=b \land D2\pr1=50}, edge label=(a1)
  [  {\textbf{b2} $\thickmuskip=0mu \mt{V2=b \land  V3>75 \land  V3<85}$} ]
  ]
  [ \newstate{s13}%
  {\,}%
  {V3>0 \land  V3<100 \land D2\pr1>0 \land V3>25+D2\pr1 \land V3<35+D2\pr1 \land V2=a \land D2\pr1>75 \land D2\pr1<85}, edge label=(a2)
  [  {\bf fail} ]
  ]
  [ \newstate{s15}%
  {\,}%
  {V3>0 \land  V3<100 \land D2\pr1>0 \land V3>25+D2\pr1 \land V3<35+D2\pr1 \land V2=b \land D2\pr1>125 \land D2\pr1<135}, edge label=(a3)
  [  {\bf fail} ]
  ]
  ]
  ]]]
  [ \newstate{s6ii}%
  { edge(b,V2,V3)}%
  { V3>0 \land  V3<100 \land  Y\pr1=V2 \land  D\pr1=V3 }
  [ \newstate{s9}%
  {\,}%
  { V3>0 \land  V3<100 \land  V2=a \land  V3>25 \land  V3<35 }
  [ {\textbf{b1} $\thickmuskip=0mu \mt{V2=a \land  V3>25 \land  V3<35}$} ]
  ]
  ]]
\end{forest}
} \\ \cline{1-1}
  \end{tabular}
  \caption{TCLP-forest of the query \texttt{?- D \#< 150,
      dist(a,Y,D)} \\ for \texttt{dist/3} with right recursion}
  \label{fig:tree04}
  \redafterfig
\end{figure}

Fig.~\ref{fig:tree04} shows the TCLP forest corresponding to querying
the right recursive \texttt{dist/3} program 
(Fig.~\ref{fig:dist-graph-right}, right).  This example is useful to
show how the algorithm works with mutually dependent
generators\footnote{I.e., generators which consume answers from each
  other.} and to see why not all the answers from a generator may be
directly used by its consumers.

Unlike the left-recursive version, which shows only one TCLP tree
(Fig.~\ref{fig:tree01-03}, bottom), Fig.~\ref{fig:tree04} has two TCLP
trees (one for each generator).
That is because the left recursive version only sought paths from the
node \texttt{a}, but the right recursive version creates a new TCLP
tree at the state \textsf{s4} to collect the paths from the node
\texttt{b}, since \texttt{edge(a,b)} had been previously evaluated at
state \textsf{s3}. As before we only explain
how we obtain some of the states: 

\begin{description}[leftmargin=!,labelwidth=\widthof{\bfseries Ans(xx)
  }, labelindent=0cm, align=right]
  
\item[s1] the TCLP tree
  $\bigsymbol{\tau}_P(\mt{dist(a,V0,V1),\, V1<150})$ is created.

\item[s4] is obtained by resolving the literal
  \texttt{edge(a,Z\pr1,D1\pr1)}.

\item[Ans(s5)] the tabled literal \texttt{dist(b,V0,D2\pr1)} is a new
  generator and a new TCLP tree
  $\bigsymbol{\tau}_P(\mt{dist(b,V2,V3),\, V3>0 \land V3<100})$ is created
  (Def.~\ref{def:call_entail}).

\item[s5] is the root node of the new TCLP tree.
  
\item[s6i/ii] are obtained by resolving the literal
  \texttt{dist(b,V2,V3)} against the clauses of the program.
  
\item[s8] is obtained by resolving the literal
  \texttt{edge(b,Z\pr1,D1\pr1)}.

In the state \textsf{s8}, the call
\goaltt{dist(a,V2,D2\pr1)}{D2\pr1>0 \land D2\pr1<75} is suspended
because it entails the former generator
\goaltt{dist(a,V0\pr1,V1\pr1)}{V1\pr1<150}.  
  
\item[Ans(s1)] the tabled literal \texttt{dist(a,V2,D2\pr1)} is
  resolved with answer resolution
  (Def.~\ref{def:call_entail}) using the
  answers from the previous TCLP tree
  $\bigsymbol{\tau}_P(\mt{dist(a,V0_1,V1_1), V1_1<150})$ because the
  renamed projection\footnote{The projection of
    $\mt{V3>0} \land \mt{V3<100} \land \mt{D1\pr1>0} \land \mt{D2\pr1>0} \land
    \mt{V3=D1\pr1+D2\pr1} \land \mt{Z\pr1=a} \land \mt{D1\pr1>25} \land \mt{D1\pr1<35}$
    onto \texttt{D2\pr1} is $\mt{D2\pr1>0} \land \mt{D2\pr1<75}$.  After
    renaming $\mt{D2\pr1=V1\pr1}$, the resulting projection is
    $\mt{V1\pr1>0} \land \mt{V1\pr1<75}$.} of the current constraint store onto
  the variable of the literal entails the projected constraint
  store of the generator:
  $(\mt{V1_1>0 \land V1_1<75}) \sqsubseteq \mt{V1_1<150}$. Since the initial
  TCLP forest is under construction and depends on itself, the current
  branch derivation is suspended.

  This suspension also causes the former generator to suspend at the
  state \textsf{s4}. 

\item[s9] is a final state obtained upon backtracking to the state
  \textsf{s6ii}.

\item[b1] is the first answer of the second generator.

  At this point the suspended calls can be resumed by consuming the
  answer \textsf{b1} or by evaluating \textsf{s2ii}. The algorithm
  first tries to evaluate \textsf{s2ii} and then it will resume
  \textsf{s4} consuming \textsf{b1}.

\item[s10] is a final state obtained upon backtracking to the state
  \textsf{s2ii}.
  
\item[a1] is the first answer of the first generator:
  $\mt{V0=b \land V1=50}$.
  
\item[s11] is a final state obtained from the state \textsf{s4} by
  consuming \textsf{b1}.
  
\item[a2] is the second answer of the first generator:
  $\mt{V0=a \land V1>75 \land V1<85}$.
  
\item[s12] is a final state obtained from the state \textsf{s8} by
  consuming \textsf{a1}.
  
\item[b2] is the second answer of the second generator.

\item[s13] is a failed derivation obtained from \textsf{s8} by consuming
  \textsf{a2}. It fails because the constraints
  $\mt{V0=a} \land \mt{V1>75} \land \mt{V1<85}$ are inconsistent with the current
  constraint store. Note that the projection of the constraint store
  of \textsf{s8} onto \texttt{V1} is $\mt{V1>0} \land \mt{V1<75}$. Its child is
  a \textsf{fail} node.
  
\item[s14] is a final state obtained from the state \textsf{s4} by
  consuming \textsf{b2}.
  
\item[a3] is the third answer of the first generator:
  $\mt{V0=b} \land \mt{V1>125} \land \mt{V1<135}$.

\item[s15] is a failed derivation obtained from \textsf{s8} by
  consuming \textsf{a3}. Its child is a \textsf{fail} node.

\end{description}

This example illustrates why left recursion reduces the execution time
and memory requirements when using tabling / TCLP: left recursion
will usually create fewer generators.  We have also seen that using
answers from a more general call, as in the answer resolution of state
\textsf{s8} (i.e., the constraint store of the consumer
$\mt{V1\pr1>0 \land V1\pr1<75}$ is more particular than the constraint
store of the generator $\mt{V1\pr1<150}$), makes it necessary to
filter the correct ones (i.e., answer resolution for \textsf{a2} and
\textsf{a3} failed). This is not required in variant tabling because
the answers from a generator are always valid for its consumers
(modulo variable renaming).

\section{Step by Step Execution of \texttt{\large dist/3} under TCLP(\Q)}
\label{sec:execution-dist3}

The trace below shows the step by step execution of the TCLP version
of the left recursive distance traversal program in
Fig.~\ref{fig:transformation}, left, with the query \mbox{\texttt{?- D
    \#< 150, dist(a,Y,D).}}  using the graph in
Fig.~\ref{fig:edges}. In this example we are using the TCLP(\Q)
interface (Section~\ref{sec:tclpq-int}). Each step is annotated with
the labels used in Fig.~\ref{fig:flow}. The execution starts with the
query \mbox{\texttt{?- D \#< 150, dist(a,Y,D)}}:

\begin{description}[leftmargin=!,labelwidth=\widthof{\bfseries 0
  }]
    
\item[ 0] the constraint \texttt{D\#<150} in the
  query is added to the current store (state \texttt{s0}). Then
  \goaltt{dist(a,Y,D)}{D<150} is called and the tabling engine takes
  the control of the execution calling
  \texttt{tabled\_call(dist\_aux(a,Y,D))}.

\item[ 1] \texttt{call\_lookup\_table/3} initializes and saves (after
  renaming) \texttt{dist\_aux(a,V0,V1)}, because it is the first
  occurrence, and returns \texttt{Vars=[D]} and \texttt{Gen=\$1},
  where \texttt{\$1} is the reference for this generator.

\item[ 2] \texttt{store\_projection([D],ProjStore)} returns
  \texttt{ProjStore=([V1],[V1\#<150])}.
  
\item[ 3] \texttt{member/2} fails because the list of projected
  constraint stores associated to \texttt{Gen=\$1} is
  empty.

\item[ 4] \texttt{save\_generator/3} saves
  \texttt{([V1],[V1\#<150])} in the list of projected constraint
  stores associated to \texttt{Gen=\$1} (state \texttt{s1}).
  
\item[ 7] \texttt{execute\_generator/2} evaluates the generator
  against the first clause of \texttt{dist\_aux/3} and adds the body
  of the clause to the resolvent of the state \texttt{s2i}. Then the
  constraints of the resolvent, \texttt{[D1\#>0, D2\#>0,
    D\#=D1+D2]}, are added to the constraint store (state \texttt{s3})
  and \goaltt{dist(a,Z,D1)}{D<150 \land D1>0 \land D2>0 \land D=D1+D2}
  is called.
  
\item[ 0] the tabling engine reenters the tabled execution with
  \texttt{tabled\_call(dist\_aux(a,Z,D1))}.
  
\item[ 1] \texttt{call\_lookup\_table(dist\_aux(a,Z,D1),Vars,Gen)}
  returns \texttt{Vars=[D1]} and \texttt{Gen=\$1}, the reference
  to the previous generator, \texttt{dist\_aux(a,V0,V1)}.
  
\item[ 2] \texttt{store\_projection([D1], \mbox{ProjStore})} returns
  \texttt{ProjStore=([V1], [V1\#>0, V1\#<150])}.  For
  clarification, the projection of the current constraint store
  $\mt{D<150} \land \mt{D1>0} \land \mt{D2>0} \land \mt{D=D1+D2}$ onto \texttt{D1} is
  $\mt{D1>0 \land D1<150}$.

\item[ 3] \texttt{member/2} retrieves the projected constraint store
  \texttt{ProjStore\_G=([V1], [V1\#<150])}.
  
\item[ 5] \texttt{call\_entail/2} succeeds
  because $(\mt{D<150} \land \mt{D1>0} \land \mt{D2>0} \land \mt{D = D1+D2})
  \sqsubseteq \mt{D1<150}$.

\item[ 6] \texttt{suspend\_consumer/1} suspends the current call
  \texttt{dist\_aux(a,Z,D1)} (state \textsf{s3}, waiting for the answer
  of the current TCLP tree, \texttt{Ans(s1)}). 
  
\item[ 7] The evaluation of the generator backtracks to evaluate the
  other clause (state \textsf{s2ii}). Now the current constraint store
  is \texttt{D\#<150} and the call \goaltt{edge(a,Y,D)}{D<150} unifies
  with \texttt{edge(a,b,50)} (state \textsf{s4}). The first answer is
  found and \texttt{new\_answer/0} is invoked to collect the answer.

\item[ 8] \texttt{answer\_lookup\_table/2} stores the
  Herbrand constraints\footnote{In solvers written in Prolog and  implemented using
    attributed variables, such as CLP(\Q) and CLP(\R), it is usual
    that variables lose their association with the constraints where
    they appeared when these variables become ground. As ground terms
    do not have attributes attached, $\mt{D=50}$ is
    handled as part of the Herbrand constraints.}  of the answer,
  $\mt{Y=b \land D=50}$, returning \texttt{Vars=[]} and
  \texttt{Ans=\$a1}, where \texttt{\$a1} is the reference for this
  answer.
  
\item[ 9] \texttt{store\_projection/2} returns
  \texttt{ProjStore=([],[])}.
  
\item[10] \texttt{member/2} fails because the list of projected
  constraint stores associated to \texttt{\$a1} is
  empty.
  
\item[11] \texttt{save\_answer/2} saves \texttt{([],[])} in the list
  of the answer constraints associated to \texttt{\$a1} (state
  \textsf{a1}). The first answer is collected.
  
\item[14] the tabling engine resumes the  goal suspended at state
  \textsf{s3} and \texttt{member/2} retrieves the Herbrand constraints
  $\mt{Y=b \land D=50}$ and the answer constraint \texttt{([],[])}.

\item[15] \texttt{apply\_answer/2} adds the answer to the current
  constraint store (state \textsf{s5}).

\item[ 7] the execution continues resolving \goaltt{edge(b,Y,D2)}{\dots \land
    D2<100} which unifies with the clause \texttt{edge(b,a,D2):-
    D2\#>25, D2\#<35} (state \textsf{s6}). The second answer is found.

\item[ 8] \texttt{answer\_lookup\_table/2} stores the Herbrand constraints
  of the answer, $\mt{V0=a}$, returning \texttt{Vars=[D]} and
  \texttt{Ans=\$a2}.
  
\item[ 9] \texttt{store\_projection/2} returns
  \texttt{ProjStore=([V1], [V1\#>75,V1\#<85])}.
  
\item[10] \texttt{member/2} fails because the list of projected
  constraint stores associated to \texttt{Ans=\$a2} is empty.
  
\item[11] \texttt{save\_answer/2} saves
  \texttt{([V1],[V1\#>75,V1\#<85])} in the list of answer constraints
  associated to \texttt{\$a2} (state \textsf{a2}). The second answer
  is collected.

\item[14, 15, 7] the tabling engine resumes the
  suspended goal at state \textsf{s3} and consumes the second answer
  following the same steps as with the first one and generating the
  states \textsf{s7} and \textsf{s8}. The third answer has been found.
  
\item[8, 9, 10, 11] the answer is collected and
  \texttt{([V1],[V1\#>125,V1\#<135])} is saved in the list of
  answer constraint associated to \texttt{\$a3} (state \textsf{a3}).

\item[14, 15] the tabling engine resumes the suspended goal at state
  \textsf{s3} and consumes the third answer.
 
\item[ 7] the execution fails resolving
  \goaltt{edge(b,V0,D2\pr1)}{V1<150 \land \dots \land D1\pr1<135}
  (states \textsf{s9} and \textsf{s10})
  
\item[14, 15] the generator has exhausted all the answers and it does not
  have any more dependencies, so \texttt{complete/0} marks the
  generator as complete.
  The query retrieves the answers from the generator one by one and
  returns them.

\end{description}

\section{Comparison of \mtclp\ using \R, \Q and \D}
\label{sec:tclpd-tclpq-tclpr}

This section highlights that the modularity of TCLP makes it possible
to choose the most adequate constraint solver for the specific problem, and
that decision should not always be based solely on the performance of
the constraint solver, but also on its expressiveness and / or
precision.  Since TCLP, unlike CLP, uses entailment checking  extensively to decide
whether to suspend and save / discard answers or not, the performance of
entailment is more relevant than in CLP.  It also makes its soundness
(which can be challenged by e.g.\ numerical accuracy) critical, as
incorrect entailment results can lead to non-termination or to
unexpected termination.

\begin{figure}
  \begin{center}
    \begin{minipage}{.37\linewidth}
      \redbeflst
\begin{lstlisting}[style=MyProlog]
fib(N,F) :-
    $\vdots$
    F #= F1 + F2,
    fib(N1, F1),
    fib(N2, F2). 
\end{lstlisting}
      \redafterlst
    \end{minipage}
    \myvrule{.1\linewidth}{5em}
    \begin{minipage}{.37\linewidth}
      \redbeflst
\begin{lstlisting}[style=MyProlog]
fib(N,F) :-
    $\vdots$
    fib(N1, F1),
    fib(N2, F2),
    F #= F1 + F2. 
\end{lstlisting}
      \redafterlst
    \end{minipage}
    \redbefcapt
    \caption{Two versions of \texttt{fib/2}: \Q and \R
      (left) vs.\ \D (right).}
    \label{fig:fibo-program}
    \redafterfig
  \end{center}
\end{figure}

We use the doubly recursive Fibonacci program
(Fig.~\ref{fig:fibo-program}).  It is well-known that tabling reduces
its complexity from exponential to linear, but, in addition, CLP makes
it possible to run exactly the same program \texttt{fib/2}
\emph{backwards} to find the index of some Fibonacci number 
by generating a system of equations whose solution is the index of the
given Fibonacci number.  Under CLP, the size of this system of
equations grows exponentially with the index of the Fibonacci number.
However,
under TCLP, entailment makes redundant equations not to be added and
solving them becomes less expensive.

We have run this benchmark using \R, \Q and \D.
Due to the characteristics of \D
(Section~\ref{sec:design-diff-constr}), the program for this
constraint system is
slightly different from the ones for \Q and \R (Section~\ref{sec:tclpq-int}).  In these two,
constraints are placed before the recursive calls.  However, \D can
have at most two variables per constraint, and therefore we had to
move the constraint \mbox{\texttt{F \#= F1 + F2}} to the end of the
clause (Fig.~\ref{fig:fibo-program}).  This can be detrimental to the
performance of \mtclp(\D), as value propagation in the constraints is
less effective.

\begin{table}[tb]
  \centering
  \setlength{\tabcolsep}{0.75em}
  \begin{tabular}{@{\extracolsep{0pt}}|l|r|r|r|}
    \cline{1-4}
    \multicolumn{1}{|c|}{}
                             & \mtclp(\R)     & \mtclp(\Q)  & \mtclp(\D)     \\[0.25em]  \cline{1-4}
    \texttt{fib(N, 832040)}  & \textbf{25}           & 61        & 147         \\ \cline{1-4}
    \texttt{fib(N, 28657)}   & \textbf{16}           & 40        & 69          \\ \cline{1-4}
    \texttt{fib(N, 610)}     &  \textbf{8}           & 19        & 24          \\ \cline{1-4}
    \texttt{fib(N, 89)}      &  \textbf{5}           & 12        & 13          \\ \cline{1-4}
  \end{tabular}
  \redbefcapt
  \caption{Run time (ms) comparison for the \texttt{fib/2} using \R, \Q and \D.} 
  \label{tab:fibo}
    \redafterfig
\end{table}

Table~\ref{tab:fibo} shows the experimental results. First, note that
the \mtclp(\D) version is slower than any of the other two.  While the
implementation of CLP(\D) is comparatively faster than CLP(\R) and
CLP(\Q), moving the \texttt{F \#= F1 + F2} to the
end of the clause (which is necessary to satisfy the instantiation
requirements of \D) reduces its usefulness to prune the generation of
redundant constraints.

Additionally, note that although the solvers for \R and \Q are
practically the same, \mtclp(\R) is fastest in all cases, since it
uses directly CPU floating point numbers while CLP(\Q) implements
rational numbers by software.  However, there is a drawback: floating
point arithmetic is not accurate, and when CLP(\R) approximates its
results, it can cause (depending on the particular program)
non-termination.  That would be the case for a query such as
\mbox{\texttt{?- fib(N, 23416728348467685)}}, which  terminates
correctly with \mtclp(\Q), but it does not (in under five minutes)
with \mtclp(\R), since the termination condition never holds.